\documentclass[twocolumn]{aastex631}

\newcommand{\etaM}{\eta_{\rm M}}

\newcommand{\etaZ}{\eta_{\rm Z}}
\newcommand{\tdep}{t_{\rm dep}}
\newcommand{\adam}{\texttt{adam}}
\newcommand{\smhm}{M_*/M_h}
\newcommand{\fgas}{f_{\rm ISM}}
\newcommand{\thetamap}{\theta_{\rm MAP}}
\newcommand{\sapphire}{\texttt{sapphire}}
\newcommand{\Tsit}{\texttt{Tsit5}}
\newcommand{\Bosh}{\texttt{Bosh3}}
\newcommand{\jax}{\texttt{JAX}}

\usepackage{amsmath}

\shorttitle{\texttt{sapphire}: Towards Hybrid Physics-Informed, Data-Driven Modeling of Galaxy Formation}
\shortauthors{Viraj Pandya et al.}

\begin{document}

\title{Introducing \texttt{sapphire}: \\ Towards Hybrid Physics-Informed, Data-Driven Modeling of Galaxy Formation}

\correspondingauthor{Viraj Pandya}
\email{vgp2108@columbia.edu}


\author[0000-0002-2499-9205]{Viraj Pandya}
\altaffiliation{Hubble Fellow}
\affiliation{Columbia Astrophysics Laboratory, Columbia University, 550 West 120th Street, New York, NY 10027, USA}

\author[0000-0003-2630-9228]{Greg L. Bryan}
\affiliation{Department of Astronomy, Columbia University, 550 West 120th Street, New York, NY 10027, USA}
\affiliation{Center for Computational Astrophysics, Flatiron Institute, 162 5th Ave, New York, NY 10010, USA}

\author[0000-0002-3795-6933]{T. Lucas Makinen}
\affiliation{Department of Applied Mathematics and Theoretical Physics, University of Cambridge, Wilberforce Rd, Cambridge, CB3 0WA, United Kingdom}
\affiliation{Imperial Centre for Inference and Cosmology (ICIC) \& Astrophysics Group, Imperial College London, Blackett Laboratory, Prince Consort Road, London SW7 2AZ, United Kingdom}

\author[0000-0003-4295-3793]{Austen Gabrielpillai}
\affiliation{Department of Astronomy, Columbia University, 550 West 120th Street, New York, NY 10027, USA}

\author[0000-0002-5840-0424]{Christopher Carr}
\affiliation{Department of Astrophysical Sciences, Princeton University, Princeton, NJ 08544, USA}
\affiliation{Department of Astronomy, Columbia University, 550 West 120th Street, New York, NY 10027, USA}

\author[0000-0003-3806-8548]{Drummond B. Fielding}
\affiliation{Department of Physics, New York University, 726 Broadway, New York, NY 10003, USA}

\author[0000-0001-6950-1629]{Lars Hernquist}
\affiliation{Center for Astrophysics | Harvard \& Smithsonian, 60 Garden Street, Cambridge, MA 02138, USA}

\author[0000-0003-3207-8868]{Matthew Ho}
\affiliation{Columbia Astrophysics Laboratory, Columbia University, 550 West 120th Street, New York, NY 10027, USA}

\author[0000-0001-9298-3523]{Kartheik Iyer}
\altaffiliation{Hubble Fellow}
\affiliation{Columbia Astrophysics Laboratory, Columbia University, 550 West 120th Street, New York, NY 10027, USA}
\affiliation{Center for Computational Astrophysics, Flatiron Institute, 162 5th Ave, New York, NY 10010, USA}

\author[0000-0002-8896-6496]{Christian Kragh Jespersen}
\affiliation{Department of Astrophysical Sciences, Princeton University, Princeton, NJ 08544, USA}

\author[0000-0002-1528-5091]{Sophie Koudmani}
\affiliation{Centre for Astrophysics Research, Department of Physics, Astronomy and Mathematics, University of Hertfordshire, College Lane, Hatfield AL10 9AB, UK}
\affiliation{Kavli Institute for Cosmology, Cambridge, University of Cambridge, Madingley Road, Cambridge, CB3 0HA, UK}

\author{Marta Laska}
\affiliation{Department of Astronomy \& Astrophysics, The Pennsylvania State University, University Park, PA 16802, USA}

\author[0000-0002-4728-8473]{Pablo Lemos}
\affiliation{Department of Physics, Universit\'e de Montr\'eal, Montr\'eal, Canada}
\affiliation{Mila - Quebec AI Institute, Montr\'eal, Canada}
\affiliation{Ciela, Montreal Institute for Astrophysics and Machine Learning, Montr\'eal, Canada}

\author[0000-0001-7964-5933]{Christopher C. Lovell}
\affiliation{Kavli Institute for Cosmology, Cambridge, University of Cambridge, Madingley Road, Cambridge, CB3 0HA, UK}
\affiliation{Institute of Astronomy, Madingley Road, Cambridge CB3 0HA, UK}

\author[0000-0002-8449-1956]{Lucia A. Perez}
\affiliation{Center for Computational Astrophysics, Flatiron Institute, 162 5th Ave, New York, NY 10010, USA}
\affiliation{Department of Astrophysical Sciences, Princeton University, Princeton, NJ 08544, USA}

\author{William F. Robinson Jr.}
\affiliation{Department of Astronomy, Columbia University, 550 West 120th Street, New York, NY 10027, USA}

\author[0000-0002-6748-6821]{Rachel S. Somerville}
\affiliation{Center for Computational Astrophysics, Flatiron Institute, 162 5th Ave, New York, NY 10010, USA}

\author[0000-0003-2539-8206]{Tjitske K. Starkenburg}
\affiliation{Center for Interdisciplinary Exploration and Research in Astrophysics (CIERA), Northwestern University, 1800 Sherman Ave, Evanston, IL 60201, USA}
\affiliation{Department of Physics and Astronomy, Northwestern University, 2145 Sheridan Rd, Evanston IL 60208, USA}
\affiliation{NSF-Simons AI Institute for the Sky (SkAI), 172 E. Chestnut St., Chicago, IL 60611, USA}

\author[0000-0002-0986-314X]{Richard Stiskalek}
\affiliation{Astrophysics, University of Oxford, Denys Wilkinson Building, Keble Road, Oxford, OX1 3RH, UK}

\author[0000-0001-5529-7305]{Bryan Terrazas}
\affiliation{Department of Physics \& Astronomy, Oberlin College, Oberlin, OH, 44074, USA}

\author[0000-0002-3514-0383]{G. Mark Voit}
\affiliation{Michigan State University, Department of Physics and Astronomy, East Lansing, MI 48824, USA}

\begin{abstract}
Semi-analytic models (SAMs) have been treating galaxy populations as dynamical systems for $\gtrsim50$ years, but their evolution equations remain poorly constrained. We introduce \sapphire, a modular, automatically differentiable, GPU-accelerated SAM written in \texttt{JAX}. For the first time, we compute exact Jacobian and Hessian matrices of a galaxy formation SAM, using the \citet{pandya23} nonlinear differential equation system as an example. These allow efficient, interpretable local and global sensitivity analyses, which reveal that supernova energy loading is the key astrophysical parameter. We use gradient descent and Hamiltonian Monte Carlo (HMC) to perform comprehensive mock parameter recovery tests. These indicate that the $z=0$ stellar-to-halo-mass relation alone does not contain enough information to infer many astrophysical parameters. Using observations of star-forming galaxies from the MaNGA survey and the \citet{behroozi19} empirical model as one baseline, we derive multiple posteriors assuming different combinations of data, including $z=0$ interstellar medium gas fractions and metallicities. The inferred physical parameters suggest that galaxies self-regulate their star formation primarily through preventative rather than ejective feedback, though this remains uncertain due to the lack of satellite galaxies, black holes and multi-phase galactic atmosphere physics. Both Fisher and HMC forecasts demonstrate the potential of \texttt{sapphire} to enable precision inference for galaxy formation and cosmology in a hybrid physics-informed, data-driven way, but more work is needed to expand its library of models and methods. We make \texttt{sapphire} publicly available at \url{https://github.com/virajpandya/sapphire}.
\end{abstract}


\section{Introduction}\label{sec:intro}

\subsection{The Cosmological Context}
Galaxies represent one of our most promising cosmological probes of fundamental physics. Their abundance, dynamics and clustering contribute a variety of evidence for dark matter, dark energy and inflation. Some examples include luminosity functions \citep{sandage88,binggeli88}, rotation curves \citep{rubin80,bosma81}, correlation functions \citep{peebles80,davis83,geller89}, absorption line studies \citep{gunn65}, lensing \citep{kaiser93}, satellite merger rates \citep{bond91,hopkins10}, intrinsic alignments \citep{kiessling15,libeskind18,pandya19,pandya25}, and maybe even morphology \citep{ostriker73,bullock17,pandya24}. These helped establish the modern ``concordance'' $\Lambda$CDM paradigm in which galaxies grow hierarchically through accretion and mergers dictated by the dark matter halos within which they are assumed to live \citep[e.g.,][]{blumenthal84}. One of the grand challenges of modern astrophysics is to develop a fully predictive theory of galaxy formation so that galaxies can become more robust cosmological probes. However, this requires overcoming five obstacles: (1) the relevant physical processes span a large dynamic range, (2) they are non-linearly coupled across scales leading to complicated emergent behavior, (3) many of those processes are not understood from first principles, (4) we cannot directly observe the evolution of individual galaxies on human timescales so must resort to statistical, population-level studies, and (5) we have noisy, incomplete data. Together, these pose a formidable challenge for the ambitious goal of connecting galaxy formation astrophysics to the fundamental physics of cosmology. 

Fortunately, the past century of observations has revealed that galaxies, despite their individual complexity, exhibit remarkable statistical regularity. For example, they appear to follow morphological ``sequences'' \citep[e.g.,][]{hubble1926}, and both local \citep[e.g.,][]{mannucci10} and global \citep[e.g.,][]{faber76,tully77} scaling relations. More generally, galaxies occupy relatively thin manifolds in the space of their observable features, which implies that physical laws drive them towards equilibrium attractor solutions. In addition, the observed clustering of galaxies agrees remarkably well with simulations of the $\Lambda$CDM cosmology in which early density fluctuations grew to become the seeds responsible for galaxy and large-scale structure formation \citep{davis85,primack12,frenk12}. Anisotropies in the cosmic microwave background (CMB) have enabled precision measurements of $\Lambda$CDM model parameters, including the relative densities of baryons, dark matter and dark energy \citep{white94,hu02,planck15}. However, the growing tensions between CMB and late-universe probes of the cosmic expansion rate (i.e., the Hubble constant $H_0$) as well as the matter density fluctuation amplitude ($\sigma_8$) may require modifications to $\Lambda$CDM such as early dark energy \citep{kamionkowski23,primack24}.

Upcoming facilities (e.g., the Nancy Grace Roman Space Telescope, Rubin Observatory, Euclid) will map tens of billions of galaxies, providing unprecedented opportunities to pursue precision cosmology. New simulation efforts are underway that aim to reconstruct our initial density field from this ``late-time'' galaxy clustering data \citep[e.g.,][]{gottloeber10,wang14,jasche19,modi21,li24,hahn24,lemos24,mcalpine25}. These have the potential to provide powerful, complementary constraints on the nature of dark matter, dark energy and inflation using galaxies as cosmological tracers. However, it is becoming increasingly clear that astrophysical uncertainties will hamper our interpretation of this forthcoming data. For example, the large-scale matter power spectrum is sensitive to the choice of small-scale baryonic physics \citep{villaescusanavarro21,ni23,delgado23,gebhardt24,schaller25}, and so far it has been challenging to robustly infer cosmology across different astrophysical models \citep{perez23,jo23,jo25}. Thus, we argue that precision cosmology with galaxies requires precision astrophysics: we need to elucidate the key astrophysical processes, identify the most sensitive observables, and forecast the uncertainties required to achieve some target precision on model parameters so that we actually believe consistency with or departures from $\Lambda$CDM in terms of galaxy properties. Fortunately, by combining these same galaxy surveys, originally intended for cosmology, with forthcoming data on the thermodynamic state of gas within and around galaxies, we can improve our understanding of the foundational astrophysics for its own sake \citep[helping to achieve the ``cosmic ecosystems'' science case of the 2020 Decadal Survey;][]{decadal23}.

\subsection{The State of Galaxy Formation}\label{sec:intro2}
Since it is not possible to perform \textit{ab initio} predictions for galaxy populations, a wide variety of approaches have been developed to model galaxy formation in a cosmological context \citep[e.g., see recent reviews by ][]{somervilledave15,naabostriker17,wechsler18}. On the one hand, ``empirical'' models are now able to fit much of the available photometric galaxy survey data by assuming relatively simple mappings between simulated dark matter halo properties and observable galaxy properties, which often happens to reproduce the observed clustering of different types of galaxies \citep{vale06,conroy06,moster10,behroozi13c,rodriguezpuebla17,behroozi19,zhang23,shuntov25}. These empirical models constrain the overall efficiency of converting baryons into stars as a function of halo mass and redshift, but remain ambiguous about the underlying astrophysics which makes it difficult to interpret their results. Furthermore, for the purposes of inferring cosmological parameters from galaxy clustering data, many empirical approaches only forward model a limited number of galaxy observables and discard information on small scales, which, for example, fails to capture halo and galaxy assembly bias \citep[][]{zentner14,hearin16,hadzhiyska20}. Like many other approaches, these empirical models are also fit assuming a single cosmological model, so great care must be taken when using them to constrain and interpret cosmological parameters.

On the other hand, modern hydrodynamical simulations can now make detailed predictions for galaxy properties down to sub-pc scales, allowing unprecedented theoretical investigations \citep[e.g., see recent reviews by][]{vogelsberger20,crain23,feldmann25}. However, galaxy-scale simulations on their own lack two essential qualities: (1) interpretability, and (2) causal identifiability. The first is due to their complexity: simple abstractions are required to extract insights from these highly nonlinear numerical experiments. The second refers to mapping noisy, observable outputs back to uncertain inputs (and vice versa), a task that requires prohibitively expensive computational resources and analysis techniques. Both \citet{pandya21} and \citet{pandya23}, for example, stressed that it would be impossible to disentangle cause-and-effect relationships from emergent galaxy properties in ``single-shot'' simulations without the ability to do parameter variations and equilibrium analysis. This is slowly changing thanks to expensive simulation suites like AGORA \citep{kim14}, CAMELS \citep{villaescusanavarro21} and FLAMINGO \citep{schaye23,kugel23} as well as the recent proliferation of machine learning techniques, which enable the training of emulators for Bayesian inference with neural networks \citep[e.g.,][]{jo23,ho24,alsing24,iyer25,lovell25}. However, as recently reviewed by \citet{primack24}, no existing simulation simultaneously reproduces all observed properties of the evolving galaxy population. Coming up with, or even choosing between, different solutions to these existing tensions is not always straightforward given the multi-scale complexity of galaxy formation. The field therefore requires a new hybrid approach that blends the strengths of both empirical and physical models. 

If we want to understand galaxy populations as dynamical systems, we must first identify their key state variables and governing equations. Similar to how hydrodynamical simulations co-evolve billions of fluid elements, semi-analytic models (SAMs) reduce galaxy evolution to the simpler problem of tracking the phase space evolution of galactic-scale summary statistics using continuity equations. It is an open question as to how many independent state variables are needed to sufficiently characterize galaxies and their cosmic ecosystems. The term ``semi-analytic'' is historical and reflects the fact that the dark matter halo formation process is so nonlinear that it must be simulated and summarized as merger trees, which then act as the backbone for ordinary differential equations (ODEs) that approximate various baryonic processes. Early works conceived SAM-like approaches to test the basic viability of different cosmological paradigms using simple physical arguments to model the competition between gas accretion, cooling, star formation, feedback and chemical evolution \citep[e.g.,][]{tinsley68,rees77,whiterees78,ostriker81,dekel86,whitefrenk91,kauffmann93,lacey93,somerville99,cole00}. In the decades since, SAMs have exploded in complexity with the inclusion of black hole growth and feedback \citep{kauffmann00,croton06,somerville08}, multi-element chemical evolution \citep{arrigoni10,yates13}, multi-phase gas partitioning \citep{lagos11,somerville15}, morphological evolution \citep{dutton07,forbes14,porter14,krumholz18,stevens24}, cosmic reionization \citep{mutch16}, and thermodynamic evolution of gaseous galactic atmospheres \citep{sharma20,pandya23}. Minimalist variants have emerged that restrict the number of state variables and impose equilibrium conditions to help extract key insights about the overall ``baryon cycling'' process \citep[these are sometimes called ``bathtub'' or ``gas regulator'' models;][]{bouche10,dave12,lilly13,dekel14,mitra15,rodriguezpuebla16,belfiore19,tacchella20,carr23,voit24a,voit24b,iyer24,burnham26}.

\subsection{A New Physics-Informed, Data-Driven Approach}
To make further progress in understanding galaxy evolution, we need to unify SAMs and hydrodynamical simulations into an interpretable Bayesian framework. This has long been a goal of the galaxy formation community \citep{benson01,yoshida02,helly03,stringer10,lu11,neistein12,hirschmann12,pandya20,mitchell22,gabrielpillai22,omoroyi26}. Previous authors have already emphasized that this will require a new standard of numerical robustness, modular code development and principled Bayesian methods  \citep{henriques09,bower10,lu11b,benson12,croton16,lagos18,forbes19,pandya21thesis}. Recently, \citet{pandya23} took a step towards this by benchmarking and improving standard SAM recipes to roughly emulate the FIRE-2 simulations, as one example testbed \citep[][and references therein]{hopkins18}. In particular, \citet{pandya23} introduced a two-zone model that tracked flows of mass, metals and energy between the interstellar medium (ISM), circumgalactic medium (CGM) and intergalactic medium (IGM). Their simple model approximately reproduced the time evolution of stellar, ISM and CGM masses and metallicities, as well as thermal and turbulent CGM pressure support, of individual galaxies in the core FIRE-2 simulation suite. They achieved this by measuring many of their free parameters in the simulations and then using the resulting fits in their SAM, noting that FIRE-2 provided only one of many possible ``priors'' for their model. \citet{carr23} and \citet{voit24a,voit24b} showed that a similar model without any calibration to FIRE-2 could also naturally explain empirical constraints on the inefficiency of star formation in Milky Way (MW) and lower mass galaxies.

Here we introduce \texttt{sapphire}, an automatically differentiable, multi-GPU-parallelized framework for modeling the dynamical phase space evolution of galaxy populations.\footnote{We make the code publicly available at \url{https://github.com/virajpandya/sapphire} and encourage community development.} We have written \sapphire\ in the lightweight AI/ML Python package \texttt{JAX} \citep{bradbury18} so that we can rapidly compute exact gradients of our ODE solutions with respect to free parameters using automatic differentiation \citep[see][for a recent review]{baydin18}. This can help address the lack of both interpretability and causal identifiability in existing approaches that we introduced earlier, and is only possible now because we are drawing on recent developments in computer science made available by open-source differentiable ODE solver packages like \texttt{diffrax} \citep[][]{kidger22}. We are following in the footsteps of \citet{hearin21,hearin23} who pioneered differentiable, probabilistic modeling of galaxy populations, except here we are numerically solving and automatically differentiating through the internals of adaptive ODE solvers for highly nonlinear systems without assuming parameterized, equilibrium solutions. Our approach is complementary to other differentiable galaxy population modeling techniques including halo occupation distributions \citep[\texttt{DiffHOD};][]{horowitz24} and hierarchical diffusion models \citep[e.g., \texttt{pop-cosmos};][]{alsing24,thorp25,deger26}. \texttt{sapphire} is designed to generalize far beyond any one dynamical model so that we can ultimately perform hybrid physics-informed, data-driven differential equation discovery using inputs from many simulations and datasets. 

This is the latest paper in a series building on \citet{pandya20,pandya21,pandya23} as well as the broader vision for multi-scale Bayesian SAMs outlined by \citet{pandya21thesis}, which grew out of the Simulating Multi-scale Astrophysics to Understand Galaxies (SMAUG) Collaboration, now part of the broader Simons Collaboration on Learning the Universe (LtU).\footnote{\url{https://learning-the-universe.org/}} Our work complements ongoing efforts to generate ever larger, higher resolution simulations with coarse parameter variations by taking a fundamentally new approach of making the entire code differentiable, inspired by the successes of BORG \citep[][see also ELUCID by \citealt{wang16} and TARDIS by \citealt{horowitz19}]{jasche19,mcalpine25}. This unlocks previously inaccessible techniques for understanding the richness of even simple dynamical systems like the ODE-based physical model of \citet{pandya23}. For example, we will use the gradients to interpret the nonlinear dynamics of our model, perform local and global parameter sensitivity analysis (i.e., understanding cause-and-effect with parameter variations), derive Fisher uncertainties and fully Bayesian posteriors, and forecast the impact of improved observational uncertainties, all of which are standard in cosmology but not galaxy formation. We will deliberately focus on a very simple initial application involving fitting three $z=0$ scaling relations where the data is reasonably well understood and complete. This extends seminal work by \citet{henriques09,bower10,henriques12,henriques13,henriques15,lu11b,forbes19} who demonstrated the potential of SAMs for Bayesian inference, as well as \citet{gomez14}, \citet{oleskiewicz20} and \citet{bose25} who performed the only other sensitivity analyses of SAM-like models that we can find.

This paper is organized as follows. In Section~\ref{sec:data}, we describe galaxy scaling relation data used for inference. In Section~\ref{sec:model}, we introduce the model and in Section~\ref{sec:methods}, we outline our methods. In Section~\ref{sec:mocks} we present sensitivity analyses and mock parameter recovery tests. In Section~\ref{sec:fitdata}, we infer model parameters by combining $z=0$ galaxy scaling relations. Finally, we discuss our results in Section~\ref{sec:discussion} and conclude in Section~\ref{sec:summary}. Throughout this paper we assume a standard \citet{planck15} cosmology with $\Omega_M=0.3075$, $\Omega_b=0.0486$ and $h=0.6774$.

\section{Observational Data}\label{sec:data}
Here we describe the observational data we use for inference. Our model does not yet include black holes or satellite galaxies, so we cannot predict total galaxy number densities or realistic group/cluster populations. However, even with this baseline model, we can already begin to probe the information content of galaxy scaling relations for constraining a subset of model parameters \citep[as was done by][]{carr23,voit24b}. For simplicity, we focus on three $z=0$ galaxy scaling relations involving ``quasi-observables'' inferred from data. We defer forward modeling of galaxy spectrophotometric observables to future work, but note that this will be necessary to disentangle astrophysics and cosmology from observational selection effects. 

First, we use the stellar-mass-halo-mass (SMHM) relation restricted to only star-forming central galaxies at $z=0$ as tabulated by \citet{behroozi19}. Our model can help provide astrophysical interpretations for the normalization and shape of this empirically-determined relation. We limit our analysis to eleven bins ranging from $\log M_{\rm vir}/M_{\odot}\sim10-12$, with the lower limit set by the resolution of their DM simulations and the upper limit chosen to avoid the contribution of black hole feedback. For each bin, they report the median $\log M_*/M_{\rm vir}$ and the $16-84$ percentile uncertainties. Appendix \ref{sec:manga} details our efforts to assign halo mass posteriors to individual MaNGA galaxies and the biased relation that results, motivating us to simply use the inferred relation from \citet{behroozi19} as-is.

We will combine the $z=0$ SMHM relation with $z=0$ ISM gas fractions and the ISM mass-metallicity relation. For this, we analyze the Mapping Nearby Galaxies at APO \citep[MaNGA;][]{bundy15} survey, which provides properties for $\sim10,000$ galaxies at $z\sim0$ that we can compute summary statistics from in the same way as we do for the model. We restrict to only star-forming galaxies and require cross-matches to both the Pipe3D stellar population catalog \citep{sanchez22} as well as Data Release 3 of HI-MaNGA \citep{masters19,stark21}, leaving 1787 galaxies in our final sample. More details are given in Appendix \ref{sec:manga}. 

Finally, to demonstrate the future generality of our framework, we will do posterior predictive checks for a few quantities that we will not directly infer in this first proof-of-concept paper. This includes all three scaling relations at one example higher redshift ($z=2$). For the SMHM relation, we again use the tabulated values for SF centrals from \citet{behroozi19} as well as new JWST-based results from \citet{shuntov25}. For the ISM mass-metallicity relation (MZR) at $z=2$, we will show both pre-JWST \citep{sanders21} and post-JWST \citep{he24,khostovan25} results, which do not always agree with each other making our approach uniquely powerful. We also predict the star-forming main sequence at $z=0$ \citep[using MaNGA SFRs from][]{sanchez22} and $z=2$ \citep{whitaker14,speagle14,pandya17,clarke24}.

\section{\texttt{sapphire} Framework Description}\label{sec:model}
\begin{figure*}
\centering
\includegraphics[width=\hsize]{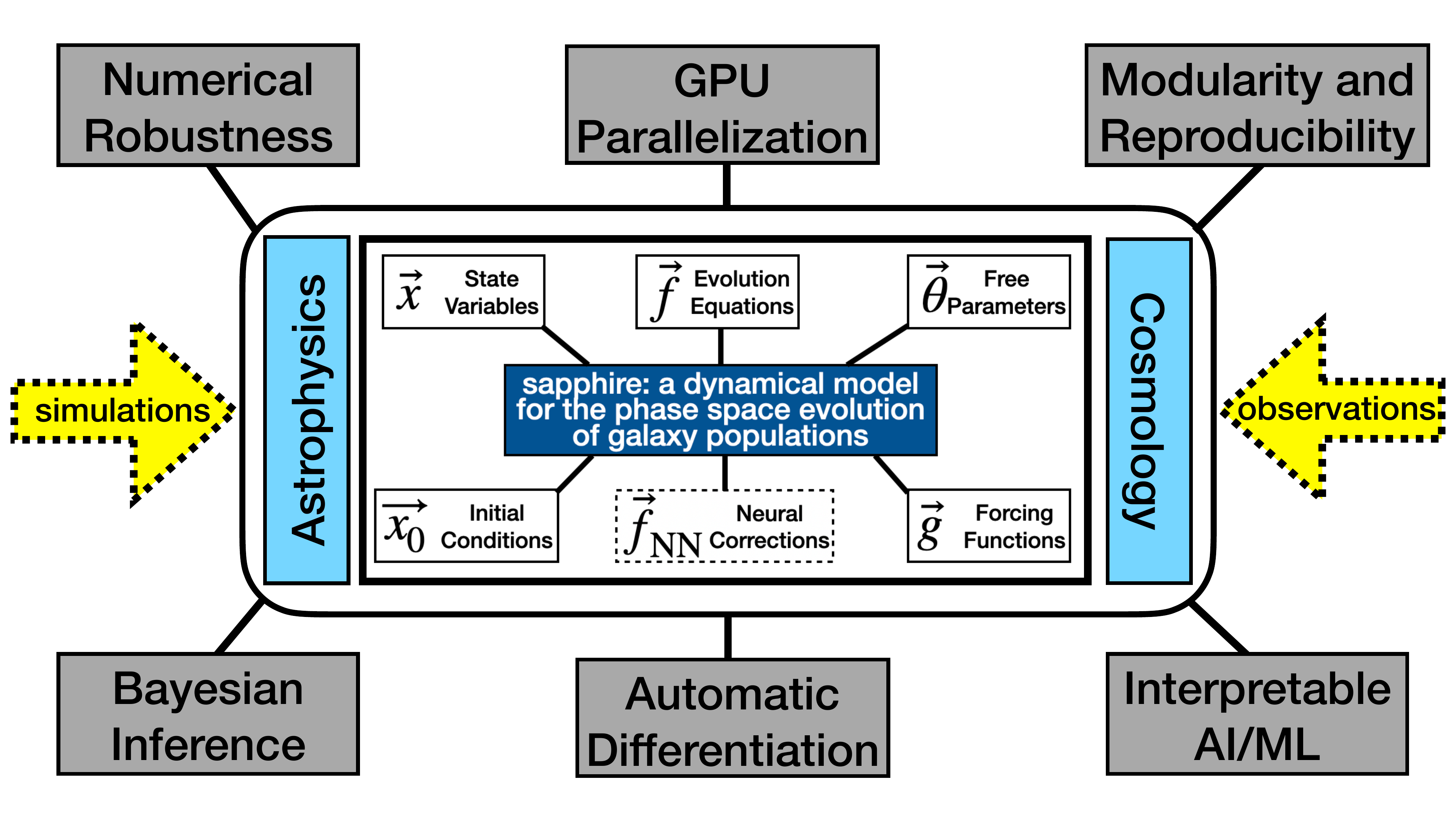}
\caption{Schematic overview of \texttt{sapphire}. The innermost boxes with solid black borders list the five universal ingredients of any dynamical system: state variables $\vec{x}$, evolution equations $\vec{f}$, free parameters $\vec{\theta}$, initial conditions $\vec{x}_0$ and forcing functions $\vec{g}$. We built \texttt{sapphire} to be differentiable so that we can add small, interpretable neural networks $\vec{f}_{\rm NN}$ to correct model mis-specification (dotted black border). The values and functional forms of these components are dictated by multiple sub-domains spanning both astrophysics and cosmology (blue boxes). The reader is directed to Figure 1 of \citet{pandya23} for a visualization of our baseline physical model. Our goal is to discover the unknown, nonlinear, time-dependent structure of this dynamical system using hybrid physics-informed, data-driven techniques, which requires input from both observations and hydrodynamical simulations (yellow arrows). Achieving this requires bridging several foundational interdisciplinary areas including robust numerical methods, GPU parallelization, modular and reproducible code, Bayesian inference techniques and automatic differentiation for training and interpreting neural corrections to baseline dynamical models (outer gray boxes).}
\label{fig:modules}
\end{figure*}

This section describes the \texttt{sapphire} framework with an emphasis on dynamics and numerics. Our baseline galaxy formation model is from \citet{pandya23} and we refer the reader to that paper for more details about it. Figure \ref{fig:modules} illustrates the core dynamical components of any baseline SAM as well as several foundational, interdisciplinary areas that we needed to bridge to build \texttt{sapphire}. Here we only review the elements most relevant for this paper. We refer the reader to Appendix \ref{sec:numerics} for essential numerical tests, including details about \texttt{diffrax} \citep{kidger22}, the \texttt{JAX}-based differential equation solver package that we use.

\subsection{State variables and evolution equations}

We implement a nearly identical version of the model from \citet{pandya23} in \texttt{JAX} with slight modifications as discussed below. Briefly, the model defines eight state variables that summarize the co-evolution of the long-lived stars, ISM and CGM of a galaxy:
\begin{align}
\vec{x} = [&M_*,M_{\rm ISM},M_{\rm CGM},E_{\rm CGM}^{\rm th},E_{\rm CGM}^{\rm turb}, \nonumber \\
& M_*^Z,M_{\rm ISM}^Z,M_{\rm CGM}^Z ]\;.
\end{align}
Here, $M_X$ refers to mass and $M_X^Z$ refers to metal mass, with the stars/ISM and CGM being treated as two separate ``zones.'' One unique aspect of the \citet[][see also \citealt{carr23,voit24a,voit24b}]{pandya23} model is that it introduces two new state variables for the thermal energy ($E_{\rm CGM}^{\rm th}$) and turbulent kinetic energy ($E_{\rm CGM}^{\rm turb}$) of the CGM. These state variables co-evolve according to this system of nonlinearly\footnote{When we say galaxy formation is nonlinear, we mean that, even in the context of this very simple model, it is impossible to rewrite the right-hand side as the product of a state-independent matrix and the state vector itself, plus some vector of constants. This makes the dynamics very hard to understand without the tools we will introduce in this and future papers. The main sources of nonlinearity in the \citet{pandya23} ODEs are the new CGM cooling, turbulence dissipation and over-pressurization terms.} coupled ODEs:

\begin{widetext}

\begin{equation}
\dot{M}_* = (1-R)\frac{M_{\rm ISM}}{t_{\rm dep}}
\end{equation}

\begin{equation}
\dot{M}_{\rm ISM} = \frac{M_{\rm CGM}}{t_{\rm cool,eff} + t_{\rm ff,eff}} - \left[1+\frac{\eta_M}{1-R}\right]\left[(1-R)\frac{M_{\rm ISM}}{t_{\rm dep}}\right] 
\end{equation}

\begin{equation}
\dot{M}_{\rm CGM} = f_{\rm prev}f_{\rm UV}f_{\rm b}\dot{M}_{\rm DM} - \frac{M_{\rm CGM}}{t_{\rm cool,eff} + t_{\rm ff,eff}} + \eta_M\frac{M_{\rm ISM}}{t_{\rm dep}} - \frac{\dot{E}_{\rm out,halo}}{E_{\rm CGM}/M_{\rm CGM}}
\end{equation}

\begin{align}
\dot{E}_{\rm CGM}^{\rm th} &= f_{\rm th}^{\rm acc} e_{\rm vir}f_{\rm prev}f_{\rm UV}f_{\rm b}\dot{M}_{\rm DM} - \int 4\pi r^2 n(r)^2\Lambda(M_{\rm CGM}, E_{\rm CGM}^{\rm th}, Z_{\rm CGM}) dr + \frac{E_{\rm CGM}^{\rm turb}}{t_{\rm turb}} \nonumber \\
& + f_{\rm th}^{\rm wind}\eta_E e_{\rm SN}\frac{M_{\rm ISM}}{t_{\rm dep}} - f_{\rm CGM}^{\rm th}\left[\frac{E_{\rm CGM}^{\rm th}+E_{\rm CGM}^{\rm turb}-e_{\rm vir} M_{\rm CGM}}{t_{\rm dyn}^{\rm halo}}\right] 
\end{align}

\begin{align}
\dot{E}_{\rm CGM}^{\rm turb} &= (1-f_{\rm th}^{\rm acc}) e_{\rm vir}f_{\rm prev}f_{\rm UV}f_{\rm b}\dot{M}_{\rm DM} - \frac{E_{\rm CGM}^{\rm turb}}{t_{\rm turb}} \nonumber \\
& + (1-f_{\rm th}^{\rm wind})\eta_E e_{\rm SN}\frac{M_{\rm ISM}}{t_{\rm dep}} - (1-f_{\rm CGM}^{\rm th})\left[\frac{E_{\rm CGM}^{\rm th}+E_{\rm CGM}^{\rm turb}-e_{\rm vir} M_{\rm CGM}}{t_{\rm dyn}^{\rm halo}}\right] 
\end{align}

\begin{equation}
\dot{M}_*^Z = (1-R)\frac{M_{\rm ISM}^Z}{t_{\rm dep}}
\end{equation}

\begin{equation}
\dot{M}_{\rm ISM}^Z = \frac{M_{\rm CGM}^Z}{t_{\rm cool,eff} + t_{\rm ff,eff}} - (1-R+\eta_M)\frac{M_{\rm ISM}^Z}{t_{\rm dep}} + (1-\eta_Z)(1-R)y_Z\frac{M_{\rm ISM}}{t_{\rm dep}}
\end{equation}

\begin{equation}
\dot{M}_{\rm CGM}^Z = Z_{\rm in,halo}f_{\rm prev}f_{\rm UV}f_{\rm b}\dot{M}_{\rm DM} - \frac{M_{\rm CGM}^Z}{t_{\rm cool,eff} + t_{\rm ff,eff}} + (1-R+\eta_M)\frac{M_{\rm ISM}^Z}{t_{\rm dep}} + \eta_Z(1-R)y_Z\frac{M_{\rm ISM}}{t_{\rm dep}}
\end{equation}

\end{widetext}

We refer the reader to Section 2 and Table 1 of \citet{pandya23} for definitions and justifications of the individual terms. In short, these ODEs approximate flows of mass, metals and energy between the galaxy and CGM zones due to cosmic accretion, gas cooling, star formation, supernova (SN) feedback, and CGM over-pressurization. They are, in a sense, ``continuity equations'' that must hold on galactic scales, but it is not clear a priori how to parameterize the individual terms describing each process. Guided by simple physical arguments, \citet{pandya23} showed that it is possible to calibrate these ODEs by extracting the time evolution of the state variables and individual source/sink terms from cosmological simulations (using the FIRE-2 suite from \citealt{hopkins18} as an initial testbed; the analysis followed what was done by \citealt{pandya20} and \citealt{pandya21}). In this way, simulations can be used to provide ``physics-informed priors'' where data may otherwise be too ambiguous.

Our goal with \texttt{sapphire} is to begin to infer the nonlinear, time-dependent structure of this differential equation system from observational data, and then extend it to more state variables in an interpretable way. As a first step towards that vision, here we stick with the baseline equations from \citet{pandya23} and infer a subset of free parameters that encode our ignorance about various astrophysical processes. For simplicity, following the Appendix of \citet{pandya23}, here we set $f_{\rm th}^{\rm acc}=f_{\rm th}^{\rm wind}=1$ which prevents driving of turbulence and puts the CGM in the purely thermal limit, negating the need to evolve the $E_{\rm CGM}^{\rm turb}$ state variable \citep[as in][]{carr23}.

\subsection{Astrophysical parameters}\label{sec:params}
In this paper, we will infer the mass, energy and metal loading factors of SN feedback as well as the ISM depletion time using galaxy scaling relations. Other parameters are fixed to what \citet{pandya23} found. These are defined as 
\begin{equation}
\eta_M \equiv \frac{\dot{M}_{\rm wind}}{\rm {SFR}}\;,
\end{equation}
\begin{equation}
\eta_E \equiv \frac{\dot{E}_{\rm wind}}{e_{\rm SN} \rm {SFR}}\;,
\end{equation}
\begin{equation}\label{eqn:etaZ}
\eta_Z \equiv \frac{\dot{M}_{\rm wind}^{Z,\rm SN}}{y_Z \rm {SFR}}\;.
\end{equation}
Here, $e_{\rm SN}=10^{51}\rm{erg}/100M_{\odot}$ and $y_Z=2M_{\odot}/100M_{\odot}\approx0.02$ are, respectively, the specific energy and metal yield from supernovae per $100M_{\odot}$ of stars formed with a canonical \citet{kroupa01} or \citet{chabrier03} initial mass function (IMF). Note that we define $\eta_Z$ with respect to only the pure SN-enriched material in the outflow, $\dot{M}_{\rm wind}^{Z,\rm SN}$. There is a separate contribution from entrained ISM metals ($Z_{\rm ISM}\eta_M \rm{SFR}$) that together determines the total outflowing metal mass $\dot{M}_{\rm wind}^Z$ \citep[as in][]{carr23}. Lastly, the ISM depletion time is defined as 
\begin{equation}
t_{\rm dep} \equiv \frac{M_{\rm ISM}}{\rm{SFR}}\;.
\end{equation}

Motivated by \citet[][see their Figures 4 and 5]{pandya23} for how these astrophysical parameters may vary with halo mass and redshift, here we introduce generalized power laws for inference: 
\begin{equation}
\theta_X = 10^{A_X} \left(\frac{V_{\rm vir}}{125 \rm{km s^{-1}}}\right)^{\alpha_X^0 + \alpha_X^z(1+z)} (1+z)^{\beta_X}\;.
\end{equation}
Here, $\theta_X$ represents any one of the four astrophysical parameters: $\eta_M$, $\eta_E$, $\eta_Z$ or $t_{\rm dep}$. These are each allowed to vary with halo virial velocity $V_{\rm vir}$ and redshift $z$ as determined by their respective power law log-amplitude $A_X$ and slope $\alpha_X^0$. The log-amplitude and slope are both allowed to vary with redshift as dictated by the corresponding $\beta_X$ and $\alpha_X^z$ parameters. In this first paper, we will only use $z=0$ galaxy scaling relations as constraints and, for simplicity, we fix all $\alpha_X^z=\beta_X=0$ except for $\beta_{\rm SF}=-0.7$ since we know empirically from observations that $t_{\rm dep}$ must decrease with redshift \citep[e.g., see review by][]{tacconi20}.

\subsection{Cosmology and halo merger trees}\label{sec:cosmology}
Cosmology provides initial conditions (ICs) and forcing functions for our ODEs in the form of halo merger trees. These merger trees determine the initial, evolving and final $z=0$ distribution of halo virial masses, radii, velocities and concentrations. For any individual halo, its time-dependent mass accretion rate acts as an external forcing function for the above ODEs. For this work, we use publicly available halo merger trees from the dark matter-only TNG100-1-Dark simulations \citep{nelson19b,gabrielpillai22}, generated with the Rockstar halo finder and consistent-trees codes \citep{behroozi13a,behroozi13b}. The $\sim10^7M_{\odot}$ DM particle mass of TNG100-1-Dark means we can resolve halos as small as $\sim10^9M_{\odot}$, assuming 100 particles are required to adequately measure halo mass and concentration.\footnote{Throughout we assume the virial overdensity mass and radius definition of \citet{bryan98}.} In order to minimally resolve the main progenitor branch, we make a factor of ten larger cut on the smallest allowed root halo mass: $\log M_{\rm vir}/M_{\odot}>10$ at $z=0$. From the full (100 Mpc)$^3$ box, we take five random (20 Mpc)$^3$ subvolumes which together contain $\sim10^5$ halos above this root mass limit. For our Bayesian inference, we only need a random subset of these halos ($\sim10^3$) which we select by assigning an inverse weight to each halo based on how many other halos we have of similar mass. This allows us to generate a uniform random root halo mass distribution without over-representing low-mass halos, which are otherwise far more abundant. Our results are not sensitive to the choice of different random subsamples.

For numerical compatibility with adaptive ODE solvers, automatic differentiation, and massive vectorization and parallelization (Appendix \ref{sec:numerics} below), we need to smoothly interpolate these merger trees. By default, the TNG100-1-Dark trees record halo properties at $\sim100$ discrete snapshots from $z\sim10$ to $z=0$. We first use finite-differencing to convert the time series of halo mass into a mass accretion rate, requiring that the halo mass can only ever increase and setting the accretion rate to zero otherwise.\footnote{In this paper, we only model central galaxies. Subhalos, which host satellite galaxies, can experience mass loss but we defer those to Gabrielpillai et al. (in prep.).} Following subsection 2.7 of \citet{pandya23}, we apply a 1D Gaussian smoothing\footnote{We arbitrarily choose a Gaussian width of 10 snapshots which corresponds to $\sim30$ Myr at $z\sim10$, increasing to $\sim200$ Myr at $z\sim0$.} to five time series: (1) halo mass accretion rate, (2) virial mass, (3) virial radius, (4) virial velocity and (5) concentration, using the \citet{klypin01} definition for the latter. Since every halo starts at a different initial redshift (depending on when its earliest main branch progenitor was first resolved/identified), we first use a cubic univariate spline to interpolate all halos onto the same common time grid, zero-padding earlier non-existing snapshots as needed. Then we use differentiable cubic Hermite splines with backward differences \citep{morrill21} to interpolate each of the five time series, resulting in a coefficient matrix encoding time-dependent forcing functions for each halo ODE system. This coefficient matrix has the same fixed shape for all halos and can be stacked for efficient vectorization and parallelization. 

Finally, we compute the ICs of our ODE state variables by assuming that every halo starts out with CGM mass $M_{\rm CGM}=f_{\rm b} M_{\rm vir}$ where $f_{\rm b}=\Omega_b/\Omega_M=0.158$ is the universal cosmic baryon fraction, CGM thermal energy $E_{\rm CGM}^{\rm th}\sim M_{\rm CGM} V_{\rm vir}^2$ and small but non-zero stellar mass, ISM mass, and metal masses corresponding to an initial metallicity $Z/Z_{\odot}=10^{-4}$ with $Z_{\odot}=0.02$. The small, non-zero initial masses are required since we solve our ODEs using logarithmic state variables for numerical stability, but our results are not sensitive to reasonable IC variations because the rapid, steep assembly phase of halos dominates early galaxy evolution in our model. Note that even though halos are initialized with their fair share of the cosmic baryon fraction, they may end up being depleted of baryons if their CGM becomes significantly over-pressurized \citep{pandya23,carr23,voit24a,voit24b,messere26} and/or if photoionization from the cosmic ultraviolet background suppresses halo accretion which can happen for the lowest mass dwarfs we consider \citep[we follow the approach of][]{gnedin00,somerville02,kravtsov04,okamoto08}.

\section{Methods}\label{sec:methods}
In this section, we describe our methods for local and global sensitivity analysis with Jacobians, parameter optimization, and Bayesian inference. We again refer the reader to Appendix \ref{sec:numerics} for important numerical tests.

\begin{figure}
\centering
\includegraphics[width=\hsize]{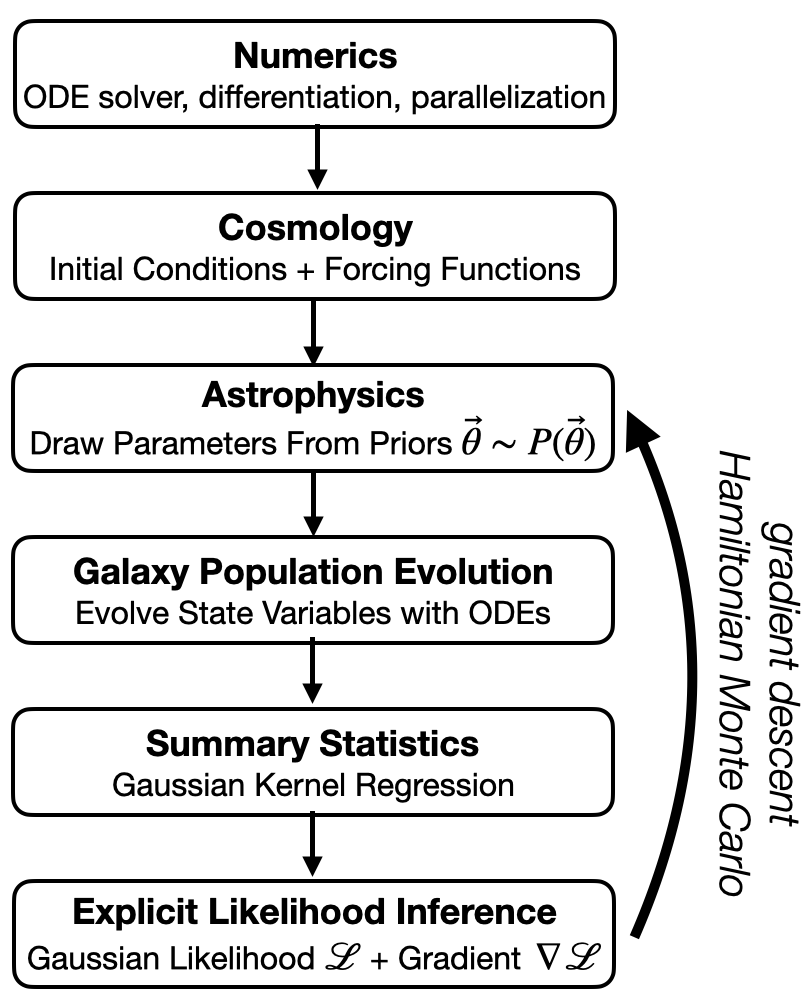}
\caption{Flowchart illustrating our galaxy population evolution model that bridges numerics, cosmology, astrophysics and Bayesian statistics. We first choose our numerical setup (ODE solver, automatic differentiation method and parallelization strategy). Next, we interpolate halo merger trees which provides initial conditions and forcing functions for our galaxy formation ODE system. In principle, this step can involve cosmology variations but we neglect that here. After drawing the astrophysical parameters from their respective priors, we evolve the ODE system for our ensemble of galaxies in parallel, compute summary statistics with differentiable Gaussian kernel regression, and finally compute the explicit assumed Gaussian likelihood and its gradient with respect to each astrophysical parameter. The latter enables efficient Bayesian MAP optimization and/or Hamiltonian Monte Carlo inference.}
\label{fig:flowchart}
\end{figure}

Figure \ref{fig:flowchart} serves as an outline for this section and summarizes our essential modeling steps. After fixing our choice of numerics (Appendix \ref{sec:numerics}) and cosmology (subsection \ref{sec:cosmology}), we draw astrophysical parameters from priors\footnote{In principle, we can also vary cosmology, but that is deferred to future work.}, evolve an ensemble of galaxy state variables through time using our ODEs, compute differentiable summary statistics, compute a likelihood and its gradient, and repeat this process until convergence using gradient descent and/or Hamiltonian Monte Carlo (HMC). We now describe each of these steps in turn, starting with Latin hypercube sampling over our choice of priors.

\subsection{Latin hypercube sampling with uniform priors}\label{sec:lhs}
Throughout this paper we will want to assess the causal identifiability\footnote{As explained in Subsection \ref{sec:intro2}, by this we mean mapping noisy observable outputs back to input model parameters.} of our model across parameter space given just a few $z=0$ galaxy scaling relations as constraints. For this, we will use mock tests and local sensitivity analyses at many different points throughout parameter space (as well as efficient parameter optimization and inference techniques). Even with only 8 free parameters, it is prohibitively expensive to do grid-based sampling so instead we adopt Latin hypercube sampling \citep[e.g., see][]{bower10,villaescusanavarro21,perez23,ho24,chaikin25}. We draw $N$ random points in our 8-dimensional parameter space by sampling each along its respective prior and then stacking the resulting parameter vectors. 

For simplicity, we adopt uniform priors for all parameters. The slope parameters are limited to $-4<\alpha_X^0<0$. For the log-amplitudes, we impose $A_M\in[-2,1]$, $A_E\in[-2,0]$, $A_{\rm SF}\in[0,1.1]$ and $A_Z\in[-2,0]$. In principle, we could have allowed for infinitely wide priors, but we already know that some combinations are unphysical. For example, energy loading from SNe alone cannot exceed one due to continuity arguments, and SN feedback alone is unlikely to cause loading factors to increase with halo mass since potential wells get deeper. Our limited (but still quite large) uniform prior ranges encode these ``implicit'' physical arguments to prevent us from wasting computation in pathological parts of solution space.

\subsection{Sensitivity Analysis with Jacobians}\label{sec:jacs}
Even without defining a loss or doing any inference, we can immediately assess the sensitivity of our dynamical model parameters to its different outputs. In this paper, we use forward-mode autodiff to compute the Jacobian matrix of partial derivatives of our $z=0$ galaxy state variables with respect to our eight free astrophysical amplitude and slope parameters:
\begin{equation}\label{eqn:jac}
\mathcal{J} = 
\begin{bmatrix}
\frac{\partial \log M_*}{\partial A_M} & \frac{\partial \log M_*}{\partial \alpha_M^0} & \dots & \frac{\partial \log M_*}{\partial A_Z} & \frac{\partial \log M_*}{\partial \alpha_Z^0} \\
\vdots & \ddots & \ddots & \ddots & \vdots\\
\frac{\partial \log M_{\rm CGM}^Z}{\partial A_M} & \frac{\partial \log M_{\rm CGM}^Z}{\partial \alpha_M^0} & \dots & \frac{\partial \log M_{\rm CGM}^Z}{\partial A_Z} & \frac{\partial \log M_{\rm CGM}^Z}{\partial \alpha_Z^0}
\end{bmatrix}
\end{equation}
The rows of $\mathcal{J}$ are called ``gradients'' of a single state variable with respect to all parameters at once. The columns are referred to as parameter sensitivities, showing the effect of perturbing a single parameter on all state variables at once. By construction, Jacobians are local measures of the nonlinear sensitivity of our galaxy state variables to free parameters. Thanks to the combination of \texttt{jit} and auto-diff, for the first time, we will be able to compute these local sensitivity metrics at any arbitrary point across the global parameter space. This provides a fundamentally new and complementary technique for model interpretability compared to traditional coarse-grained parameter variations. Note that $\mathcal{J}$ is time-dependent since the ODE system is subject to forcing functions from cosmology-dependent merger trees as well as redshift-dependent astrophysical parameters. In Appendix \ref{sec:numerics}, we compare the accuracy of our auto-diff Jacobians to those computed with expensive finite-differencing.

\subsection{Differentiable Gaussian kernel regression}\label{sec:nadaraya}
In this paper, we will employ explicit likelihood inference so we need to compute summary statistics from the set of discrete data points representing our individual evolved model galaxy properties.\footnote{Makinen et al. (in prep.) will present implicit likelihood inference which can non-parametrically learn the full, joint distribution of \texttt{sapphire} model outputs.} Traditional approaches involving hard, fixed bins suppress the flow of gradient information, preventing the application of differentiable inference techniques. As a simple alternative, here we use univariate Gaussian kernel regression for each of the three scaling relations individually (essentially soft binning). Briefly, each point is convolved with a Gaussian along the independent variable axis to provide a smooth, non-parametric, locally-weighted average of the dependent variable. The normalized weight of the $i$th point at location $x$ is
\begin{equation}
\mathcal{W}_i(x) = \frac{\mathcal{N}(x-x_i,\sigma_{\mathcal{W}}^2)}{\sum_j \mathcal{N}(x-x_j,\sigma_{\mathcal{W}}^2)}
\end{equation}
where $\mathcal{N}(\ldots)$ denotes a Gaussian. The Gaussian kernel bandwidth $\sigma_{\mathcal{W}}$ is assumed to be the same for every point, and it depends on both the actual covariance of the data as well as the sample size following the empirical ``Scott's rule'' \citep{scott10}: 
\begin{equation}
\sigma_{\mathcal{W}} = \sigma_x \cdot N^{-1/(d+4)}
\end{equation}
with $d=1$ for our univariate regression case. Here we take the zeroth order average of the Gaussian contributions at a given location, which is known as Nadaraya-Watson regression \citep{nadaraya63,watson64}. As an example, the kernel-weighted average SMHM ratio $y\equiv \log M_*/M_{\rm vir}$ at an arbitrary $x\equiv \log M_{\rm vir}/M_{\odot}$ is: 
\begin{equation}
\bar{y}(x) = \sum\mathcal{W}_i(x-x_i)\cdot y_i
\end{equation}
where the sum runs over all data points $(x_i,y_i)$. The intrinsic standard error on $\bar{y}(x)$ is:
\begin{equation}
\sigma_{\bar{y}}(x) = \left[\frac{\sum\mathcal{W}_i(x-x_i)\cdot (y_i - \bar{y}(x))^2}{N_{\rm eff}(x)}\right]^{1/2}
\end{equation}
where 
\begin{equation}
N_{\rm eff}(x) = \frac{1}{\sum \mathcal{W}_i^2(x)}
\end{equation}
is the local effective sample size at $x$ based on the squared sum of normalized Gaussian weights.

\begin{figure*}
\centering
\includegraphics[width=\hsize]{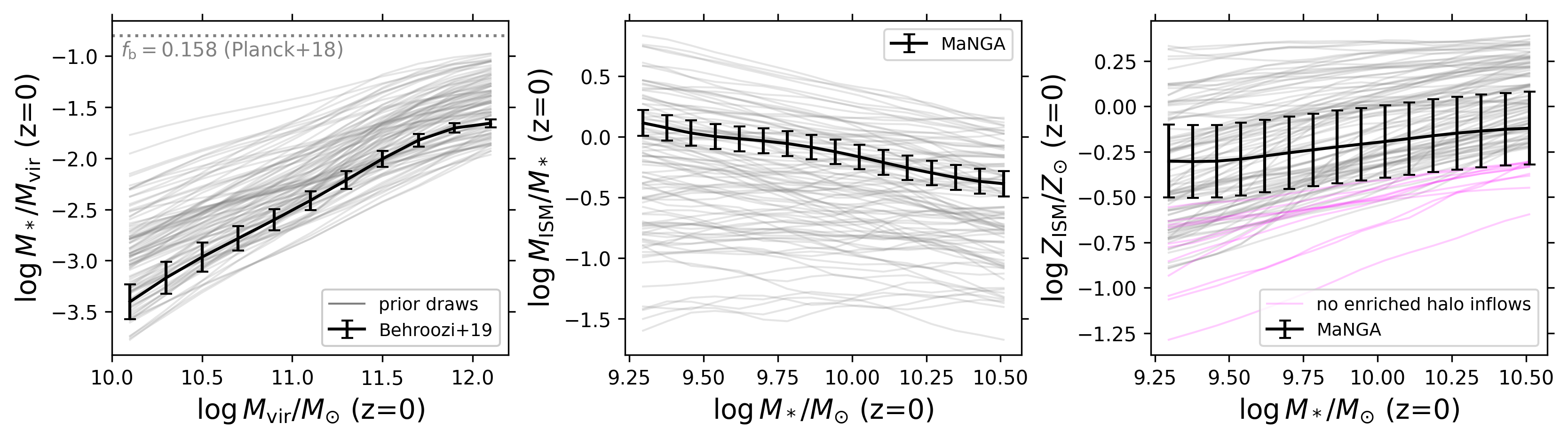}
\caption{Prior predictive checks for the three $z=0$ galaxy scaling relations we focus on in this paper. In each panel, the black line comes from data and the gray lines are draws from uniform priors for model parameters. Left: stellar-mass-halo-mass relation. Middle: total ISM-to-stellar-mass ratios as a function of stellar mass. Right: ISM mass-metallicity relation. The prior predictive checks span the range of the data. In the right panel, we further show the effect of turning off pre-enriched halo inflows (magenta lines) which causes the ISM MZRs to reach a lower floor compared to our fiducial model. This will be used in Subsection~\ref{sec:perturbs} to interpret our baseline posteriors and illustrate the impact of MZR systematics.}
\label{fig:priorpreds}
\end{figure*}

Figure \ref{fig:priorpreds} shows prior predictive checks for the three $z=0$ galaxy scaling relations that we will focus on in this paper. The priors correspond to the same uniform ranges for the eight parameters discussed in Subsection~\ref{sec:lhs}. For all three quasi-observables, the prior predictive checks encompass the data, which means we can pursue inference in Section~\ref{sec:fitdata}. Interestingly, the $f_{\rm ISM}\equiv M_{\rm ISM}/M_*$ relations span a wide range above and below the data but the distribution of SMHM and MZR relations tend to be on the higher side compared to the data. This is telling us something about both the baseline model and data. For example, if we turn off pre-enrichment of halo inflows, this allows us to predict ISM MZRs with even lower normalizations, which will be useful when we interpret our posterior predictive checks in Subsection~\ref{sec:perturbs}.

\subsection{Loss function and posterior}\label{sec:loss}
We assume an independent Gaussian log-likelihood for every individual kernel-weighted average across all three $z=0$ scaling relations:
\begin{equation}
\small
\log\mathcal{L}(\vec{\theta}) = -\frac{1}{2} \sum_i \left[\left( \frac{\bar{y}_i - \bar{y}^{\rm obs}_i}{\sigma_{i,\rm tot}} \right)^2 + \log(2\pi \sigma_{i,\rm tot}^2) \right]
\end{equation}
where the subscript $i$ runs over regression kernels and
\begin{equation}
\sigma_{i,\rm tot}(\vec{\theta}) = \sqrt{\sigma_{\mathcal{W}}^2+\sigma_{i,\rm obs}^2}
\end{equation}
is the total uncertainty from quadrature summing the intrinsic model scatter and measurement errors on $y$. To account for measurement errors in both $x$ and $y$, we use the approach of ``multiple imputations'' as described in Appendix \ref{sec:manga}. The dependence of the log-likelihood on parameters $\vec{\theta}$ comes through both $\bar{y}_i$ and $\sigma_{i,\rm tot}$. For simplicity, we assume no covariance between the three $z=0$ scaling relations so we can further just sum their individual log-likelihood contributions. We know this is not correct because of the existence of multi-dimensional galaxy scaling relations and that it erases some information \citep[e.g.,][]{jo26}, but it is sufficient for our proof-of-concept demonstration \citep[we can also address this with implicit likelihood inference;][and Makinen et al., in prep.]{ho24}. Since we assume uniform priors for all parameters, our likelihood is equivalent to the posterior. Taking its negative gives us a loss function that we can minimize with gradient descent (subsection \ref{sec:map}) and/or explore in a fully Bayesian way with Hamiltonian Monte Carlo (subsection \ref{sec:hmc}).

\subsection{Fisher/Laplace analysis}\label{sec:fisher}
The curvature of the loss landscape encodes the sensitivity of the model to our astrophysical parameters $\vec{\theta}$. This loss landscape remains largely uncharted for complicated, nonlinear and historically expensive dynamical galaxy formation models such as ours. One way to probe this sensitivity both locally and globally is to compute the Hessian of our loss function throughout parameter space. This is routinely done in precision cosmology \citep[e.g.,][]{turner22,lanzieri23} and sometimes for empirical abundance-matching models \citep{wechsler18}, but never for physically-grounded dynamical models. The Hessian is a symmetric, positive-definite matrix that encodes the second-order gradients of the loss:
\begin{equation}\label{eqn:hessian}
\mathcal{H} = 
\begin{bmatrix}
-\frac{\partial^2 \log \mathcal{L}}{\partial^2 A_M} & \cdots & \cdots & \cdots & \cdots \\
-\frac{\partial^2 \log \mathcal{L}}{\partial A_M \partial\alpha_M^0} & -\frac{\partial^2 \log \mathcal{L}}{\partial^2 \alpha_M^0} & \cdots & \cdots & \cdots \\
\vdots & \ddots & \ddots & \ddots & \vdots\\
-\frac{\partial^2 \log \mathcal{L}}{\partial A_M \partial A_Z} & -\frac{\partial^2 \log \mathcal{L}}{\partial \alpha_M^0 \partial A_Z} & \cdots & -\frac{\partial^2 \log \mathcal{L}}{\partial^2 A_Z} & \cdots \\
-\frac{\partial^2 \log\mathcal{L}}{\partial A_M \partial \alpha_0^Z} & -\frac{\partial^2 \log \mathcal{L}}{\partial \alpha_M^0 \partial \alpha_0^Z} & \cdots & -\frac{\partial^2 \log \mathcal{L}}{\partial A_Z \partial \alpha_Z^0} & -\frac{\partial^2 \log \mathcal{L}}{\partial^2 \alpha_Z^0}
\end{bmatrix}
\end{equation}
The diagonal entries give the marginal parameter sensitivities whereas the off-diagonals encode degeneracies between different combinations of two parameters. Just like the Jacobian (equation \ref{eqn:jac}), the Hessian varies across parameter space. At minima, we can use the Fisher/Laplace approximation to invert $\mathcal{H}$, which provides errorbars (i.e., a full covariance matrix) for parameters assuming the posterior was locally Gaussian. At saddle points, the Hessian is non-positive-definite so an inverse does not exist and local parameter uncertainties cannot be computed. In this paper, we will run multiple gradient descent trajectories (subsection \ref{sec:map}) to identify the maximum a posteriori (MAP) point estimate of parameters and compute the covariance matrix there. This is a useful, efficient method that complements fully Bayesian HMC, which is much more expensive (subsection \ref{sec:hmc}).

\subsection{MAP optimization with gradient descent}\label{sec:map}

To efficiently find MAP point estimates, we use gradient descent with the \adam\ optimizer \citep{kingma17} as implemented in the \texttt{JAX} \texttt{optax} library \citep{optax}. Briefly, \adam\ applies a series of transformations to our ``raw'' first-order gradient of the loss function.\footnote{Since the loss is a scalar, its Jacobian is a single-row matrix.} These transformations include averaging over the history of past gradients (``momentum'') to smooth over stochastic noise and identify directions in parameter space where the loss consistently decreases. To prevent large parameter jumps, we first clip the raw gradient vector so that its norm is one. At each iteration, the parameters are updated in the direction opposite to the gradient so that the loss decreases:
\begin{equation}
\vec{\theta}_{\rm new} = \vec{\theta} - \alpha_{\rm lr} f_{\rm adam}[\nabla (-\log\mathcal{L})]
\end{equation}
where $f_{\rm adam}(\dots)$ denotes that \adam\ transformations are applied to the raw-clipped gradient of the loss (negative log-likelihood). $\alpha_{\rm lr}$ is the learning rate (effectively a step size) which determines the update for each parameter. Normally, the learning rate is considered a hyper-parameter that must be tuned but we forgo that for the simple proof-of-concept demonstrations in this paper. We assume an exponentially decaying learning rate schedule with an initial value of $0.01$, a decay rate of 0.95 every 50 steps, and a floor of $10^{-4}$. Our convergence/stopping criteria are that both the norm of the parameter update vector and the relative change in the loss must be less than $10^{-3}$. We find that this generally suffices to get near the central bottom of minima. 

\subsection{Hamiltonian Monte Carlo}\label{sec:hmc}

To complement our local Fisher/Laplace analysis (subsection \ref{sec:fisher}), we utilize HMC which performs fully Bayesian exploration of the global parameter space guided by gradients. We use the No U-Turn Sampler variant \citep[NUTS;][]{hoffman14} implemented in the \texttt{JAX}-compatible \texttt{numpyro} probabilistic programming language \citep{numpyro}. We refer the reader to \citet{betancourt17} for a conceptual introduction to HMC. Briefly, at each point in parameter space, the loss is analogous to a potential energy that we want to minimize and the gradient of the loss is akin to a force that determines parameter update directions. Given some starting position in parameter space, a random momentum vector is drawn from a multivariate normal with zero mean and a covariance (``mass'') matrix that is typically adapted during warm-up based on the average curvature of the loss landscape. The joint position–momentum system is then evolved according to Hamilton’s equations, where gradients of the loss act as forces that update the momentum. Unlike \adam, a single iteration of HMC can involve hundreds or thousands of loss and gradient evaluations to generate a long, correlated trajectory through parameter space. In exact arithmetic the Hamiltonian (the sum of potential and kinetic energies) would be conserved; in practice, numerical integration errors are corrected by a Metropolis–Hastings accept/reject step at the end of each trajectory. The NUTS variant of HMC further automates this process by adaptively selecting: (1) the step size to target a desired acceptance probability during warm-up, and (2) the number of integration steps by terminating trajectories that begin to turn back on themselves. 

Although HMC can be faster to converge than standard MCMC (see our comparison against an affine-invariant ensemble sampler in Appendix \ref{sec:numerics}), it is still extremely expensive, especially for dynamical population evolution models like ours where we need to solve ODE systems in parallel for $\sim10^3$ galaxies per iteration. Since the Hamiltonian needs to be conserved, we do not have the flexibility to clip and arbitrarily transform gradients like for \adam. Thus, HMC/NUTS is susceptible to choosing impractically small step sizes if it detects regions of very high curvature, as can happen near steep minima. In practice, by including measurement uncertainties for our Gaussian kernel regression, we effectively smear and smooth the loss landscape, enabling efficient HMC. We will show that the loss landscape is slightly multimodal depending on what constraints are used, but different HMC chains are able to switch between the modes. We do not find evidence that our fiducial choice of ODE solver and tolerances affect convergence across parameter space, which would otherwise require these to be fine-tuned. 

Here we impose a target acceptance probability of 0.8 with a maximum allowed $2^{10}$ steps per trajectory, as recommended by \texttt{numpyro}. We generally only need $\sim1000$ warmup iterations with a dense mass matrix and draw $\gtrsim1000$ samples to map the posterior. We run four chains in parallel and verify that we are converged by requiring that the \citet{gelman92} statistic is $\approx1$. We will use HMC primarily for fitting the observed data and forecasting the impact of different statistical and systematic uncertainties.

\begin{figure*}
\centering
\includegraphics[width=\hsize]{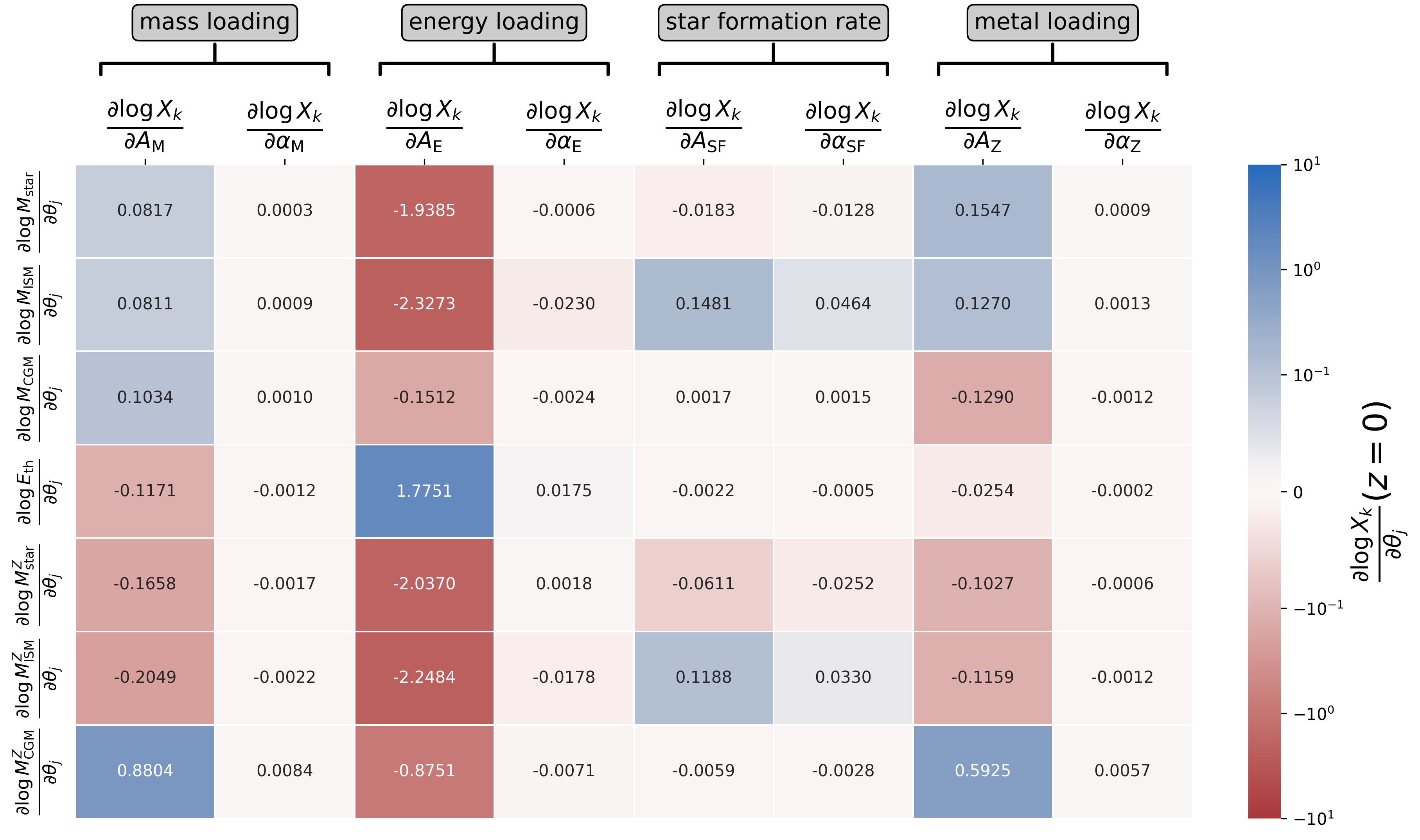}
\caption{Example Jacobian at a random point in parameter space for a single MW-mass halo showing the exact $z=0$ sensitivity of the ODE state variables to parameter variations. The non-random structure of this Jacobian encodes the astrophysics and locally linearized dynamics of our galaxy formation model. Note that the colorbar uses a symmetric log normalization. The greatest sensitivity is to the energy loading normalization (third column). Mass loading sensitivity is weaker and has the opposite effect of energy loading for many state variables. Metal loading only affects the metal mass state variables (lower right). There is only mild sensitivity to the ISM depletion time parameters.}
\label{fig:jacobian}
\end{figure*}

\section{Sensitivity Analyses and Mock Tests}\label{sec:mocks}
In this section, we show that the Jacobians of individual and ensembles of halos have non-random, interpretable structure. We then use the information provided by these gradients to perform mock parameter recovery tests.

\subsection{Example Jacobian of a single MW galaxy at $z=0$}

Figure \ref{fig:jacobian} illustrates an example Jacobian sensitivity matrix for a single $z=0$ MW-mass halo at a random point in parameter space. We can immediately see that the energy loading power law amplitude $A_E$ dominates the sensitivity for every state variable. This is generally the case throughout parameter space and for other halos (see next subsection). The signs of the gradients make sense: as we increase energy loading, the thermal energy of the CGM increases which reduces ISM accretion and ultimately the SFR. We also find a much weaker sensitivity to mass loading amplitude, which in some cases has the opposite sign (e.g., increasing $A_M$ increases $z=0$ stellar mass, unlike increasing $A_E$; see also \citealt{carr23,voit24a,voit24b}). There is only mild sensitivity to the ISM depletion time and metal loading parameters. The $z=0$ metal mass state variables show an order of magnitude greater sensitivity to $A_E$ than $A_Z$, but the sensitivity to the latter may help break degeneracies between $\eta_M$ and $\eta_Z$.

\subsection{Jacobians across parameter space for many halos}

\begin{figure*}
\centering
\includegraphics[width=\hsize]{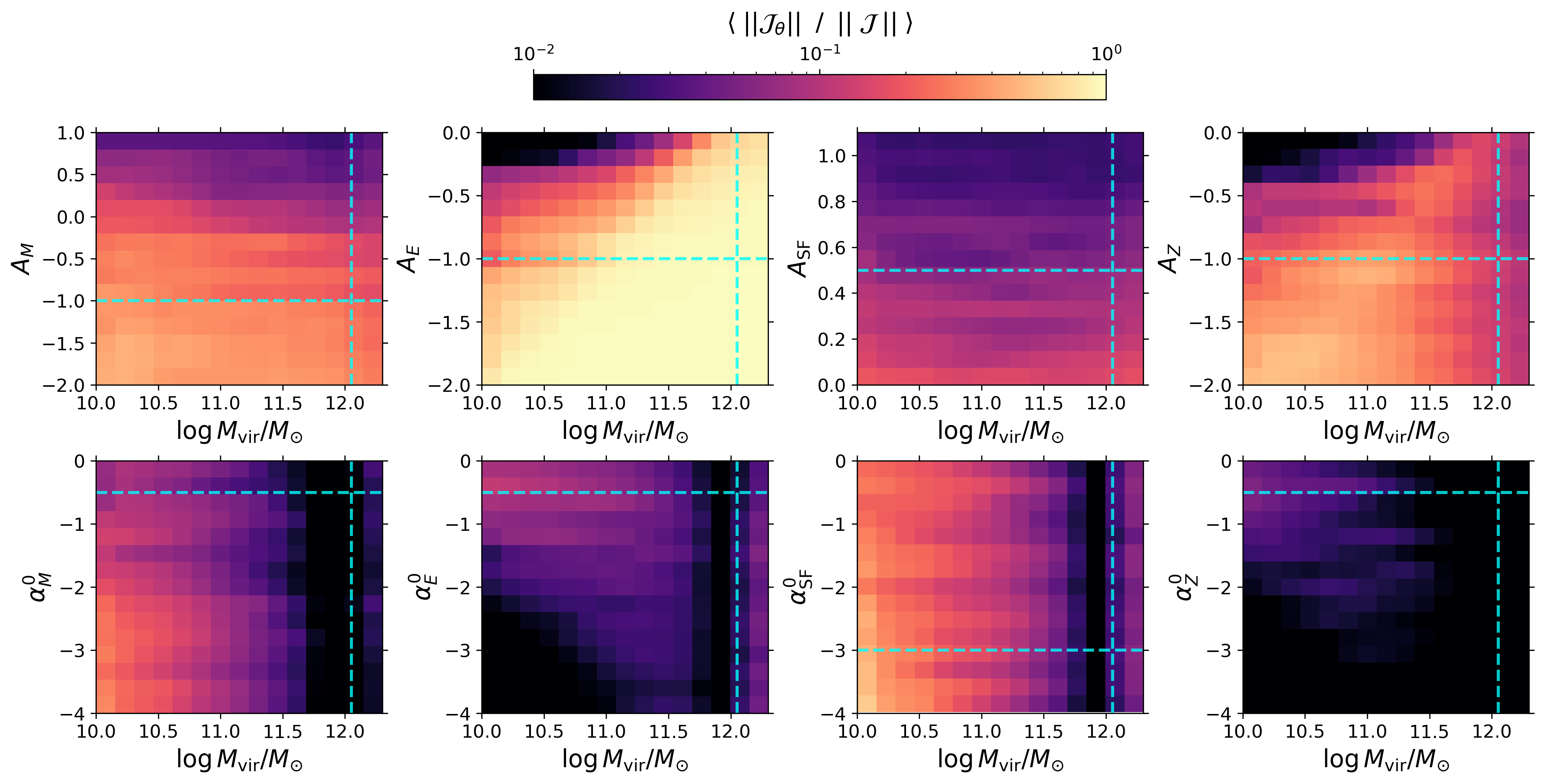}
\caption{Fractional sensitivity across parameter space for a range of halo masses. Brighter colors correspond to higher sensitivities. The 2D histogram panels show the ratio of the norm of each Jacobian column (parameter sensitivity) to the norm of the entire Jacobian matrix as a function of parameter value and halo mass. The combined sensitivity of all state variables to $A_E$ dominates over their sensitivity to the other parameters. The $A_E$ sensitivity drops for low-mass halos in high $A_E$ and/or very steep $\alpha_E^0$ realizations because we clip $\eta_E\leq1$. The cyan dotted lines mark the parameter values of the example Jacobian for the single MW-mass halo from Figure \ref{fig:jacobian}. The lack of sensitivity for the four slope parameters near $\log M_{\rm vir}/M_{\odot}\sim11.8$ is due to reference point $V_{\rm vir}\approx125$ km s$^{-1}$ in our power law functional form where the gradient with respect to $\alpha$ naturally goes to zero.}
\label{fig:jacall}
\end{figure*}

It is remarkable that we were able to identify and interpret systematic patterns in the Jacobian of a single random MW-mass halo given the nonlinearity of our ODE system. But does that clarity translate to ensembles of halos? Figure \ref{fig:jacall} extends the previous analysis to 1000 halos spanning a range of halo masses (and formation histories) for which we compute the $z=0$ Jacobian in each of 1000 random latin hypercube realizations (for a total of 1 million Jacobians).\footnote{This expensive application demands multi-GPU parallelization.} We confirm that $A_E$ dominates the sensitivity regardless of halo mass or location in parameter space. The only exception is low-mass halos for high $A_E$ which show decreased sensitivity because $\eta_E$ gets clipped to be $\leq 1$ (similar behavior is seen for high values of $A_M$ and $A_Z$ as well as very steep $\alpha_E^0$). Another interesting feature is a low-sensitivity ``stripe'' at $\log M_{\rm vir}/M_{\odot}\sim11.8$ for all four slope parameters. This happens because that mass corresponds to the reference point $V_{\rm vir}\approx125$ km s$^{-1}$ in our assumed power law functional forms (see Subsection~\ref{sec:params}). The gradient of those functions with respect to the slope goes to zero exactly at the reference point and remains $\lesssim10^{-2}$ in a region whose width depends on the value of the power law normalization. This just means that we need more halos of other masses to constrain slope parameters (though of course the reference point is arbitrary and could be moved).

\subsection{Mock parameter recovery tests with \adam}\label{sec:adam}

\begin{figure*}
\centering
\includegraphics[width=\hsize]{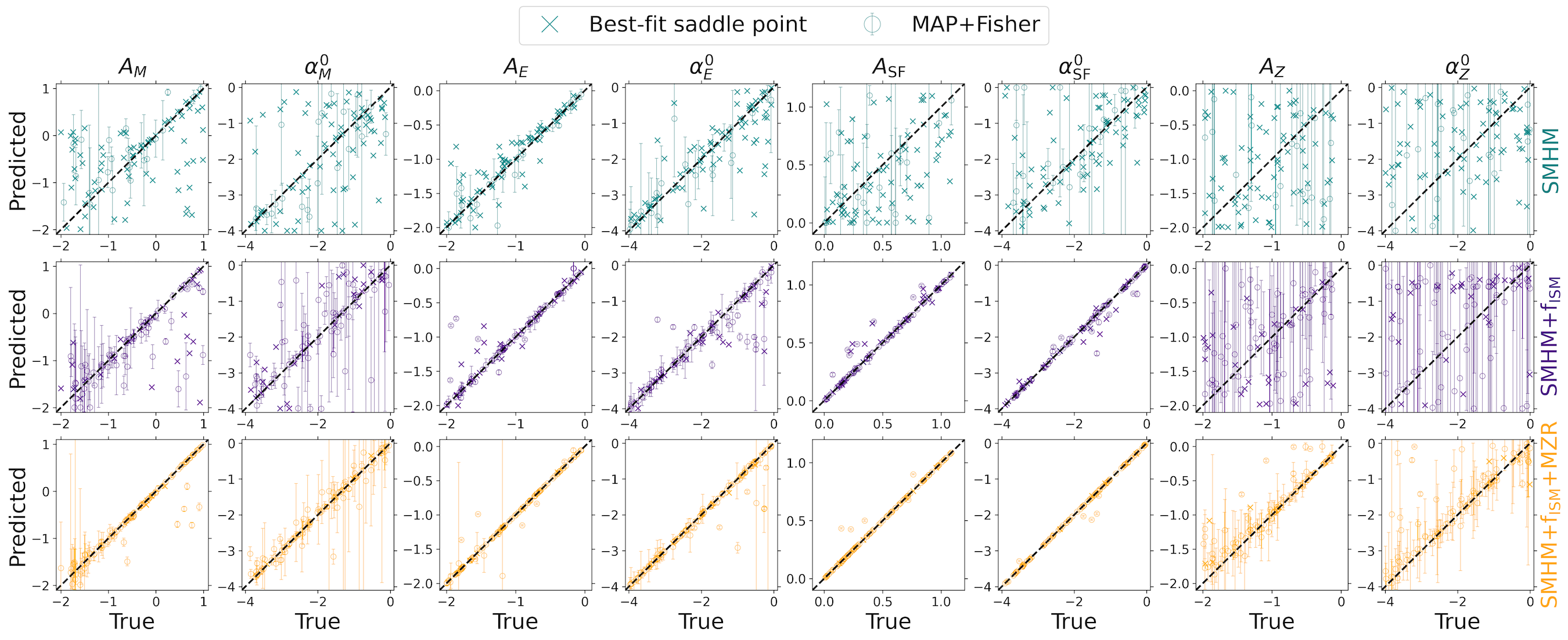}
\caption{Mock parameter recovery tests using different combinations of $z=0$ quasi-observable scaling relations in the best case scenario of no measurement uncertainties (see Appendix \ref{sec:adamerrs} for the latter). Circles and errorbars show the MAP and Fisher uncertainty from \adam. Crosses indicate that all three \adam\ trajectories ended at a saddle point and we report only the lowest-loss solution. From top to bottom, we show our default progression for the rest of this paper: starting with SMHM as the only constraint (teal), then adding $f_{\rm ISM}$ (indigo), and subsequently adding ISM MZR (orange). The best precision and accuracy are achieved when all three constraints are combined. In other cases, $A_E$ is well recovered but there are often saddle points leading to poor convergence for the other parameters. Note how combining SMHM and $f_{\rm ISM}$ pins down the ISM depletion time and energy loading, but strong degeneracies remain between mass and metal loading. Incorporating the ISM MZR distinguishes mass and metal loading parameters with smaller uncertainties.}
\label{fig:adamcoverage8}
\end{figure*}

Now we turn to mock parameter recovery tests using gradient descent. Following Section~\ref{sec:methods}, we generate 100 random points in parameter space, evolve 1000 halos for each realization, summarize the three $z=0$ scaling relations, define a loss involving combinations of the mock relations, run \adam\ from three different initial guesses to find the MAP, and finally compute the local Fisher matrix. Our default progression in this paper will be to start with the SMHM only, then add ISM gas fractions, and then incorporate the ISM MZR. We tested other combinations of one or two relations and find similar behavior. In this subsection, we start with the best case scenario of no measurement uncertainties, exploring those in Appendix \ref{sec:adamerrs}.

Figure \ref{fig:adamcoverage8} summarizes the identifiability of our input model parameters given different combinations of output $z=0$ scaling relations. With the $z=0$ SMHM relation as the only constraint, $A_E$ is often converged to the true value. However, the loss landscape contains many saddle points and long, flat plateaus that \adam\ gets stuck in, so many of the other parameters are not well constrained and Fisher uncertainties cannot be computed due to the absence of a local minimum where \adam\ ended.\footnote{Later in section \ref{sec:fitdata}, when fitting actual data, we will get around this by also running HMC to map the full posterior, not just the MAP with \texttt{adam}.} This is also the case if we use only $f_{\rm ISM}$ or the ISM MZR as single constraints. If we combine the $z=0$ SMHM relation with $\fgas$, many of these false minima and saddle points go away because the ISM depletion time parameters get pinned down precisely, in turn enabling tight constraints on $A_E$, $\alpha_E$, $A_{\rm SF}$, $\alpha_{\rm SF}$ and cases with high $A_M\gtrsim0$. But strong degeneracies remain between mass and metal loading parameters with $A_Z$ and $\alpha_Z$ for the latter showing large scatter. Adding ISM MZR information to SMHM and $f_{\rm ISM}$ helps break the $\eta_M-\eta_Z$ degeneracy leading to smaller uncertainties and more accurate parameter recovery. Naturally, the best precision and accuracy are achieved with all three constraints combined.

\subsection{Precision and bias summary statistics}\label{sec:summarystats}

\begin{figure}
\centering
\includegraphics[width=\hsize]{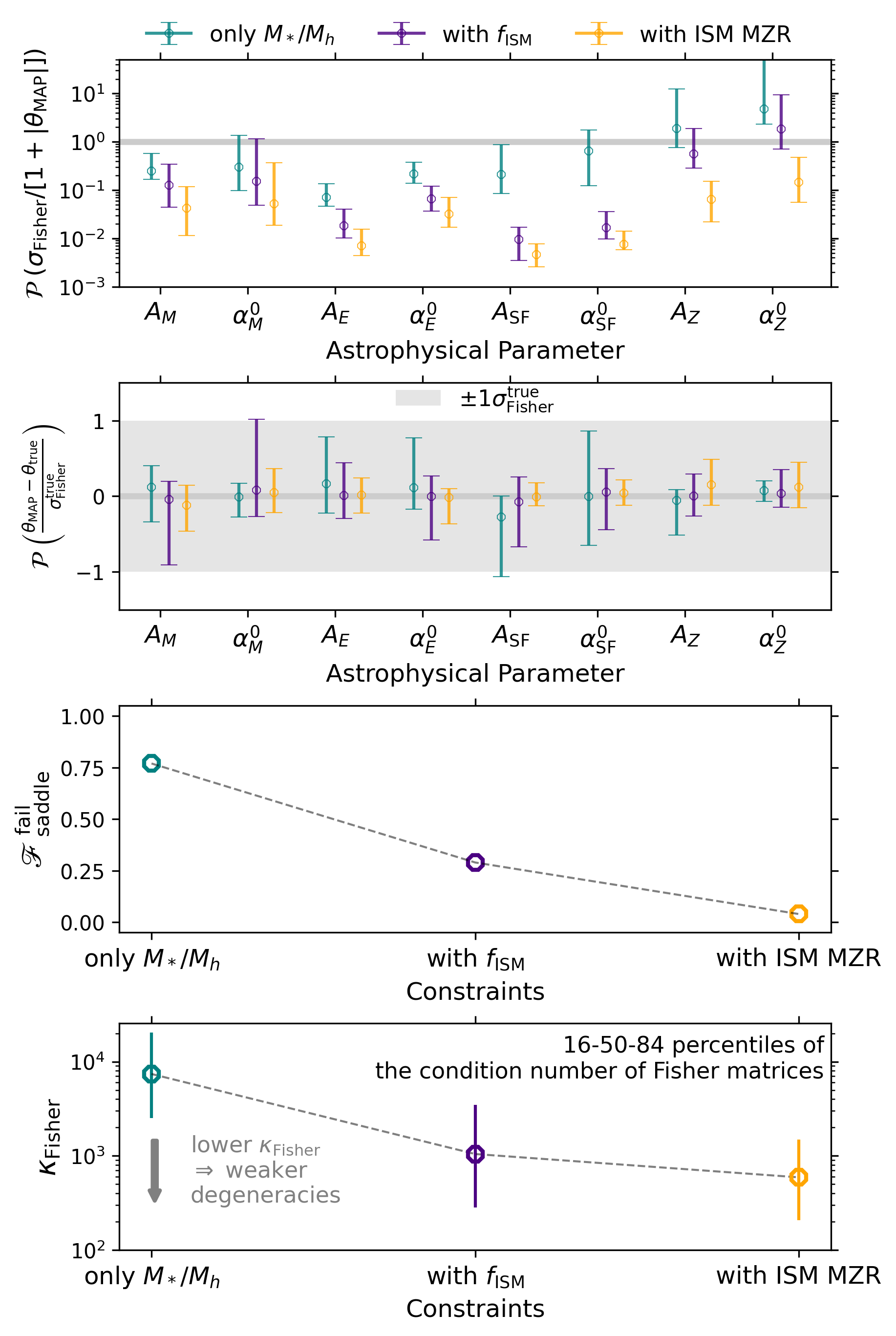}
\caption{Mock parameter recovery summary statistics for our default progression of combining SMHM, $f_{\rm ISM}$ and ISM MZR (i.e., the top three rows from Figure \ref{fig:adamcoverage8}). This is the best case with no measurement uncertainties (we defer those to Appendix \ref{sec:adamerrs}). \textit{Top row:} Precision as quantified by Fisher parameter uncertainties normalized by the MAP parameter value. The errorbars denote the 16, 50 and 84th percentiles from all mock realizations. Naturally, uncertainties improve as more constraints are included: teal for only $\smhm$, purple when adding $\fgas$, and orange when further including the ISM MZR.\textit{Second row:} Bias defined as the deviation of the MAP from the true parameter value normalized by the true minimum Fisher curvature. We are generally within $\pm1\sigma$ of the true minimum and hence unbiased. \textit{Third row:} optimization failure rate (fraction of mocks where all 3 adam optimizers ended at a saddle point) drops as we add more constraints. With $\smhm$ alone, there are many more false saddle points for gradient descent to get stuck in. \textit{Fourth row:} condition number of the Fisher matrix (proxy for parameter degeneracy strength) also drops as we add more constraints.}
\label{fig:summaryfisher}
\end{figure}

Figure \ref{fig:summaryfisher} summarizes our maximum achievable precision in the best case scenario of no measurement uncertainties (see Appendix \ref{sec:adamerrs} for noisy mock tests). We focus only on the default progression corresponding to the top three rows of Figure \ref{fig:adamcoverage8}. With $\smhm$ as the only constraint, $A_E$ is pinned down relatively well but many of the other parameters have large Fisher uncertainties. This happens because \adam\ frequently cannot pin down the true $\tdep$ parameters. Progressively adding $\fgas$ and ISM MZR decreases the uncertainties around $\thetamap$ as given by the diagonals of the Fisher matrix. 

We also consider the bias (accuracy) which we define as the difference between $\theta_{\rm MAP}$ and $\theta_{\rm true}$ normalized by $\sigma_{\rm Fisher}^{\rm true}$. The distribution of these normalized Fisher residuals is centered on zero and generally falls within $\pm1\sigma$ of the true minimum. Note that it is difficult to get exactly to the bottom of the true minimum, especially when the loss landscape is steep, and this requires fine-tuning of the optimizer which is beyond the scope of (and unnecessary for) this paper. Overall, we are unbiased in the noise-free case and, as we will show in the next subsection, also when including observational uncertainties.

The third row of Figure \ref{fig:summaryfisher} investigates the optimization failure rate as a function of included constraints. We define failure as \adam\ ending at a saddle point from all three initializations. Thus the Hessian would not be invertible and the Fisher matrix cannot be computed to give parameter uncertainties and degeneracies. The failure rate defined in this way is surprisingly high at $\sim90\%$ when $\smhm$ is the only constraint. Again, this happens because without additional constraints from $\fgas$, \adam\ cannot identify the true $\tdep$ parameters so even if it has found the correct value of, e.g., $A_E$, we are still likely at a saddle point in the broader high-dimensional landscape. As we add more constraints, the failure rate drops dramatically and is nearly zero when including both $\fgas$ and the ISM MZR. This emphasizes the need to supplement \texttt{adam} with methods like HMC that map the full posterior.

Lastly, in the bottom row of Figure \ref{fig:summaryfisher}, we compute the condition number $\kappa$ of the Fisher matrix, which quantifies how ``stretched out'' the Fisher ellipsoid is locally around the $\thetamap$ found by \adam. A smaller $\kappa$ denotes a more spherical, less degenerate local posterior whereas a large $\kappa$ indicates that the posterior is strongly stretched out along certain parameters compared to others.\footnote{Specifically, $\kappa$ is the ratio of the largest to smallest eigenvalue of the Fisher matrix. The individual off-diagonal entries of the Fisher matrix encode detailed information about correlations between specific parameters whereas $\kappa$ is a simple summary statistic of the overall ellipticity of the $N$-dimensional Fisher ellipsoid.} Note that we deliberately defined our free parameters and state variables to all be of order unity which simplifies the interpretation of $\kappa$. Figure \ref{fig:summaryfisher} reveals that $\kappa$ decreases by a factor of $\sim10$ as we add more constraints. This can be understood as degeneracies being weakened or even broken as additional quasi-observables provide more information about model parameters. There is not much difference in $\kappa$ between including or excluding the ISM MZR at least in this best case of assuming no measurement uncertainties. This plateau likely reflects the strong degeneracy between parameters like $\etaM$ and $\etaZ$ that may require additional data beyond these three $z=0$ constraints alone to break.

\section{Application to data}\label{sec:fitdata}
In this section, we combine the different quasi-observables described in Section~\ref{sec:data} to infer \texttt{sapphire} parameter posteriors and interpret the underlying physics.

\subsection{Posterior predictive checks}

\begin{figure*}
\centering
\includegraphics[width=\hsize]{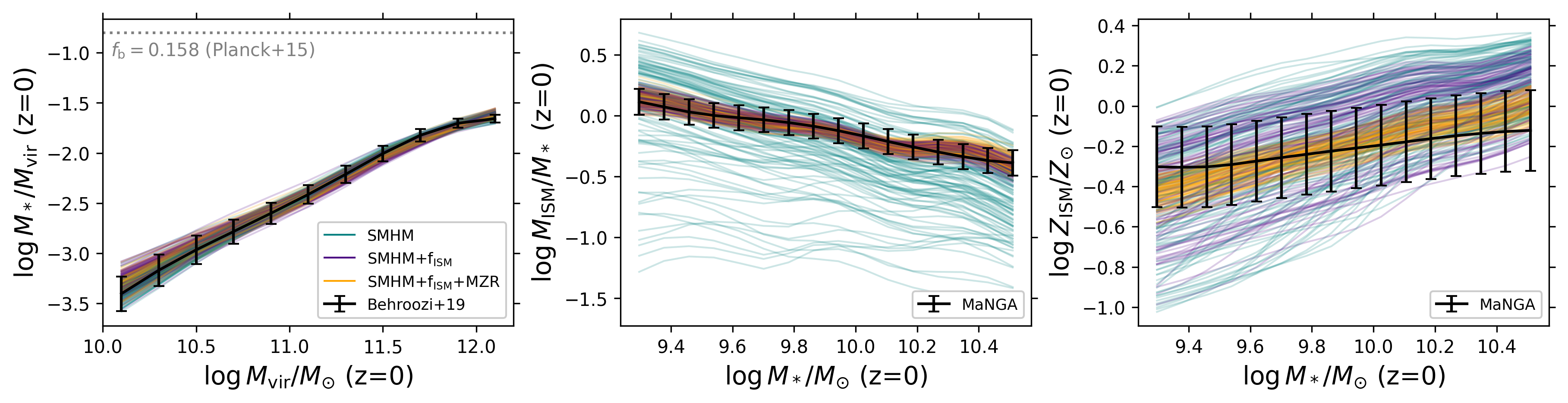}
\caption{Posterior predictive checks from fitting different combinations of the $z=0$ SMHM relation (left panel), ISM gas fractions (middle panel) and ISM MZR (right panel). In every panel, the black lines are for the data and there are three different sets of 100 colored lines which represent random draws from the posterior of a particular data combination (see legend). Note how the posterior draws are much tighter than the prior predictive checks from Figure \ref{fig:priorpreds}. When fitting to the SMHM relation alone (teal lines), large variations are allowed for the model $f_{\rm ISM}$ and MZR. Combining SMHM and $f_{\rm ISM}$ (indigo lines) leads to tight fits for those two, but posterior predictive checks for the MZR tend to be higher than observed. It is possible for \texttt{sapphire} to simultaneously fit all three relations together (orange).}
\label{fig:obspost}
\end{figure*}

Figure \ref{fig:obspost} starts off by showing posterior predictive checks from fitting different combinations of the data. Whereas the prior predictive checks from Figure \ref{fig:priorpreds} showed much larger variations in the model relations compared to the data, the posterior predictive checks are more tightly constrained. It is not surprising that, when we leave one or more constraints out, the posterior predictives for those observables are wider than allowed by the data. For example, fitting SMHM alone does not automatically constrain the $f_{\rm ISM}$ and the ISM MZR predictions to the range of the data. This implies that there are multiple star formation and stellar feedback scenarios consistent with the same $z=0$ SMHM relation. Combining SMHM with $f_{\rm ISM}$ leads to many MZR realizations higher than observed. Nevertheless, it is encouraging that when we fit to all three constraints together, \texttt{sapphire} is able to simultaneously match those relations.\footnote{Our predicted MZR seems steeper than observed. However, as described in Appendix \ref{sec:manga}, systematic errors from different metallicity indicators make the observed slope highly uncertain and motivate our additional HMC forecasts with controlled data perturbations below.} The agreement with the data is essential because it means that we can analyze the parameter posteriors to interpret the physics implied by our dynamical model, which we turn to next.

\subsection{Joint posteriors for astrophysical parameters}

\begin{figure*}
\centering
\includegraphics[width=\hsize]{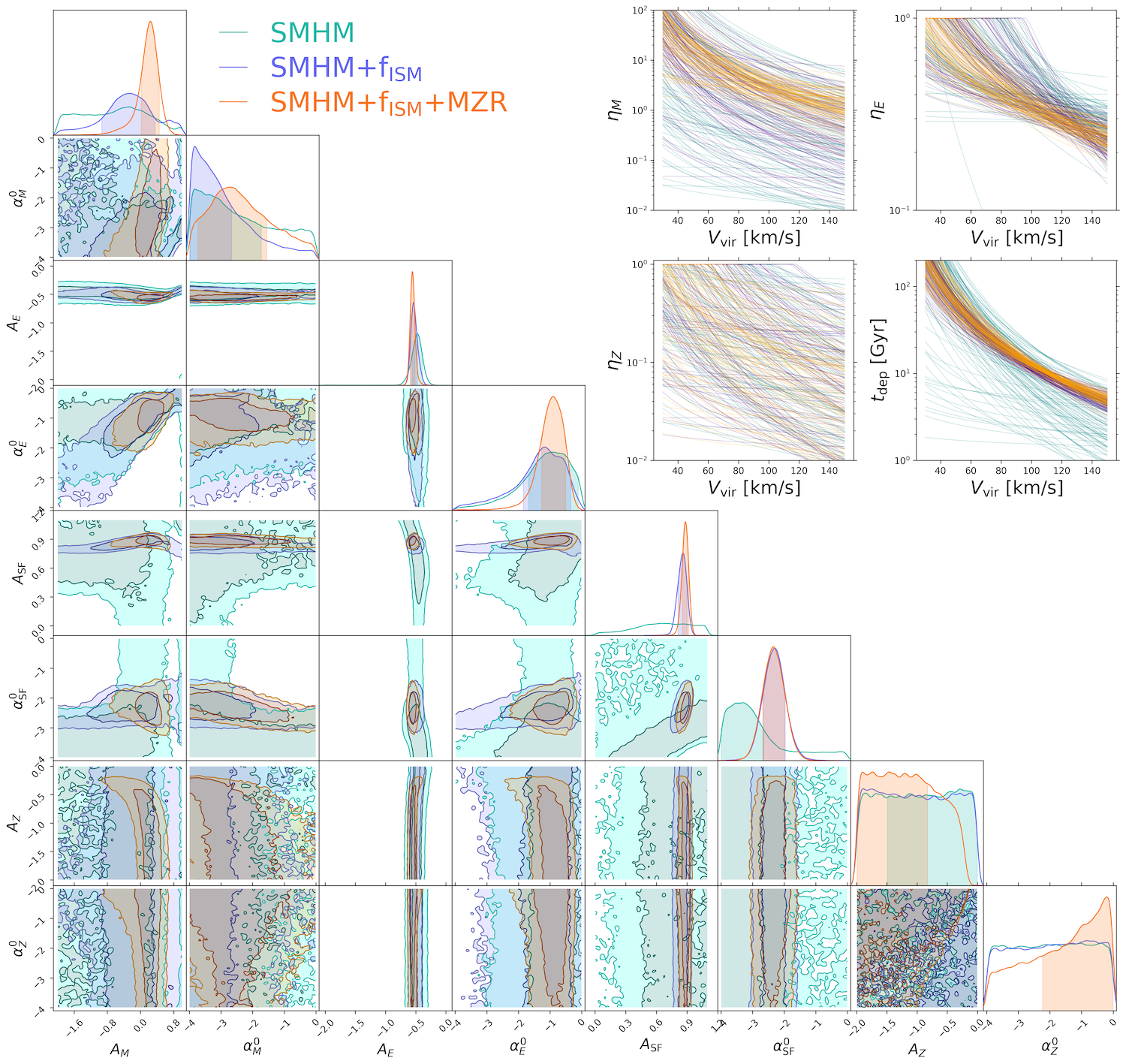}
\caption{Joint posterior for astrophysical parameters when fitting to different combinations of data. The color scheme is the same as in the previous Figure \ref{fig:obspost}: teal for SMHM only, indigo for SMHM and $f_{\rm ISM}$, and orange for all three together. The top-right panels show power laws corresponding to 100 random draws from each posterior. The most tightly constrained parameter is the energy loading power law amplitude $A_E$. When fitting to only SMHM, many other parameters are poorly constrained. Adding $f_{\rm ISM}$ and ISM MZR progressively tightens the posteriors, though many still remain broad.}
\label{fig:obscorner}
\end{figure*}

\begin{table}[h!]
\begin{tabular}{|c|c|c|c|}
$\theta$ & $p_{16}$ & $p_{50}$ & $p_{84}$ \\\hline
$A_M$ & -0.013 & 0.218 & 0.404 \\
$\alpha_M^0$ & -3.313 & -2.420 & -1.279 \\
$A_E$ & -0.590 & -0.554 & -0.515 \\
$\alpha_E^0$ & -1.325 & -0.934 & -0.588 \\
$A_{\rm SF}$ & 0.851 & 0.884 & 0.914 \\
$\alpha_{\rm SF}^0$ & -2.672 & -2.342 & -1.989 \\
$A_Z$ & -1.768 & -1.214 & -0.626 \\
$\alpha_Z^0$ & -2.992 & -1.441 & -0.402 \\
\end{tabular}
\caption{Parameter posteriors from fitting all three $z=0$ scaling relations together (corresponding to the orange curves in Figures \ref{fig:obspost} and \ref{fig:obscorner}). We provide the 16, 50 and 84 percentiles to summarize each marginal parameter posterior. The HMC chains used to create this table are available on the \texttt{sapphire} GitHub repository.}
\label{tab:post}
\end{table}

Figure \ref{fig:obscorner} shows joint posteriors for astrophysical parameters when fitting different combinations of data (corresponding to the previous Figure \ref{fig:obspost}). One immediate takeaway is that the amplitude of the energy loading power law is tightly constrained to $A_E\sim-0.5$, implying $\eta_E\sim0.3$ for MW-mass halos and rising to $\eta_E\sim0.4-1.0$ for lower-mass halos, depending on the slope $\alpha_E$ which is less well constrained. The tight posterior on $A_E$ is not surprising since our Jacobian analysis revealed that it is the parameter that state variables have the highest sensitivity to, including $M_*$ alone (Figures \ref{fig:jacobian} and \ref{fig:jacall}). 

However, when fitting to SMHM alone, most other parameters are poorly constrained. For example, the metal loading parameter posteriors are effectively the same as our assumed uniform priors. The mass loading power law amplitude posterior is also unconstrained, though there is a slight preference for a very steep slope $\alpha_M<-2$. Some weakly constrained parameters show complicated degeneracies. For instance, the amplitude and slope of the $t_{\rm dep}$ power law are positively correlated: higher $A_{\rm SF}$ corresponds to a shallower slope. This makes sense since the $z=0$ SMHM relation encodes only the integrated star formation history, and many different $t_{\rm dep}$ power laws can reproduce this depending on whether star formation is efficient at early or late times. Breaking that degeneracy between $A_{\rm SF}$ and $\alpha_{\rm SF}$ requires adding an independent constraint such as SFRs or ISM gas fractions.

Indeed, when we add $f_{\rm ISM}$ as a second constraint, the $t_{\rm dep}$ power law amplitude and slope posteriors become tighter and more localized. The posterior of the mass loading power law amplitude also begins to take shape: it is multimodal with a large, broad peak at $A_M\sim0$ (corresponding to a weak mass loading solution) as well as a second mode at $A_M\sim0.8$ (corresponding to more ejective feedback).\footnote{As explained in Subsection~\ref{sec:hmc}, we verified that our \citet{gelman92} statistic is $\sim1$ so this multimodality is not due to different chains getting stuck in separate minima. As described in Appendix \ref{sec:numerics}, we also find the same posteriors using a GPU-native gradient-free affine-invariant ensemble sampler \citep[from \texttt{numpyro};][]{numpyro}.} The energy loading and metal loading posteriors remain unchanged. 

When we add the ISM MZR as a third constraint, the mass loading amplitude posterior becomes tighter at $A_M\sim0.2$, consistent with relatively weak ejective feedback. The metal loading is still largely unconstrained, likely due to the large uncertainties on the ISM MZR, though there is a slight preference for low amplitude and shallow slope. The other posteriors remain similar to the case of jointly fitting SMHM and $f_{\rm ISM}$. Table \ref{tab:post} summarizes our posteriors for this case of combining all three observables.

This leaves us wondering: what do these posteriors mean? Why do our observational constraints correspond to the posterior distributions that they do? By how much would the posteriors improve if the data uncertainties were lower? Understanding this requires additional inference runs with controlled data variations, which we turn to next.

\begin{figure*}
\includegraphics[width=\hsize]{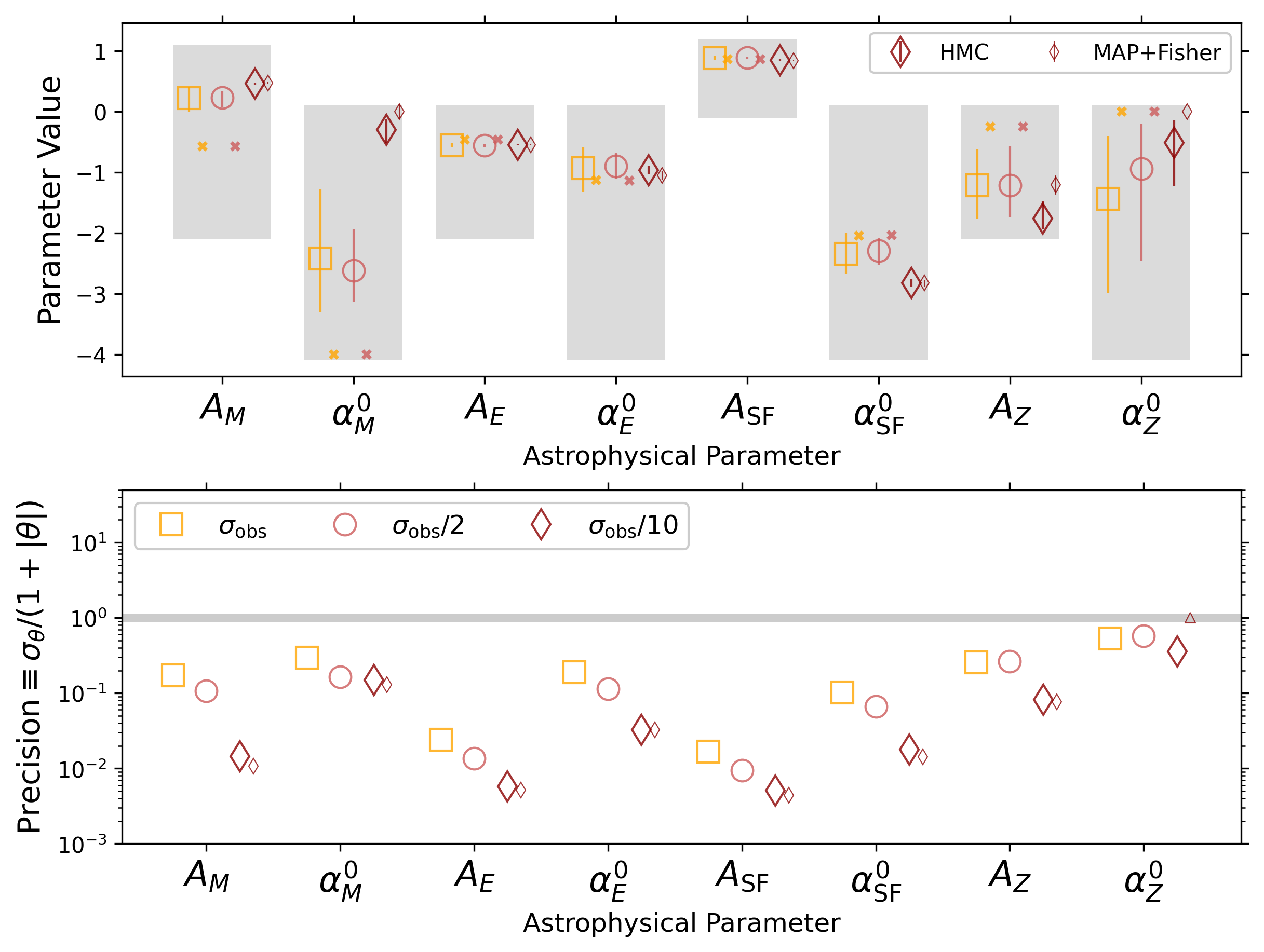}
\caption{Forecasts for improved precision on our baseline model parameters assuming the observational errors are cut by a factor of two or ten including systematics. Top panel: 16-50-84 percentiles of the HMC posterior (large symbols) and the lowest-loss MAP and local Fisher uncertainty out of ten gradient descent trajectories (small symbols). For the latter, we show a cross if all ten \adam\ trajectories ended at a saddle point, which happens for the fiducial and half uncertainty cases. The gray shaded area shows the uniform prior for each parameter. Bottom panel: precision defined as the ratio of the posterior width to the median HMC posterior or MAP value. The precision generally improves with decreasing observational error as expected. Some parameters benefit enormously, e.g., mass loading precision increases by $\sim10$ if the data has ten times lower uncertainties including systematics. We use an upward triangle for the \adam\ precision on $\alpha_Z^0$ since the MAP is near the prior boundary and the Fisher uncertainty is unreliable. Of course, this is a lower limit on the precision since the baseline model has many other parameters that we have not yet allowed to vary, our baseline model is just one of many that must be explored, and there are many more observables left to incorporate. But this gives a preview of what \texttt{sapphire} will enable in a uniquely scalable way.}
\label{fig:forecasts}
\end{figure*}

\begin{figure*}
\centering
\includegraphics[width=\hsize]{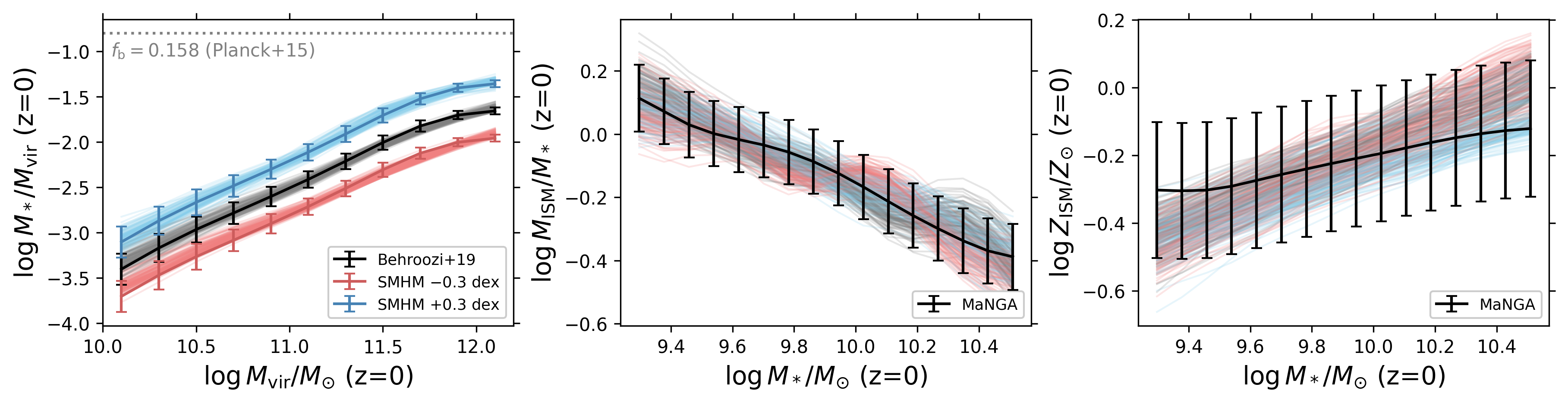}
\includegraphics[width=\hsize]{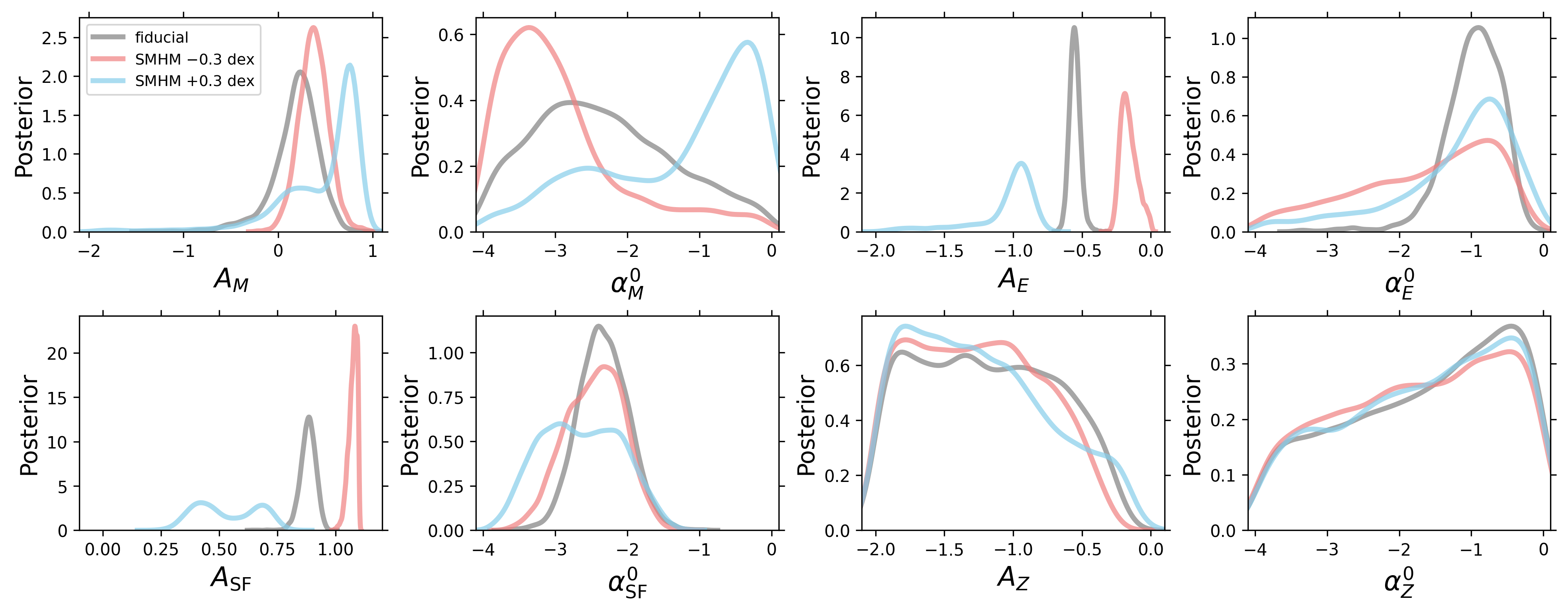}
\caption{Posterior predictive checks (top panels) and marginal parameter posteriors (bottom panels) illustrating the effect of shifting the normalization of the SMHM relation. Gray curves reflect the original data whereas blue (red) curves are from shifting the SMHM relation by $+0.3$ ($-0.3$) dex. With all else fixed, increasing the SMHM relation corresponds to lower energy loading, faster depletion time, more mass loading and shallower $\alpha_M$. In other words, producing more stars requires less feedback and/or higher SFRs per unit gas mass, and the opposite is true if the SMHM normalization were decreased.}
\label{fig:shiftsmhm}
\end{figure*}

\subsection{Forecasting the impact of smaller errors}\label{sec:forecasts}
One way to appreciate the power of the \texttt{sapphire} approach is to forecast how the precision on physical parameters improves with tighter observational errorbars. This can quickly become unwieldy so for simplicity, here we restrict to scaling the existing observational errorbars down by a constant factor of two or ten, including systematics. We only consider the case of fitting all three datasets together since we showed earlier that is necessary. We also compare HMC and gradient descent, where for the latter we take the lowest-loss MAP from ten \adam\ trajectories. Given the flatness of the loss landscape along the mass and especially metal loading parameter directions, \adam\ always converges to a saddle point so we cannot obtain a local Fisher uncertainty, except in the case where we drop the observational errors by a factor of ten. For these and more generally, the global HMC posterior should be taken more seriously. 

Figure \ref{fig:forecasts} shows the precision, defined as the HMC posterior width (half of the 16-84 percentile range) or Fisher uncertainty normalized by the median or MAP value. Naturally, this precision improves as the observational error decreases. For some parameters like $A_M$, the boost is large: ten times lower uncertainties on our three observed $z=0$ scaling relations corresponds to $\sim10\times$ smaller uncertainties on $A_M$. For others like $\alpha_M^0$, the precision decreases only by a factor of a few. In general, percent-level precision on the data seems to correspond to percent-level precision on some parameters, with sub-percent precision possible for $A_E$ and $A_{\rm SF}$. In the tightest observational error forecasts, \adam\ and HMC parameter values and their precision agree remarkably well, increasing our confidence in using the much more cost-effective gradient descent method for future \texttt{sapphire} applications. Of course, these precisions should only be taken as illustrative lower limits. Once we start leaving more parameters free and/or transitioning between different, more complex models, this precision will get worse but how that happens has not been charted before in galaxy formation. With the unified \texttt{sapphire} framework we will be in a position to rigorously explore these questions.

\subsection{Interpreting the posterior with data perturbations}\label{sec:perturbs}
Instead of changing the errorbars, let us now apply systematic shifts to one $z=0$ relation at a time, leaving the other relations and all uncertainties fixed. By re-running HMC multiple times, we can derive even more intuition about our fiducial posterior and the robustness of our interpretation of it, at least within the context of this one baseline model. Again in this subsection, we will always fit all three observational constraints together while perturbing only one relation at a time. Since this is a proof of concept demonstration, we will restrict ourselves to simple, constant normalization shifts.

\subsubsection{Perturbing the SMHM relation}

Figure \ref{fig:shiftsmhm} compares the fiducial posterior based on the original data to posteriors obtained after shifting the SMHM relation either up or down by $0.3$ dex. With the shifted SMHM relations, we are still able to match the other two relations reasonably well but with different parameter posterior distributions. If the SMHM relation had a higher normalization, that requires a lower energy loading amplitude, faster depletion time amplitude, higher mass loading amplitude or shallower mass loading slope. The other parameters remain largely unchanged. This makes sense: feedback has to be less efficient and/or the SFR per unit gas mass must be higher. The former can be achieved by launching lower specific energy winds, i.e., lower $A_E$ and higher $A_M$. If instead the SMHM relation had a lower normalization, feedback would have to be stronger (higher $A_E$) and/or the star formation efficiency would need to be lower (longer $A_{\rm SF}$). The fact that we can make intuitive sense of the changes in the posteriors demonstrates the potential of \texttt{sapphire} for interpretable precision inference under different astrophysical scenarios.

\subsubsection{Perturbing the $f_{\rm ISM}-M_*$ relation}

\begin{figure*}
\centering
\includegraphics[width=\hsize]{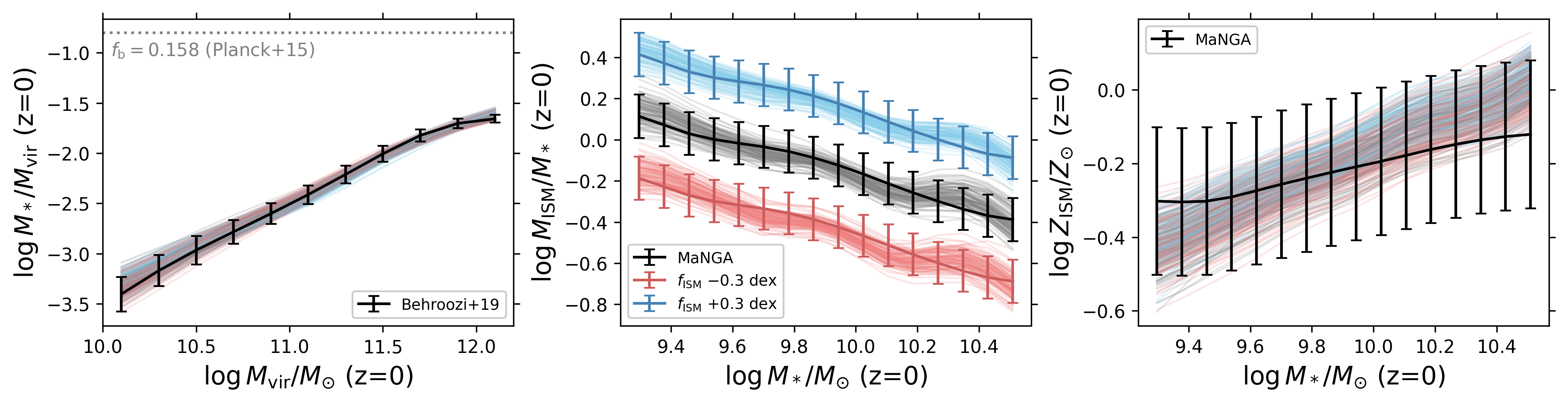}
\includegraphics[width=\hsize]{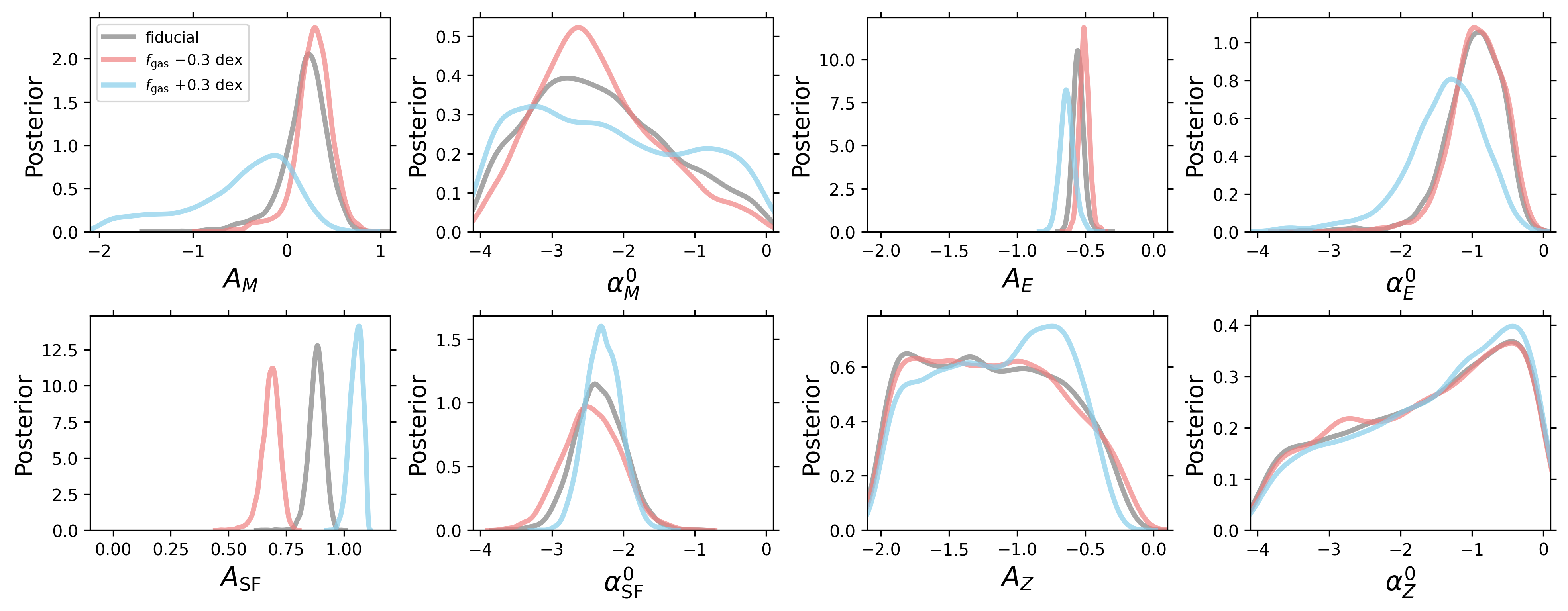}
\caption{Same as Figure \ref{fig:shiftsmhm} but now shifting only the $f_{\rm ISM}-M_*$ scaling relation. Higher (lower) gas fractions map primarily to higher (lower) depletion time power law amplitudes. Higher gas fractions also correspond to lower mass loading (less gas ejection) and lower energy loading (enhanced CGM cooling and gas accretion).}
\label{fig:shiftfgas}
\end{figure*}

Figure \ref{fig:shiftfgas} similarly shows the effect of perturbing the $f_{\rm ISM}-M_*$ relation up or down, leaving the other two fixed. Again, we are able to reproduce all three relations simultaneously as evidenced by our HMC posterior predictive checks. If gas fractions were overall higher, that primarily maps to longer depletion times and lower mass and energy loadings. This implies the gas takes longer to be converted into stars, less gas is ejected, and there is less CGM heating (and hence more gas accretion). Since we enforced a fit to the SMHM relation, the energy loading slope becomes somewhat steeper to prevent excess CGM cooling and gas accretion in dwarfs. If instead we decrease the gas fractions relative to the original observed relation, all parameters remain largely unchanged except for $A_{\rm SF}$, which shifts to lower values, i.e., faster depletion times which consume more of the gas.

\subsubsection{Perturbing the ISM MZR}

\begin{figure*}
\centering
\includegraphics[width=\hsize]{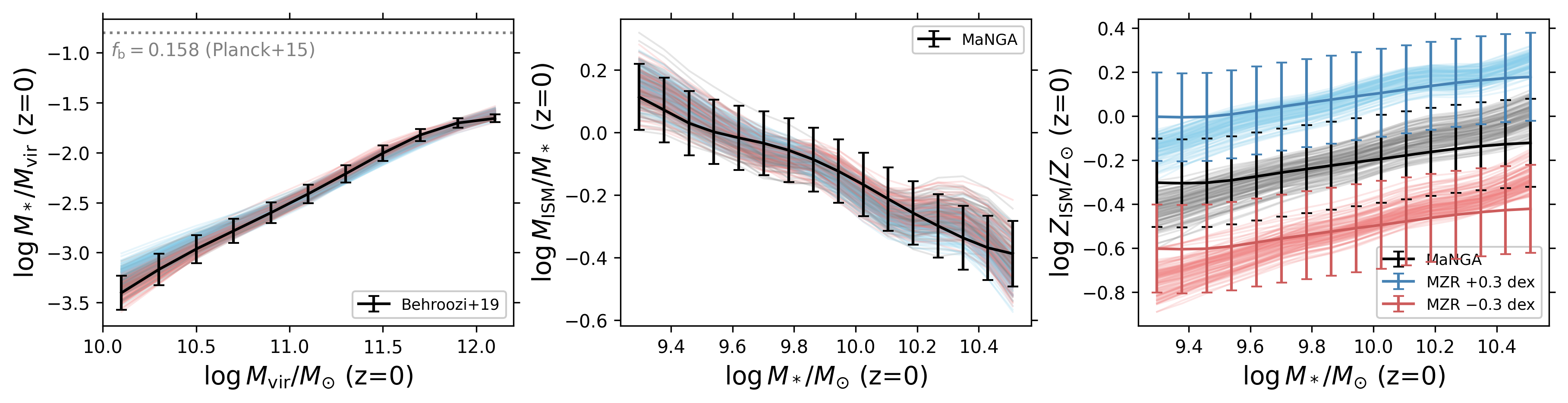}
\includegraphics[width=\hsize]{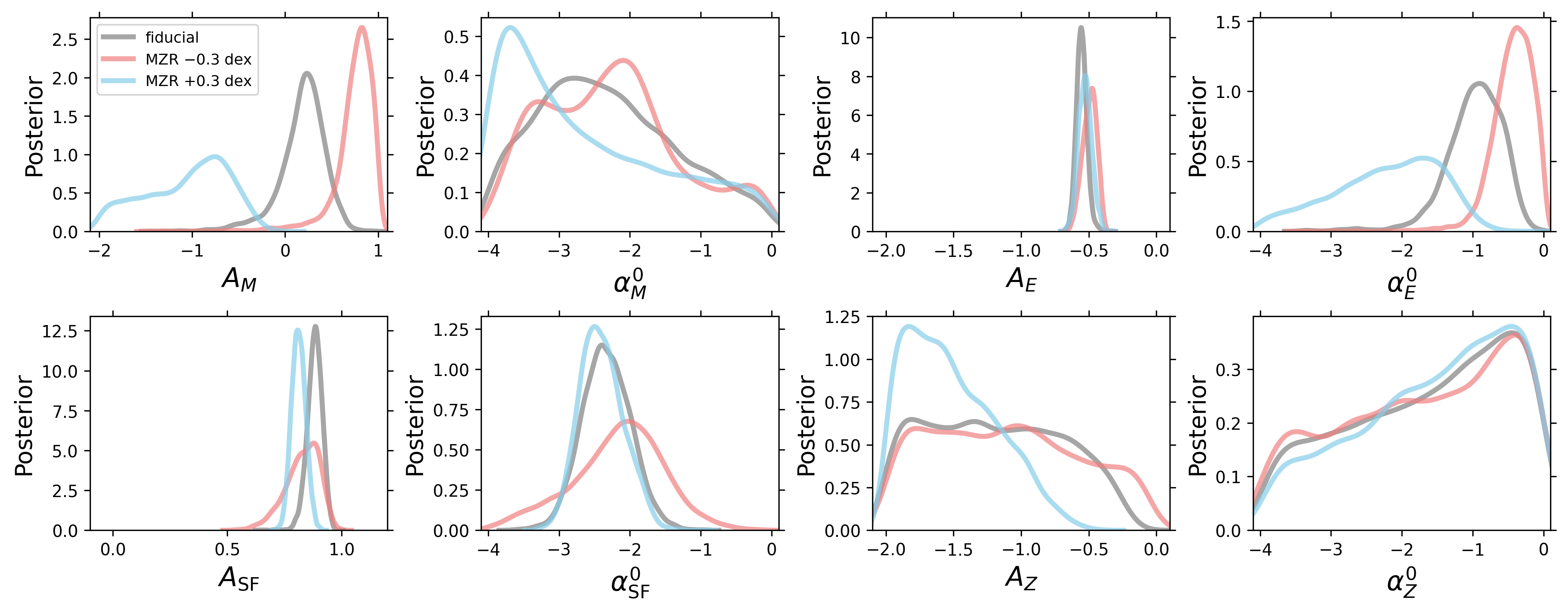}
\caption{Same as Figure \ref{fig:shiftsmhm} but now shifting only the ISM MZR. Increasing the MZR normalization causes a substantial decrease in the mass and metal loading since galaxies need to retain more of their metals. The energy loading power law also becomes steeper to prevent excess CGM cooling in lower mass halos which now retain more of their ISM gas due to $\eta_M$ decreasing. For the downward shift, we also turn off pre-enrichment of halo inflows since otherwise the prior predictive checks in Figure \ref{fig:priorpreds} do not extend much below MaNGA. This scenario primarily maps to a larger mass loading, which ejects more metals via entrainment in ISM outflows.}
\label{fig:shiftmzr}
\end{figure*}

Lastly, Figure \ref{fig:shiftmzr} shows the effect of shifting the ISM MZR on the posterior. Relative to the fiducial observed relation, the higher normalization MZR induces a significant decrease in the mass and metal loading power law amplitudes. This makes sense: if galaxies have higher metallicities, then they must retain more of their metals which constrains both the metals entrained from the ISM in mass-loaded outflows as well as the metal loading of pure supernova ejecta (see Equation \ref{eqn:etaZ}). The energy loading power law also has a steeper slope. This follows from the decreased mass loading: low-mass halos retain more of their ISM gas, so the energy loading needs to increase more steeply in the dwarf regime to prevent excess CGM cooling and further gas accretion. The gas depletion time also becomes slightly faster perhaps indicating a degeneracy with the metal yield. The other parameter posteriors are largely unchanged. For the downward shift, we further turn off pre-enrichment of halo inflows since without those, our prior predictive checks in Figure \ref{fig:priorpreds} do not extend much below MaNGA at the MW-scale. This primarily maps to an increase in the mass loading amplitude, causing more metals to be ejected via entrainment in ISM outflows. The metal loading is unchanged perhaps because we forced the overall system to now also receive less metals from halo inflows. The energy loading slope also becomes shallower. Since MZRs suffer from large systematic uncertainties, this exercise shows the utility of \texttt{sapphire} not only for interpreting baseline posteriors, but also assessing the impact of those systematic errors.

\subsection{Additional Posterior Predictive Distributions}

\begin{figure*}[h!]
\centering
\includegraphics[width=0.95\hsize]{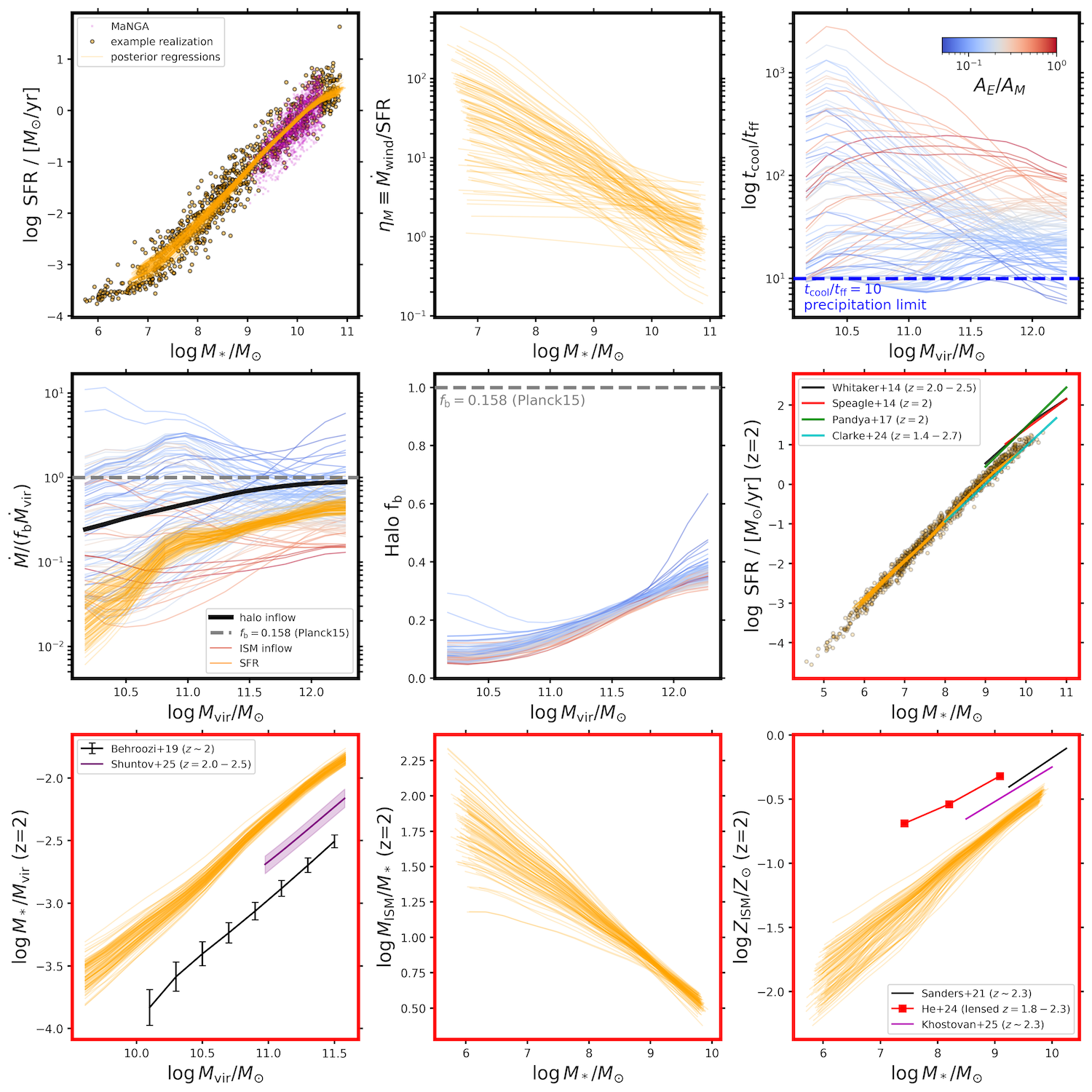}
\caption{Posterior predictive distributions for a few additional quantities that we did not fit to. Panels with black (red) borders are for $z=0$ ($z=2$). As in previous figures, orange curves show Nadaraya-Watson regressions for 100 random draws from the posterior with fitting all three $z=0$ scaling relations. In some panels, we color-code these curves by the ratio of $A_E/A_M$ (colorbar range shown in top-right panel). \textit{Top-left:} the $z=0$ SFMS predicted by the model agrees with MaNGA (magenta points). The scatter also roughly matches as indicated by one example model realization (orange points). \textit{Top-center:} mass loading factor decreases as a function of stellar mass but shows a large spread between realizations. \textit{Top-right:} the ratio of cooling time to free-fall time in the inner halo spans a large range but tends to decrease with halo mass. On average, this ratio is higher in realizations with larger $A_E/A_M$, and almost always exceeds the $t_{\rm cool}/t_{\rm ff}=10$ precipitation limit. \textit{Middle-left:} halo gas inflow (black), ISM accretion (colored) and SFR (orange) all normalized by DM halo growth rate. Dwarfs experience preventative feedback for halo inflows. ISM inflows can exceed the cosmic halo baryon accretion rate (gray dashed) if SN feedback is highly mass-loaded. \textit{Middle-center:} halo baryon fractions are below the cosmic value due to CGM over-pressurization. \textit{Middle-right:} $z=2$ model SFMS tends to be lower than observed \citep{whitaker14,speagle14,pandya17}, though systematics can be large \citep{clarke24}. \textit{Lower-left:} $z=2$ model SMHM is higher than empirically inferred \citep{behroozi19,shuntov25}. \textit{Lower-center:} $z=2$ model ISM gas fractions vary by $\sim1$ dex on average for low-mass halos between posterior realizations. \textit{Lower-right:} the $z=2$ model ISM MZR tends to be lower than observed \citep{sanders21,he24,khostovan25}. Many of these discrepancies suggest the model complexity needs to be increased.}
\label{fig:postextras}
\end{figure*}

In this paper, we have only fit to various combinations of the $z=0$ SMHM relation, ISM gas fractions and ISM MZR as a proof of concept, but the baseline time-dependent model from \citet{pandya23} makes self-consistent predictions for many other quantities. Figure \ref{fig:postextras} shows predictions for a few additional example quantities at both $z=0$ and $z=2$ using random draws from the posterior based on fitting all three $z=0$ scaling relations together. In a way, we are ending where we began: similar to the prior predictive checks from Figure \ref{fig:priorpreds}, the agreements and discrepancies we show here reflect the starting point for future iterations of model improvement and inference.

\subsubsection{$z=0$ SFMS, feedback and CGM thermodynamics}
Both the mean and scatter of the $z=0$ model star-forming main sequence (SFMS) agree quite well with the H$\alpha$-derived SFMS for MaNGA galaxies \citep{sanchez22}. This is perhaps not surprising since we included $z=0$ ISM gas fractions as a constraint, but since SFRs are easier to obtain than multi-phase gas observations, the two may be interchangeable. We do not attempt to compile SFRs at lower masses and/or quenched fractions but that is an obvious direction for future work. Closely related is the mass loading factor which decreases as a function of stellar mass but shows a large spread between posterior realizations. At the MW-scale, $\eta_M$ is at most of order unity whereas for $\log M_*/M_{\odot}\sim8$ dwarfs, it can vary by more than an order of magnitude from $\eta_M\sim10-100$.

As emphasized by \citet[][see also \citealt{pandya20,faerman22,carr23,voit24a,voit24b,messere26}]{pandya23}, there is additional information about model parameters in both gas flows and CGM observables. As an example, we show how various model quantities depend on the ratio $A_E/A_M$ as a proxy for SN wind specific energy. On average, the ratio of the cooling time to freefall time in the inner halo tends to be larger for lower-mass halos and is correlated with $A_E/A_M$ such that higher specific energy winds increase this ratio. Note that we do not include non-thermal pressure support such as turbulence driven by the full \citet{pandya23} model, which would affect this ratio.

Figure \ref{fig:postextras} also shows the ratio of halo inflow, ISM accretion and SFR normalized by the DM halo growth rate. The halo inflow rate drops below the cosmic value \citep[$f_{\rm b}=0.158$;][]{planck15} for dwarfs due to preventative feedback \citep{pandya20,pandya23,carr23}. The star formation rate is only a few percent of the instantaneous DM halo growth rate for dwarfs but increases towards $\sim50\%$ for MW-mass halos. The ISM accretion rate naturally correlates with $A_E/A_M$ and can exceed the halo accretion rate for highly mass-loaded realizations. Lastly, the significant over-pressurization of our purely thermally-supported CGM causes the halo baryon fractions to be significantly lower than the cosmic value, even at MW scales. \citet{pandya23} showed that it is possible to construct MW-mass halos containing their fair share of baryons with a roughly virial temperature CGM, but more work is needed to reconcile our posterior results with those. 

\subsubsection{$z=2$ scaling relations}
Figure \ref{fig:postextras} additionally shows posterior predictions for the SFMS, SMHM relation, ISM gas fractions and ISM MZR at $z=2$ as one other example redshift. Our SFMS tends to be lower than observed \citep{whitaker14,speagle14,pandya17} but there can be systematics of order this discrepancy due to different selection functions and calibrations even for the same SFR indicator, dust attenuation laws, etc. \citep[e.g.,][]{shivaei15,leja22,clarke24}. Our $z=2$ SFMS also has noticeably less scatter than the $z=0$ one; e.g., for galaxies with $\log M_*/M_{\odot}=9-9.5$, the standard deviation of our model SFRs is $\sim0.33$ dex at $z=0$ and $\sim0.18$ dex at $z=2$. This is likely lower than and in conflict with estimates at high-redshift and may indicate that we need additional sources of scatter beyond halo mass accretion histories. 

Our $z=2$ SMHM relation is systematically higher than both \citet{behroozi19} and the more recent JWST-based \citet{shuntov25} by $\sim0.5$ dex and $\sim0.25$ dex, respectively. This likely indicates that our model galaxies are forming their stars too fast and too early. We show predictions for ISM gas fractions at $z=2$ even though current facilities lack the sensitivity to detect the bulk of the neutral atomic ISM in individual galaxies beyond $z\sim0.5$ \citep{fernandez16,tacconi20}. Our model predicts $\sim1$ dex variation in the mean gas fraction of low-mass dwarfs at $z=2$. While this may not be pinned down any time soon in terms of HI, there are deep surveys with, e.g., ALMA for molecular gas and other datasets that could be incorporated for \texttt{sapphire} calibration purposes \citep[e.g.,][]{popping15,guo23}. Lastly, our ISM MZR at $z=2$ is lower than observed by several tenths of a dex. This includes comparisons to both pre-JWST data \citep[e.g.,][]{sanders21} as well as JWST data which go to lower masses \citep[e.g.,][]{he24,khostovan25}. 

Of course, we did not fit to any of these extra quantities but it is interesting that fitting to $z=0$ does not automatically reproduce the high redshift data. Our model has additional redshift dependent terms that may need to take on non-zero values to fit this data. The fact that some of the observed relations are different from each other underscores the need for our approach to map systematics back to interpretable model variations as in Subsection~\ref{sec:perturbs}. It is possible that our baseline model cannot simultaneously accommodate the data, in which case \texttt{sapphire} is uniquely positioned to discover interpretable corrections to our assumed equations (Figure \ref{fig:modules}).

\section{Discussion}\label{sec:discussion}
Here we discuss the astrophysical implications of our results, place our work in a greater context, and outline current limitations and caveats.

\subsection{Implications for galaxy population evolution}
Our results are consistent with the idea that galaxies self-regulate their star formation via preventative rather than ejective feedback \citep[e.g.,][]{lu15,lu17,pandya20,carr23,voit24a,voit24b}. There are at least two channels for this: (1) reducing halo gas accretion below the cosmic baryon fraction, and (2) over-pressurizing the CGM so that it cools and accretes more slowly into galaxies. On the first point, our model explicitly suppresses some fraction of halo accretion preferentially in dwarfs (middle-left panel of Figure \ref{fig:postextras}). Although we fixed that parameter in this paper following \citet{pandya23}, in principle it can and should be inferred from data, perhaps folding in IGM observables. As for CGM over-pressurization, we infer relatively low mass loading ($\eta_M\lesssim1$ at the MW-scale and $\eta_M\sim10$ for classical dwarfs; Figures \ref{fig:obscorner} and \ref{fig:postextras}) and high energy loading ($\eta_E\sim0.2$ for MW-mass halos and $\eta_E\approx0.5-1$ for dwarfs; Figure \ref{fig:obscorner}). Together, these imply high specific energy winds, which increase the CGM cooling time, decrease the ISM accretion rate, evacuate CGM baryons and lead to low halo baryon fractions (top-right, middle-left and middle-center panels of Figure \ref{fig:postextras}). The sensitivity of the scaling relations to this ratio of energy to mass loading is made clear by our inference with controlled normalization shifts applied to the data; e.g., a higher SMHM relation requires higher $A_M$ and lower $A_E$ (in other words, lower specific energy feedback; Figure \ref{fig:shiftsmhm}). Our low mass loadings are in line with recent observational determinations \citep[e.g.,][]{heckman15,mcquinn19,kadofong24} but comparison to CGM data is needed to thoroughly test the preventative feedback interpretation.\footnote{One future direction is to incorporate a new module for \texttt{ExpCGM}, which generalizes the \citet{pandya23,carr23} model to keep track of the baryons that feedback has pushed beyond the virial radius of a halo. See \citet{voit24a,voit24b} and \url{https://gmvoit.github.io/ExpCGM/}} Interestingly, if the SMHM relation was the only constraint, the high mass loading solution is allowed (teal posterior in Figure \ref{fig:obscorner}), as is assumed or required by some large-volume simulations \citep[see Figure 13 of][]{mitchell20}, but this would predict much lower gas fractions than observed. The fact that including the ISM MZR, despite its large uncertainties, strongly localizes the mass loading posterior reaffirms that observables tracing chemical evolution can help break astrophysical parameter degeneracies. 

Our approach has the potential to provide additional constraints on assumed and/or emergent properties of galaxies in sophisticated hydrodynamical simulations. Many if not all simulations are currently benchmarked against the SMHM relation inferred from simple empirical approaches \citep{wechsler18}. The forthcoming inclusion of satellite and black hole feedback processes within \texttt{sapphire} will enable us to add large-scale galaxy clustering as a constraint and therefore derive plausible SMHM relations independently of subhalo abundance matching or halo occupation distribution models. Beyond that, we can also provide constraints on interpretable effective parameters such as mass, energy, metal loading and ISM depletion time. These constraints can then inform more refined numerical experiments to map out plausible physical scenarios consistent with the phenomenological constraints. For example, what freedom do we have with supernova clustering, thermalization efficiency, multi-phase gas partitioning, CGM and cosmic accretion geometry, etc. to reproduce the SMHM ratios of $M_{\rm vir}\sim10^{11}M_{\odot}$ dwarfs while enforcing $\eta_M\sim1$ and $\eta_E\sim0.5$? Addressing questions like these requires the ability to systematically explore parameter and model variations as well as mapping the stability of equilibria \citep[as argued by][]{pandya21,pandya23}. Fortunately, at least with the simple baseline model considered here, it seems that combining even a few $z=0$ scaling relations leads to relatively tight constraints on the assumed parameterizations of input physics.

Taking our results at face value (but see caveats in Subsection~\ref{sec:caveats} below), our inferred mass, energy and metal loadings are broadly consistent with what \citet{pandya21} and \citet{pandya23} found for the FIRE-2 simulations \citep[see also][]{muratov15,muratov17,anglesalcazar17,hafen19}. This agreement is not trivial since although we did fix a few other parameters to our FIRE-2 ``priors,'' our inference for loading factors and $\tdep$ are based on wide uniform priors. The functional forms for the loading factors and $\tdep$ in FIRE-2 required additional redshift dependence for both the amplitude and slope that we neglect here, but we mostly agree at the order of magnitude level. The lowest mass FIRE-2 dwarfs have $\eta_M$ approaching $\sim100$ if all outflowing material is considered rather than just the high specific energy wind component, so this is potentially a point of tension with our lower inferred $\eta_M\lesssim10$ for $M_{\rm vir}\sim10^{10}M_{\odot}$ ultrafaint halos. However, the top-middle panel of Figure \ref{fig:postextras} shows that $\eta_M\sim100$ is not ruled out for dwarfs, though it is unlikely. We would need to explore a wider variety of models including non-thermal partitioning \citep[e.g.,][]{martinalvarez26} before drawing stronger conclusions about consistency with different simulations. 

Switching to large-volume simulations, \citet{nelson19} report $\eta_M\sim3$ with no outflow velocity cut for MW-mass galaxies in the TNG50 simulations \citep{pillepich19}, which falls within our posterior range. Their $\log M_*/M_{\odot}\sim9$ dwarfs have $\eta_M\gtrsim10$ which is at the higher end of our posterior (top-middle panel of Figure \ref{fig:postextras}). \citet{oren25} recently presented measurements of the emergent energy loading for TNG100 \citep{nelson19b}. Their Figure 13 shows a different dependence of $\eta_E$ on halo mass than we infer in our Figure \ref{fig:obscorner}, though the flatness of their relation may in part be driven by black hole feedback which we do not yet include in \texttt{sapphire}. \citet{bennett24} also emphasize that the energy loading assumed by TNG at injection at the MW-scale is $\eta_E\sim1$, far above our inferred $\eta_E\sim0.3$. For EAGLE, \citet[][their Figure 13]{mitchell20} measured $\eta_M\sim1$ at the MW-scale and $\eta_M\lesssim10$ at the dwarf scale, which is consistent with our posteriors \citep[see also][]{wright24}. They also report $\eta_E\approx0.1-0.3$ without a strong dependence on halo mass (their Figure 5); this may be consistent with our $\eta_E$ posterior although we find a steep dependence on $V_{\rm vir}$. 

Returning to the SMAUG vision for multi-scale Bayesian SAMs originally laid out in \citet{pandya21thesis}, our \texttt{sapphire} approach may eventually be able to connect simulations and observations across scales. For example, how do we reconcile the lower mass loading factors predicted by small-scale suites of idealized simulations like TIGRESS \citep{kim20} with the larger loading factors required by traditional cosmological simulations and SAMs \citep[e.g.,][]{pandya20,pandya21,oren25}? Can we use our inference framework to bridge both global and local scaling relations \citep[e.g.,][]{motwani22}? The answers to at least some of these questions may lie in how the mass, energy and metals carried by multiphase winds couple to the CGM on larger scales \citep[e.g.,][]{fielding17,fielding20,smith24a,smith24b,bennett24,messere26}. Previous SAMs have already hinted that predictions may be very sensitive to the assumed fate of ejected gas \citep[e.g.,][]{henriques13,white15}. This motivated \citet{pandya21thesis} and \citet[][see also \citealt{carr23,voit24a,voit24b}]{pandya23} to try to self-consistently couple energy flows between galaxies and their CGM so that observables of the latter could be incorporated as additional constraints. With \texttt{sapphire} we are now in a position to accelerate the exploration and discovery of more sophisticated, yet still interpretable, multi-scale models of galaxy evolution beyond the simple \citet{pandya23} baseline.

\subsection{Differentiable GPU-accelerated galaxy evolution}
We are not the first to pursue numerical robustness, modular code and Bayesian inference for SAMs \citep[e.g., see][]{henriques09,bower10,lu11,benson12,croton16,lagos18,forbes19}. We are also not the first to explore the usage of automatic differentiation \citep[e.g.,][]{hearin21,hearin23,horowitz24,horowitz25,pandey25} and GPU parallelization \citep[e.g.,][]{schneider15cholla,ocvirk16,villasenor21,wibking22,alarcon25} for modeling galaxy formation. However, to the best of our knowledge, our work is the first to combine all of the above elements into a single framework for modeling galaxy populations as dynamical systems with interpretable equations and parameters (see again our Figure \ref{fig:modules}). This allows us to perform unprecedented local and global sensitivity analysis across parameter space, which helps reveal the most important model parameters and quasi-observables in a principled way. The ability to compute parameter gradients through the internals of our numerical model offers a fundamentally new way to assess the sensitivity of galaxy formation to uncertain astrophysics. Our work complements coarse parameter variations of existing simulation and SAM codes to generate training data for emulators \citep{bower10,villaescusanavarro21,jespersen22,perez23} and related generative modeling approaches \citep[e.g.,][]{alsing24,thorp25,nguyen25,pandey25b,pandey25c,nguyen26,deger26}. \texttt{sapphire} also sets the stage for data-driven corrections to the baseline dynamics using small neural networks to address model mis-specification in an interpretable way \citep[these are the $f_{\rm NN}$ terms in the center of Figure \ref{fig:modules} that we defer to a future paper; e.g.,][]{chen18,rackauckas20,kidger22,lanzieri22,horowitz25b}.

There will likely be challenges with continuing to maintain differentiability and interpretability as additional physical processes are added to \sapphire, but we argue that those difficulties will teach us about important, poorly understood regimes of galaxy formation. For example, we did not allow SNe to drive CGM turbulence in this paper but \citet{pandya23} showed that the dynamical system experiences a ``bifurcation'' (sudden change in equilibrium) when radiative cooling fails to balance heating. Should the model remain differentiable through the resulting phase transition from a cool, turbulent, early CGM to a warm-hot corona at low-redshift? Similarly, during galaxy mergers, would parameter gradients explode, and might the small scatter of scaling relations contain information to preserve gradient flow through such chaotic events \citep{genel19,keller19}? These cases all raise deep questions about the search for a single, effective, coarse-grained model that can describe galaxies through various transformations across cosmic time. On top of this, there is the fact that we do not observe the intrinsic physical state of galaxies but rather noisy, incomplete observables which requires forward modeling spectra \citep[e.g.,][]{conroy13,hearin23,synthesizer,synthesizer2}.

The combination of differentiability and GPU parallelization will be tremendously helpful for identifying a minimal set of state variables, evolution equations and astrophysical parameters to summarize galaxy evolution. For example, here we assumed simple deterministic power laws for our astrophysical parameters, but there may be additional dependencies beyond $V_{\rm vir}$ and/or other functional forms may have been equally capable of fitting the $z=0$ scaling relations. One way to get at this is to continue to build out a ``library'' of priors for \texttt{sapphire} by analyzing different simulations in the same way, perhaps using symbolic regression \citep[e.g.,][]{cranmer23,salim25,iyer25}. Another way is hierarchical Bayesian inference where we fit for scatter in the coarse-grained parameters assigned to each galaxy. This would introduce more hyper-parameters to be inferred but gradients can help us scale, together with new sampling methods \citep[e.g.,][]{behroozi25}. Together, these can help us map the most uncertain aspects of the state space and parameter space in which galaxies evolve. This interpretable SAM-guided analysis of simulations was started by \cite{pandya20,pandya21,pandya21thesis,pandya23} using the FIRE-2 suite \citep{hopkins18} as one proof of concept example and has recently been extended to TNG \citep{oren25}. With the full \texttt{sapphire} framework, we will now be able to generalize and bridge to data as well.

It is widely believed that galaxy formation modeling approaches span a continuum with an underlying dichotomy between empirical and physically-motivated methods \citep[e.g.,][]{wechsler18}. However, given the five grand challenges of galaxy formation that we outlined in the first paragraph of Section~\ref{sec:intro}, clearly every galaxy-scale simulation and model (including ours) must make choices for numerics, physics and statistics to confront noisy, incomplete observations of the evolving galaxy population. Even if we had infinite resolution, that may not be enough because we do not fully understand the microphysics of many relevant phenomena from first principles (e.g., cosmic rays, turbulence, dust, multi-phase mixing). On macroscopic scales, this is compounded because so many variables are at play, and often we do not even know what all the variables are. The vast freedom allowed for these choices therefore necessitates re-framing the problem of galaxy evolution as the empirical exercise that it is: repeatedly assess, in a Bayesian way, whether and how different descriptions of various physical processes fit together to match multi-scale datasets and simulations. We argue that the overall challenge of galaxy formation then is to map and interpret the space of plausible dynamical models for how galaxy populations assemble into the $n$-dimensional manifolds that summarize their observable properties, accounting for variations in both astrophysics and cosmology. This could mean exploring the space of functional forms for our different ODE terms and/or identifying when more sophisticated sets of state variables and evolution equations are needed to maximally extract information from both simulations and data, which is exactly what \texttt{sapphire} was designed to do.

\subsection{Limitations, caveats and future prospects}\label{sec:caveats}

Our study, like any other, is subject to limitations and caveats. Here we discuss some as avenues for future work.

First, we have only allowed eight parameters to vary, with the remaining $\sim20$ fixed based on simple physical arguments or ``priors'' from the FIRE-2 simulations following \citet[][]{pandya23}. The limited-scope dynamical model considered here is capable of reproducing the $z=0$ quasi-observable scaling relations so we are justified in trying to interpret the parameter posteriors. However, if we had allowed more parameters to be free, that would likely result in broader posteriors. The three constraints we chose almost certainly cannot, by themselves, inform the choice of functional forms and parameter values for CGM structure, turbulence driving and dissipation, large-scale halo gas inflow suppression, redshift dependent structure growth and other elements of our full model. Constraining these and other physical processes likely requires additional quasi-observables describing large-scale CGM/IGM properties, archaeological star formation histories, and the redshift evolution of many more scaling relations. In parallel, by incorporating priors from a wider range of multi-scale simulations, \texttt{sapphire} can serve as a unified framework for comparing and testing the emergent predictions of different subgrid recipes.

Second, we are missing several core physical processes in the baseline model used here, namely satellite galaxy evolution, mergers and black hole feedback. This prevents us from predicting total galaxy number counts, understanding environmental processes, and extending to group/cluster scales where SN feedback alone cannot limit excess star formation. We are actively implementing these and defer describing them to future work (Gabrielpillai et al., in prep.; Terrazas et al., in prep.). Capturing galaxy morphology and structure will require adding more zones, more state variables and partial differential equations to model spatially-dependent physical processes \citep[e.g.,][]{dekel09,krumholz10,elmegreen11,forbes12,forbes14,porter14,bialy20,sharda21,ginzburg22}. This will be challenging to do while retaining interpretability and causal identifiability in \texttt{sapphire}, but will be important for leveraging spatially-resolved scaling relations \citep[e.g.,][]{romanowsky12,romeo20}. Relatedly, \texttt{sapphire} will need to be extended to model the back-reaction of baryons on DM halo properties \citep{dutton07,schneider15,pandey25} and account for variations in cosmology \citep[e.g., by changing the input merger trees;][Robinson et al., in prep.]{zhao09,benson13,perez23,stiskalek26}. On the latter point, it may be fruitful to connect our \texttt{sapphire} dynamical systems approach with recent advances in differentiable halo finding \citep[\texttt{jFoF};][]{horowitzbayer26}, merger tree generation \citep[\texttt{JZ-TREE};][]{stucker26}, cosmological summary statistics \citep[\texttt{jax-cosmo};][]{campagne23}, and cosmological perturbation evolution \citep[\texttt{DISCO-DJ};][]{hahn24dj}.

As for our inference, there are numerous things that we could have improved. For example, we chose a very simple Gaussian log-likelihood that neglected covariance between the different galaxy scaling relations, which erases information. In principle, this covariance could be recovered with more sophisticated summary statistics like a multivariate kernel density estimate after dimensionality reduction and/or with a neural network that learns the implicit likelihood \citep[][Makinen et al., in prep.]{makinen23,ho24}. On a longer timescale, we will take advantage of the modular nature of our code to enable Bayesian evidence calculations across different galaxy formation models with fixed numerics. Nested sampling \citep{handley15} and/or evidence networks \citep{jeffrey24} may be useful for this. This is of paramount importance because without a unified framework like \texttt{sapphire}, the field will not be able to clearly attribute similarities and discrepancies between different galaxy evolution scenarios to physics, numerics, statistics or some combination thereof.

\section{Summary}\label{sec:summary}
We introduced \sapphire, which is a modular, automatically differentiable, multi-GPU-parallelized, publicly available framework for evolving and understanding galaxy populations as dynamical systems. \sapphire\ is written in \texttt{JAX} \citep{bradbury18} and leverages the open-source \texttt{diffrax} package for solving and computing gradients through nonlinear, physically-motivated differential equations \citep{kidger22}. Building on work started by \citet{pandya20,pandya21thesis,pandya23} as part of the SMAUG and LtU collaborations, \sapphire\ bridges astrophysics, cosmology, numerics, dynamics and statistics in new ways (Figure \ref{fig:modules}) to provide much-needed interpretability and causal identifiability for both simulations and observations. Our key takeaways are:

\begin{itemize}

\item For the first time, we used exact Jacobians to perform both local and global sensitivity analysis across parameter space of a SAM, using the \citet{pandya23} physical model as an example baseline. The Jacobians of our model galaxies contain non-random, interpretable structure. Notably, SN wind energy loading dominates the sensitivity of all $z=0$ state variables whereas mass loading has much weaker, often opposite-sign, gradients. This is the case across a large range in parameter space and halo mass accretion histories. (Figures \ref{fig:jacobian}, \ref{fig:jacall})

\item We use gradient descent for comprehensive mock parameter recovery tests to show that the $z=0$ SMHM relation alone does not contain enough information to infer model parameters related to star formation and stellar feedback. Adding $z=0$ ISM gas fractions and the ISM MZR breaks degeneracies and improves parameter recovery. Fisher forecasts suggest that percent-level precision can be achieved on some parameters by combining these $z=0$ galaxy scaling relations, though systematics and broader model degeneracies may be a problem. (Figures \ref{fig:adamcoverage8}, \ref{fig:summaryfisher}, \ref{fig:adamerrs})

\item We simultaneously fit the $z=0$ SMHM relation from \citet{behroozi19} as well as ISM gas fractions and the ISM MZR for star-forming galaxies from the MaNGA survey. We infer relatively low mass loading ($\eta_M\lesssim1$ at the MW-scale and $\eta_M\sim10$ for classical dwarfs) and high energy loading ($\eta_E\sim0.2$ at the MW-scale and $\eta_E\sim0.5-1$ for dwarfs). Together these argue against ejective feedback as the dominant mechanism for regulating galaxy evolution and instead are consistent with preventative feedback. (Figures \ref{fig:obspost},\ref{fig:obscorner})

\item We use both gradient descent and HMC to forecast how the precision on model parameters would improve if the data uncertainties were cut by a factor of two or ten, including systematics. Consistent with our mocks, we can achieve percent-level precision on some parameters though more work is needed to see how this degrades as we allow more parameters to vary. We also apply systematic shifts in the normalizations of the three $z=0$ scaling relations to interpret our fiducial posteriors. The SMHM relation is most sensitive to the amplitude of $\eta_E$ and $t_{\rm dep}$. ISM gas fractions are primarily controlled by the amplitude of $\eta_M$ and $t_{\rm dep}$. The ISM MZR strongly constrains the amplitude of mass and metal loading as well as slope of energy loading. (Figures \ref{fig:forecasts}, \ref{fig:shiftsmhm}, \ref{fig:shiftfgas}, \ref{fig:shiftmzr})

\item We present posterior predictive checks for a few example $z=0$ and $z=2$ quantities we did not fit to but which are obvious avenues for future work. We are in good agreement with the normalization and scatter of the $z=0$ SFMS from MaNGA. Our mass loading factors show a large spread at fixed stellar mass for dwarfs. The ratio of cooling time to free-fall time in the inner halo, ISM accretion rates and halo baryon fractions all correlate with the specific energy of SN feedback across different posterior realizations. At $z=2$, our model SFMS tends to be below the data with smaller scatter. Our $z=2$ SMHM posterior draws are systematically higher than empirically determined both pre- and post-JWST. ISM gas fractions at $z=2$ show a large scatter between posterior draws. Finally, our ISM MZR at $z=2$ is systematically lower than inferred from pre- and post-JWST datasets. Some of these discrepancies may require additional redshift dependent parameters which \texttt{sapphire} is uniquely positioned to help constrain. (Figure \ref{fig:postextras})
\end{itemize}

\sapphire\ represents a step towards building an interpretable dynamical model that can reveal the key state variables, evolution equations and astrophysical parameters of galaxies directly from observational data, using simulations and simple arguments as ``prior'' inputs. In future papers, we will introduce more physical processes and numerical techniques for hybrid physics-informed, data-driven Bayesian modeling of galaxy population evolution in a cosmological context.

\begin{acknowledgements}
We thank the anonymous referee for comments that improved the clarity of the manuscript. VP thanks the Simons Collaboration on Learning the Universe and specifically credits Jens Jasche, Guilhem Lavaux, Ben Wandelt and Chirag Modi for first making him aware of differentiable programming. VP dedicates this paper to the memory of Joel Primack who taught him much about galaxy formation and the history and purpose of SAMs during his PhD at UC Santa Cruz. VP also thanks Sandy Faber, Shy Genel, Doug Hellinger, Patrick Kidger, Avery Kim, David Koo, Lachlan Lancaster, Hui Li, Shivam Pandey, Hiranya Peiris, Volker Springel and others for helpful discussions. Support for VP was provided by: (1) NASA Hubble Fellowship Grant HST-HF2-51489 awarded by the Space Telescope Science Institute, which is operated by the Association of Universities for Research in Astronomy, Inc., for NASA, under contract NAS5-26555, and (2) NSF Astronomy \& Astrophysics Grant No. 2307419. GLB acknowledges support from the NSF (AST-2108470, AST-2307419), NASA TCAN award 80NSSC21K1053, and the Simons Foundation through the Learning the Universe Collaboration. D.B.F. gratefully acknowledges support from NSF through grants AST-2407387 and from NASA through grants HST-AR-17859.015-A and HST-AR-17559.009-A. This work was supported by a grant from the Simons Foundation (Grant Award ID BD-Targeted-00017375, DBF). TS gratefully acknowledges the support of the NSF-Simons AI-Institute for the Sky (SkAI) via grants NSF AST-2421845 and Simons Foundation MPS-AI-00010513. TS is supported by NSF through grant AST-2510183 and by NASA through grants 22-ROMAN22-0055 and 22-ROMAN22-0013. This work would not be possible without the open-source development efforts of the \texttt{JAX} team at Google, and without the significant computational resources provided by the Scientific Computing Core at the Flatiron Institute. We also thank the MaNGA team and Peter Behroozi for making their data publicly available in tabular form.

\end{acknowledgements}

\software{\texttt{sapphire} \citep{sapphire}, \texttt{JAX} \citep{bradbury18}, \texttt{diffrax} \citep{kidger22}, \texttt{optax} \citep{optax}, \texttt{numpyro} \citep{numpyro}, \texttt{interpax} \citep{interpax}, \texttt{Astropy} \citep{astropy13,astropy18,astropy22}, \texttt{ytree} \citep{ytree}, \texttt{matplotlib} \citep{matplotlib}, \texttt{seaborn} \citep{seaborn}, \texttt{chainconsumer} \citep{chainconsumer}, \texttt{numpy} \citep{numpy}, \texttt{pandas} \citep{pandas}, \texttt{disBatch} (\url{https://github.com/flatironinstitute/disBatch})}\\

\noindent Here we use the Contributor Roles Taxonomy (CRediT) system\footnote{\url{https://credit.niso.org/}} to list the roles and contributions of the co-authors: 
\textit{\textbf{Conceptualization:}} VP 
\textit{\textbf{Data curation:}} VP 
\textit{\textbf{Formal analysis:}} VP 
\textit{\textbf{Investigation:}} VP 
\textit{\textbf{Methodology:}} VP, GB, LM, MH, PL
\textit{\textbf{Project administration:}} VP 
\textit{\textbf{Resources:}} RSS 
\textit{\textbf{Software:}} VP 
\textit{\textbf{Supervision:}} GB 
\textit{\textbf{Validation:}} VP, GB, LM, AG, DF, KI, PL, WR
\textit{\textbf{Visualization:}} VP 
\textit{\textbf{Writing -- original draft:}} VP 
\textit{\textbf{Writing -- review \& editing:}} VP, GB, LM, AG, CC, DF, LH, MH, KI, CJ, SK, ML, PL, CL, LP, WR, RSS, TS, RS, BT, MV

\bibliographystyle{aasjournal}
\bibliography{references}

\appendix 

\section{MaNGA Summary Statistics}\label{sec:manga}

Here we describe our analysis of star-forming galaxies in the final Data Release 17 (DR17) of the MaNGA survey \citep{bundy15}, which is part of the Sloan Digital Sky Survey IV \citep[SDSS-IV;][]{blanton17}. We begin with a description of the sample selection and then describe each of our three summary statistics in order of increasing complexity. The \texttt{sapphire} github repository contains a Jupyter notebook to reproduce our calculations.

\subsection{Sample selection}
MaNGA obtained integral field unit (IFU) spectroscopy of $\sim10,000$ galaxies. These were selected from the NASA Sloan Atlas \citep[NSA;][]{blanton11} using simple cuts on redshift, $i$-band luminosity and color designed to mitigate selection biases. Here we combine the Primary, Secondary and Color-Enhanced Samples. The Primary Sample selects galaxies at $z\sim0.03$ for IFU coverage out to $1.5$ times the half-light radius. The Secondary Sample includes more distant galaxies at $z\sim0.06$ to enable IFU coverage out to $2.5$ times the half-light radius. Finally, the Color-Enhanced Sample includes rarer classes of galaxies. We will refer to this combination of Primary, Secondary and Color-Enhanced Samples as our Main Galaxy Sample.

\begin{figure}
\centering
\includegraphics[width=\hsize]{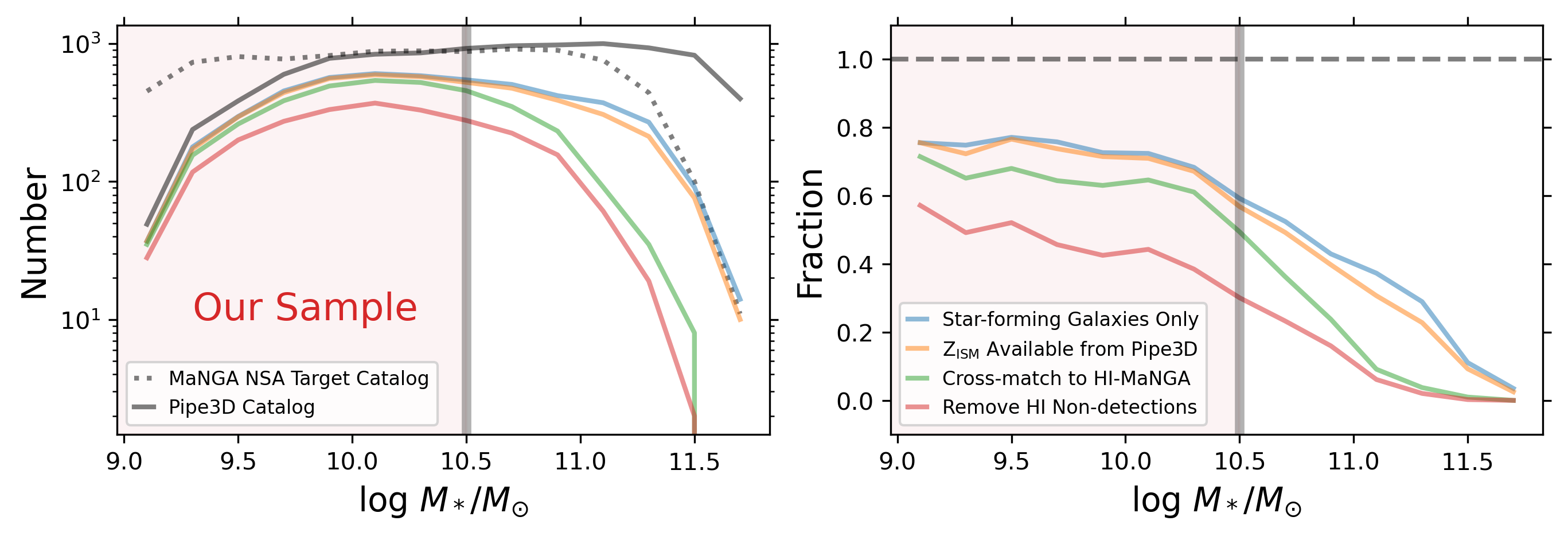}
\caption{Effect of our selection cuts on the stellar mass distribution of the MaNGA sample. The left panel shows the number of galaxies whereas the right panel shows the fraction relative to the Pipe3D catalog. We use 0.2 dex wide bins between $\log M_*/M_{\odot}=9-12$. Note how the Pipe3D catalog (solid black line) has fewer low-mass and more high-mass galaxies than the parent NSA catalog (dotted black line), which was originally used for MaNGA target selection and has a flat stellar mass distribution. The blue curves show the effect of limiting to star-forming galaxies only. The orange curves show the impact of requiring $Z_{\rm ISM}$ measurements and uncertainties. The green curves shows the drop after cross-matching to the HI-MaNGA catalog. Finally, the red curves denote the effect of requiring HI detections. The vertical gray lines  mark our upper stellar mass limit ($\log M_*/M_{\odot}=10.5$) corresponding roughly to the maximum halo masses we model ($\log M_{\rm vir}/M_{\odot}\sim12$).}
\label{fig:manga_sample}
\end{figure}

Figure \ref{fig:manga_sample} illustrates how our sequence of selection cuts affects the MaNGA galaxy stellar mass distribution. First, we start with the 9853 objects in the Main Galaxy Sample using standard flags provided by the Data Reduction Pipeline \citep[DRP;][]{westfall19}. We cross-match these to the Pipe3D value-added catalog of gas and stellar population properties \citep{sanchez22}. In this paper, we only require stellar masses, gas-phase metallicities, star formation rates and their measurement uncertainties. These are available for 9833 galaxies. Although this is essentially the full Main Galaxy Sample, the left panel of Figure \ref{fig:manga_sample} shows a deficit of low-mass and excess of high-mass galaxies relative to the parent NSA catalog used for original target selection, with the NSA catalog being roughly uniform between $\log M_*/M_{\odot}=9-11$ as is often claimed for MaNGA.

Since the version of \texttt{sapphire} we use is only appropriate for star-forming galaxies, we follow \citet{pandya17} and impose a specific SFR cut of $\rm{SFR}/M_*>10^{-11}$ yr$^{-1}$ to keep only star-forming galaxies. The SFR comes from integrating H$\alpha$ emission over all spaxels using the \citet{kennicutt98} calibration. As explained by \citet{sanchez22}, the Balmer decrement was used to correct for dust attenuation using a MW-like \citet{cardelli89} curve. This cut preferentially removes higher mass galaxies, leaving 4943 objects. We also require $Z_{\rm ISM}$ measurements and uncertainties from Pipe3D, which drops our sample slightly to 4665 galaxies.

Next, we merge with Data Release 3 of the HI-MaNGA value-added catalog \citep{stark21}. As detailed in \citet{masters19}, HI-MaNGA targets a subset of MaNGA galaxies with the Green Bank Telescope down to comparable sensitivity limits as the ALFALFA survey \citep{haynes11} while avoiding overlapping targets. They imposed a maximum distance cut which preferentially removed high-mass galaxies as confirmed by our Figure \ref{fig:manga_sample}. We are able to cross-match our star-forming sample above to 3567 galaxies in the HI-MaNGA survey. Of these, only 2393 galaxies have HI detections. Lastly, because we only model halos up to $\log M_{\rm vir}/M_{\odot}\sim12$, we cut off the sample at $\log M_*/M_{\odot}=9-10.5$, leaving 1787 galaxies in our final sample. 

\subsection{ISM mass-metallicity relation}
\begin{figure}
\includegraphics[width=0.5\hsize]{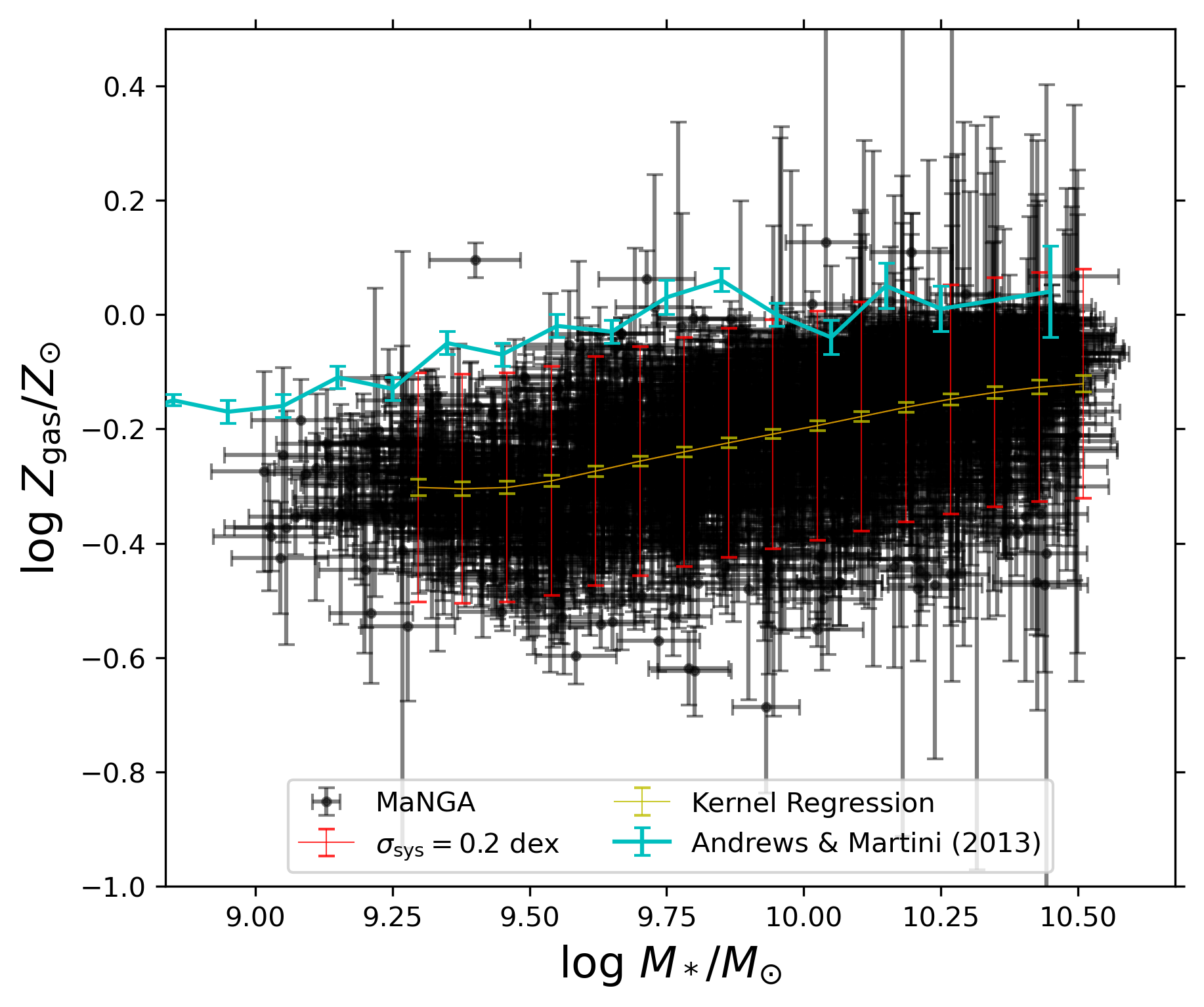}\includegraphics[width=0.5\hsize]{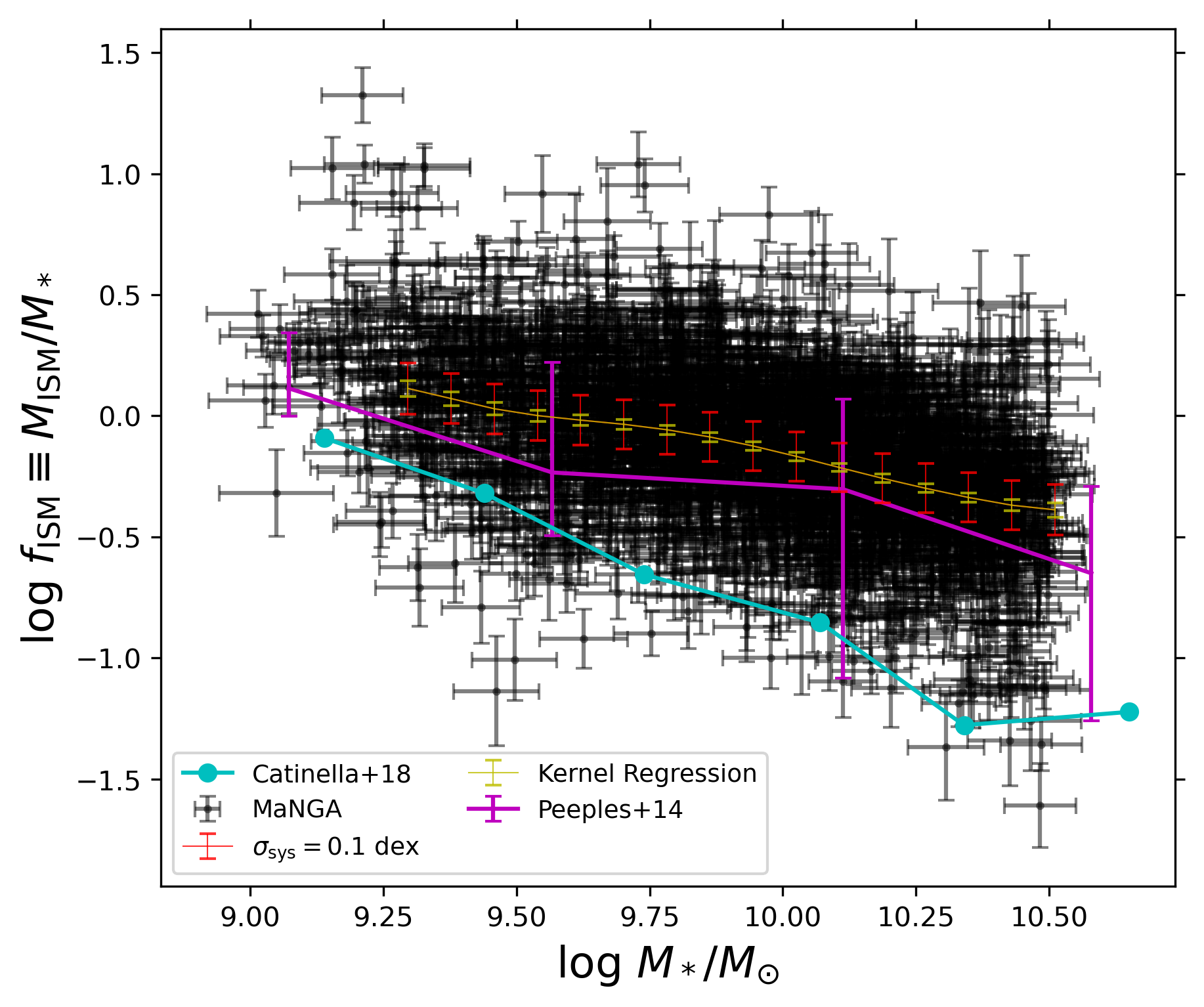}
\caption{ISM MZR (left) and ISM gas fractions (right) as a function of stellar mass for our star-forming, HI-detected MaNGA sample. In both panels, black points show individual galaxies with their measurement uncertainties. Yellow lines are from our Gaussian kernel regression and denote purely statistical uncertainties. Red lines add our assumed systematic uncertainty in quadrature: 0.2 dex for the MZR and 0.1 dex for the gas fractions. Cyan and magenta lines show relations from the literature but for these the errorbars reflect the dispersion, not standard error on the mean like for our regression. Our MZR is $\sim0.1$ dex lower than \citet{andrews13} but consistent within systematics. Our $f_{\rm ISM}-M_*$ relation roughly agrees with \citet{peeples14} whereas \citet{catinella18} is lower since they included non-detections in their averaging. In Subsection~\ref{sec:perturbs} we show how systematic shifts affects our parameter posteriors.}
\label{fig:manga_mzr_fgas}
\end{figure}

Figure \ref{fig:manga_mzr_fgas} shows our fit to the ISM MZR, which is arguably the least complicated of our three summary statistics. The gas-phase metallicities come from Pipe3D fits to various emission lines, which are then combined for multiple oxygen abundance ($12+\log(O/H)$) indicators as detailed in \citet{sanchez22}. For simplicity, here we use the common $R_{23}$ indicator involving the [O II] 3727$\AA$, [O III] 4959,5007$\AA$ and H$\beta$ emission lines in a central 2.5 arcsec diameter aperture, based on the \citet{kobulnicky04} calibration. We convert these oxygen abundances to overall $Z_{\rm ISM}$ assuming solar abundances from \citet{asplund09}. 

We use the approach of ``multiple imputations'' \citep[e.g.,][]{littlerubin,vanbuuren18} to generate 100 realizations of the data, drawing each galaxy's stellar mass and ISM metallicity from two independent Gaussians with means equal to the measured values and standard deviations set to the measurement uncertainties. For each realization of points, we perform Nadaraya-Watson regression as detailed in Subsection~\ref{sec:nadaraya}. Then, we interpolate all regressions onto a common stellar mass grid and use Rubin's rules \citep{rubin87} to compute the average regression and its standard error. The typical uncertainty on individual ISM metallicities is of order 0.05 dex, but this is likely too small and neglects systematics from, e.g., using different indicators. Thus we add, in quadrature, an arbitrary systematic uncertainty of 0.2 dex to the baseline standard errors. This reflects our ignorance on the exact normalization and slope of the ISM MZR. Our regression is relatively consistent with others in the literature \citep[e.g.,][]{andrews13} and in Subsection~\ref{sec:perturbs} we assess the impact of systematic shifts in the ISM MZR on model parameter posteriors.

\subsection{ISM gas fractions}
Figure \ref{fig:manga_mzr_fgas} also shows our regression of the ISM gas fraction, which is more involved since we start with only HI masses from HI-MaNGA \citep{masters19,stark21}. We assign every MaNGA galaxy a value of $R_{\rm mol}\equiv M_{\rm H_2}/M_{\rm HI}$ based on the average $M_*$-dependent relation given in Table 3 of \citet{catinella18} and their constant reported Gaussian scatter of 0.41 dex. Summing the random assigned H$_2$ mass with the observed $M_{\rm HI}$ and multiplying by $1.3$ to account for helium gives us an estimate of the total ISM mass which we can use to compute our gas ``fraction'' (ratio) $f_{\rm ISM}\equiv M_{\rm ISM}/M_*$. Then we use the same multiple imputation approach as described above for the MZR to derive an average regression and its standard error. The measurement uncertainties on $M_{\rm HI}$ are typically very small, and although the uncertainty on $M_{\rm H_2}$ is much higher at $\sim0.3$ dex \citep{saintonge11}, since H I dominates over H$_2$, the formal standard errors on our average $f_{\rm ISM}$ are $\lesssim0.1$ dex. We add 0.1 dex in quadrature to account for any extra systematics. Our average $f_{\rm ISM}-M_*$ regression is significantly higher than \citet{catinella18}, by up to $\sim1$ dex at high mass. This is likely because \citet{catinella18} included HI non-detections in their average. Indeed, we are in relatively good agreement with \citet{peeples14} who restricted their analysis to star-forming galaxies.

\subsection{Stellar-mass-halo-mass relation}
We investigated the possibility of assigning halo mass posteriors to individual MaNGA galaxies using the \citet{behroozi19} SMHM relation. We built a simple hierarchical Bayesian model that accounts for the scatter in the assumed relation and measurement uncertainty on stellar mass: 
\begin{equation}
\mathcal{P}(M_{\rm halo}|M_*) \propto \mathcal{P}(M_*|M_{\rm halo})\mathcal{P}(M_{\rm halo})
\end{equation}
where the left-hand side is the posterior, the first-term on the right-hand side is the likelihood, and the second term is the prior. The likelihood is our assumed SMHM relation from \citet{behroozi19} and the prior is the halo mass function, for which we assume the \citet{tinker08} function implemented in the \texttt{colossus} package \citep{diemer18}. In practice, the likelihood dominates over the prior. Since the \citet{behroozi19} SMHM relation has asymmetric uncertainties on the mean and scatter, we assume a split-normal distribution to draw 100 random values for both. For each of those SMHM realizations, we evaluate a Gaussian likelihood on a halo mass grid spanning $\log M_{\rm halo}/M_{\odot}=10-13$ for each galaxy using its measured stellar mass. The measurement uncertainty on stellar mass is convolved with the SMHM relation scatter for the Gaussian likelihood calculation. For every SMHM realization, we also compute the fraction of galaxies that would satisfy our MaNGA selection of $\log M_*/M_{\odot}=9-10.5$, assuming a Gaussian distribution for stellar mass at a given halo mass. This provides a way to estimate the completeness.

\begin{figure}
\centering
\includegraphics[width=\hsize]{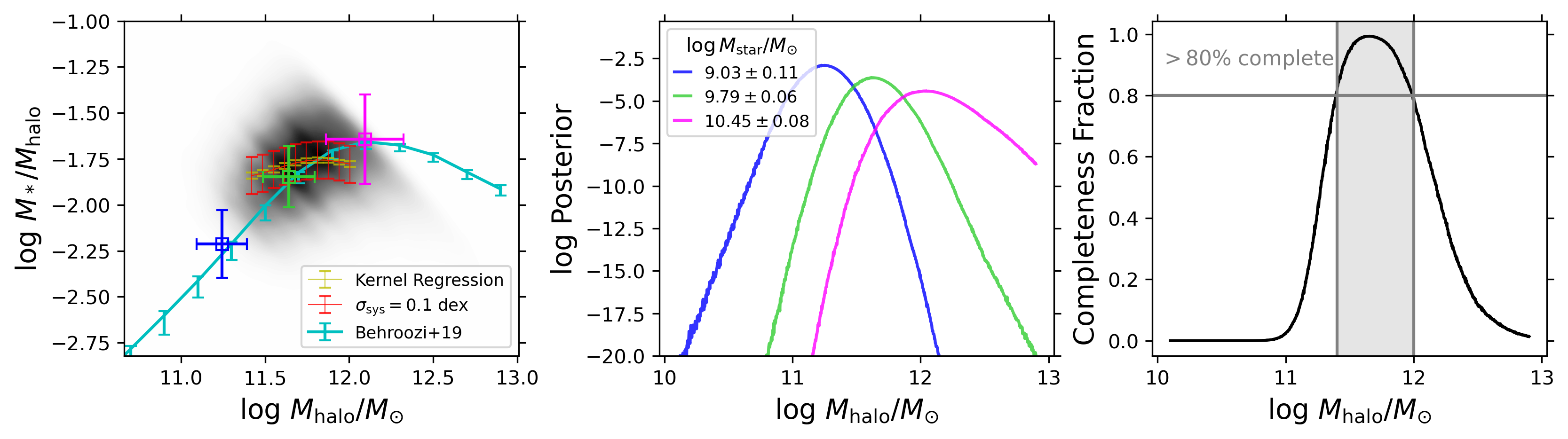}
\caption{Simple hierarchical Bayesian model for SMHM relation of MaNGA galaxies. Left: the SMHM ratio and halo mass for three example galaxies (blue, green and magenta errorbars) follow the assumed \citet{behroozi19} relation. However, the posterior density from pooling all random realizations together reveals a bias in the average SMHM relation. The negative diagonal striping pattern is due to the hard $\log M_*/M_{\odot}=9-10.5$ selection cut. Since low-$M_*$ galaxies are missing, the average SMHM ratio is biased high at low-$M_{\rm halo}$, and vice versa at high halo mass. Middle: halo mass posteriors for the same example low-mass, intermediate-mass and high-mass galaxies marked in the left panel. These individual posteriors look reasonable, suggesting that it is the hard stellar mass cut that is biasing the shape of the MaNGA SMHM relation. Right: the fraction of galaxies at each halo mass that would fall within the MaNGA $\log M_*/M_{\odot}=9-10.5$ window. Only for a small halo mass range of $\log M_{\rm halo}/M_{\odot}\approx11.4-12$ do we surpass 80\% completeness.}
\label{fig:manga_smhm}
\end{figure}

Figure \ref{fig:manga_smhm} shows the resulting SMHM relation for MaNGA. Our Bayesian procedure works well for individual galaxies: their posteriors for halo mass place them squarely on the assumed \citet{behroozi19} relation. However, the hard selection window of $\log M_*/M_{\odot}=9-10.5$ results in a biased SMHM relation for MaNGA. The average SMHM ratio for $\log M_{\rm vir}\sim11.5$ is biased high since only galaxies with $\log M_*/M_{\odot}>9$ are in our sample and lower-$M_*$ galaxies are missing. Similarly, at $\log M_{\rm vir}/M_{\odot}\sim12$, the average is biased low since we are missing $\log M_*/M_{\odot}>10.5$ galaxies that would be assigned to that halo mass. This is immediately apparent from the right-most panel which shows that we reach $>80\%$ completeness only in a small $\log M_{\rm vir}/M_{\odot}\approx11.4-12$ window where mostly only $\log M_*/M_{\odot}=9-10.5$ galaxies live. 

\begin{figure*}
\centering
\includegraphics[width=\hsize]{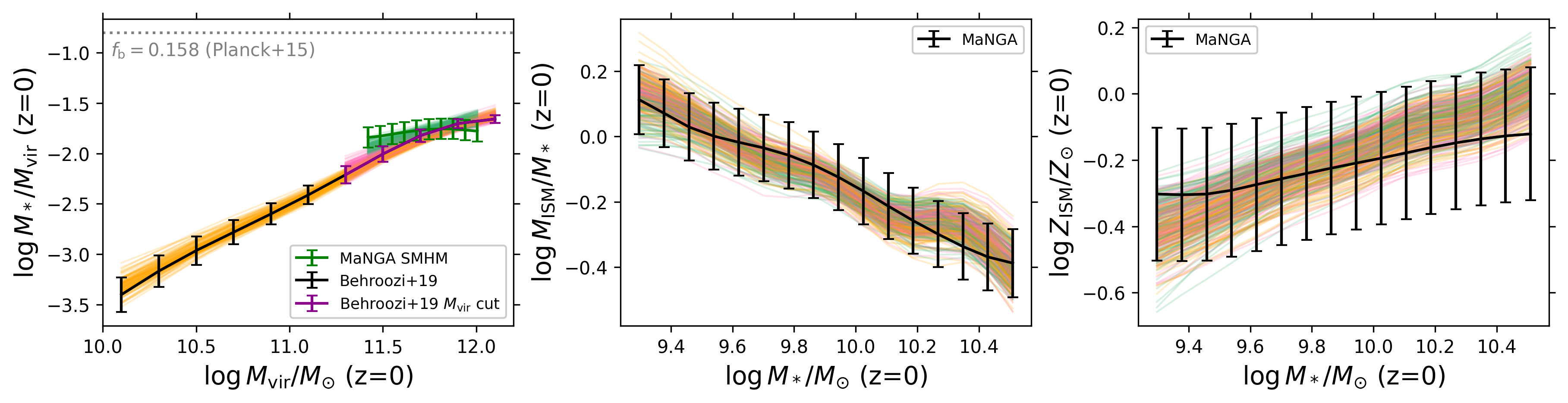}
\includegraphics[width=\hsize]{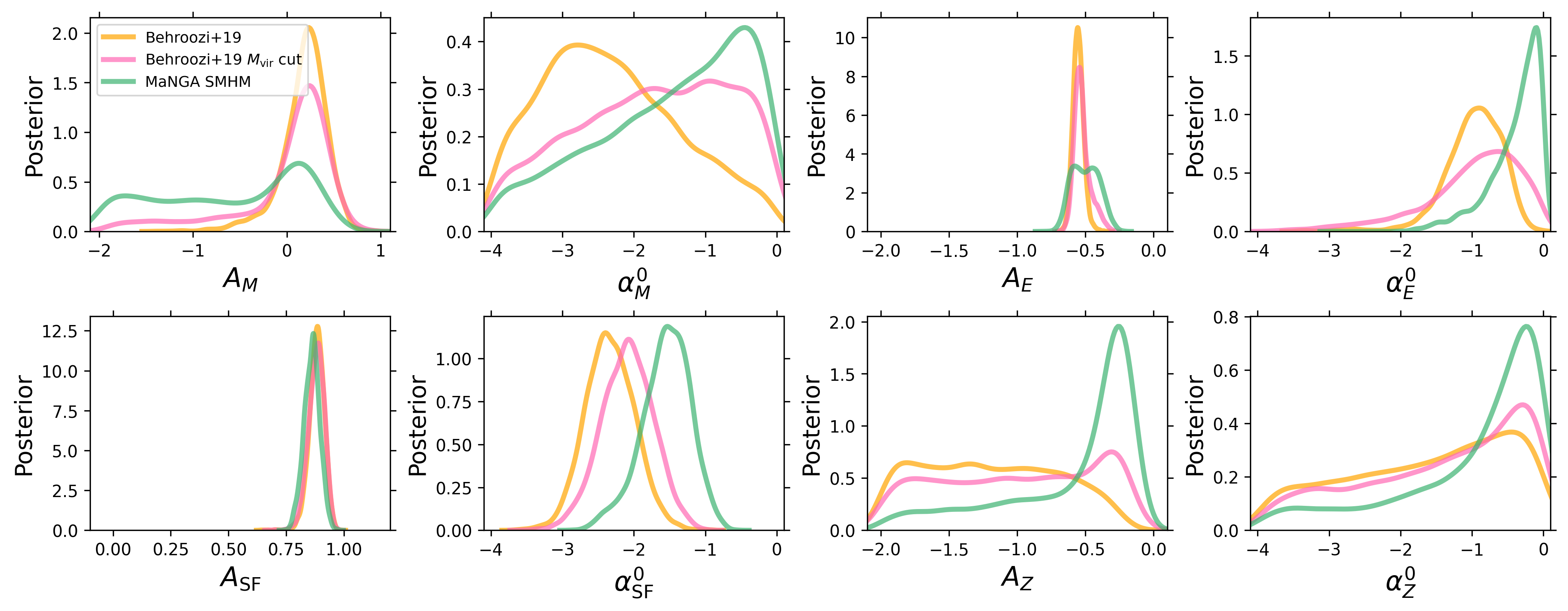}
\caption{Analogous to Figure \ref{fig:shiftsmhm} but now trimming the \citet{behroozi19} SMHM relation to the limited halo mass range where MaNGA is complete (pink) or using our biased MaNGA SMHM relation (green). In both cases we are still able to fit the SMHM relation while simultaneously matching the ISM gas fractions and ISM MZR. The bottom panels compare the two resulting parameter posteriors to the fiducial case (orange). The main change is that the ``degraded'' SMHM relations both lead to a shallower mass loading slope and higher $A_Z$ preference. The flatter biased MaNGA SMHM relation leads to a systematically shallower preferred ISM depletion time slope. This is a simple precursor to using \texttt{sapphire} to disentangle astrophysics and cosmology from observational selection effects.}
\label{fig:smhmbias}
\end{figure*}

Since the shape of the MaNGA SMHM relation is biased relative to the underlying population one from \citet{behroozi19}, it would be hard to trust any inference with it which is why we opted to ``mix and match'' the MaNGA $f_{\rm ISM}$ and ISM MZR data with the full SMHM relation from \citet{behroozi19}. Nevertheless, Figure \ref{fig:smhmbias} illustrates the impact of limiting the \citet{behroozi19} halo mass range to $\log M_{\rm vir}/M_{\odot}\sim11.4-12$ where MaNGA is complete, or using the biased, flatter MaNGA SMHM relation. We are able to still fit either of those along with the ISM gas fractions and ISM MZR. Most parameter posteriors remain relatively unchanged except both degraded SMHM relations prefer an opposite, shallower $\eta_M$ slope and higher $A_Z$. The flattening of the biased MaNGA SMHM relation also leads to a preference for a flatter $t_{\rm dep}$ slope. This example demonstrates the potential power of \texttt{sapphire} for disentangling astrophysics from observational selection effects and should remind the reader of similar systematic normalization shifts we did for the other two relations in Figures \ref{fig:shiftsmhm}, \ref{fig:shiftfgas} and \ref{fig:shiftmzr}.

\section{Numerics}\label{sec:numerics}

Here we briefly describe numerical details which provide the foundation for this paper. Note that many of the tests in this Appendix would be borderline impossible if we did not use multiple GPUs given the large numbers of ODEs that we need to evolve for different solver settings.

\subsection{Adaptive ODE solver}\label{sec:solver}
This paper deals with first- and second-order gradients of dynamical systems so first we need to establish that we can actually solve our nonlinear galaxy formation ODEs accurately and efficiently. We use the \texttt{diffrax} Python package which provides a suite of adaptive ODE solvers compatible with \texttt{JAX} \citep{kidger22}.\footnote{We originally wrote our own adaptive 2/3-order Runge-Kutta solver from \citet{bogacki89} in \texttt{JAX} using the Butcher tableau approach described in \citet{hairer93}, but the extra features of \texttt{diffrax} made it the natural choice for our work.} To ensure numerically stable adaptive timestepping, we solve our ODEs using (base-10) logarithmic state variables since they otherwise evolve over orders of magnitude (both for a single galaxy and across different systems). Next we justify our choice of adaptive timestepping algorithm, and absolute and relative error tolerances \citep[$a_{\rm tol}$ and $r_{\rm tol}$, respectively; see][for more details]{corless13}.

\begin{figure}
\centering
\includegraphics[width=\hsize]{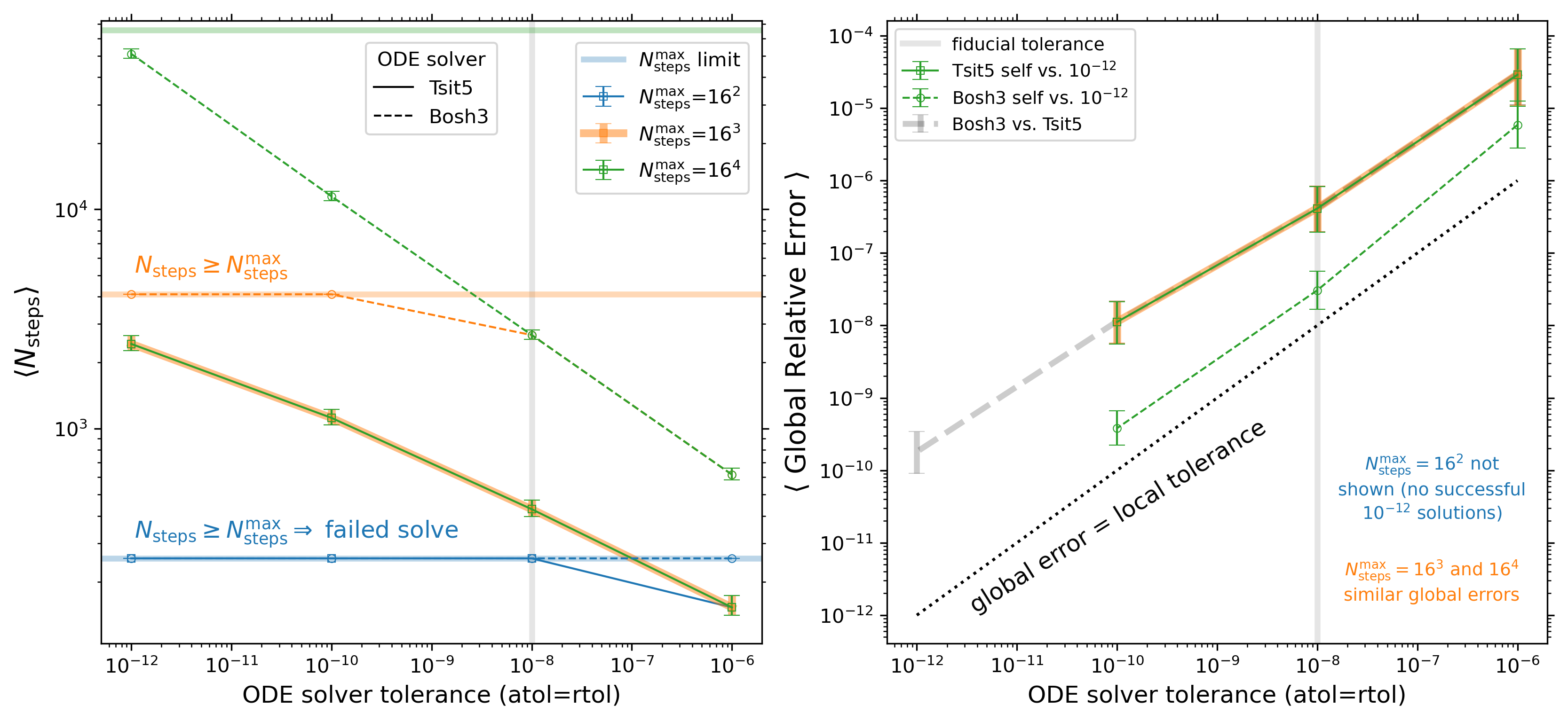}
\caption{Numerical convergence rate (left) and global relative errors (right) for different ODE solvers (solid lines for \Tsit, dashed lines for \Bosh) as a function of local tolerance and maximum allowed steps (blue/orange/green). Errorbars reflect the 16/50/84 percentiles of the distribution for 1000 random halos each evaluated at 1000 parameter sets uniformly sampled over a latin hypercube. \textit{Left:} Both solvers require more steps for lower tolerance settings but \Tsit\ is more efficient and less likely to hit the pre-allocated $N_{\rm steps}^{\rm max}$ limit. Note that \Tsit5\ finishes in $<16^3$ steps so increasing the max step limit to $16^4$ has no effect whereas \Bosh\ would take $>10\times$ as many steps at $a_{\rm atol}=r_{\rm tol}=10^{-12}$. \textit{Right:} For a given solver, the global error in the $z=0$ state vector goes down as the local tolerance is decreased reflecting the internal consistency of the solution. The lower order \Bosh\ solution also agrees with the higher order \Tsit\ solution at a given local tolerance but uses more steps. This justifies our choice to use \Tsit\ with $a_{\rm tol}=r_{\rm tol}=10^{-8}$ and $N_{\rm step}^{\rm max}=16^4$.}
\label{fig:nfail}
\end{figure}

Figure \ref{fig:nfail} compares the convergence and global error rates of a low 2/3-order \citep[][hereafter \Bosh]{bogacki89} and high 4/5-order \citep[][hereafter \Tsit]{tsit5} adaptive ODE solver. We evolve 1000 random halos at 1000 points in parameter space uniformly sampled from a latin hypercube \citep[][as described in Subsection~\ref{sec:lhs}]{mckay79}. We solve each galaxy ODE system 12 times on a grid of $a_{\rm tol}=r_{\rm tol}=(10^{-6},10^{-8},10^{-10},10^{-12})$ and $N_{\rm steps}^{\rm max}=(16^2,16^3,16^4)$. The latter is required since \jax\ pre-allocates memory so array shapes must be known at runtime. This means we have 12 million ODE solves which we vectorize over four Nvidia A100-80GB GPUs in parallel (following Subsection~\ref{sec:parallel} below). As we decrease the tolerances, both solvers require more steps to reach $z=0$ which makes sense because smaller timesteps are needed to control local errors. Clearly, \Tsit\ is more efficient than \Bosh\ and less likely to hit the $N_{\rm steps}^{\rm max}$ limit. 

Our galaxy formation ODEs do not have analytic solutions so we must resort to numerical tests to assess convergence. We define the global relative error for a given solver and tolerance by computing the norm of the residuals of the $z=0$ state vector $\vec{x}_{\rm tol}$ relative to the $10^{-12}$ ``best'' tolerance setting state vector $\vec{x}_{12}$:
\begin{equation}
\Delta (\vec{x}_{\rm tol},\vec{x}_{12}) = \left\|\frac{\vec{x}_{\rm tol}-\vec{x}_{12}}{\vert \vec{x}_{12} \vert}\right\|\;.
\end{equation}
Similarly, for each tolerance setting, we compare \Bosh\ to \Tsit\ assuming the latter is the ``truth'' since it is a higher order method. We mask failed ODE solves (i.e., those where $N_{\rm steps}$ hit the $N_{\rm steps}^{\rm max}$ limit). We find that this global relative error falls as the local tolerance decreases which is encouraging because it means both solvers are internally self-consistent. The global error is $\sim10-100\times$ larger than the requested local tolerance because local errors accumulate over the numerical integration history depending on the stability properties of our particular ODE system. This global-local error discrepancy is lower for \Bosh\ likely because it is a lower order method so the solution may not change as much when varying the local tolerance. Crucially, \Bosh\ agrees with \Tsit\ for a given tolerance with the same global relative error that \Tsit\ has compared to its own $10^{-12}$ tolerance solution.

The combined better efficiency (fewer total steps) and adequate numerical accuracy of \Tsit\ makes it our fiducial solver choice with $a_{\rm tol}=r_{\rm tol}=10^{-8}$ and $N_{\rm steps}^{\rm max}=16^4$ by default. We will show below in Subsection~\ref{sec:autodiff} that these tolerances are also sufficient for accurately computing Jacobian and Hessian sensitivity matrices.

\subsection{Automatic compilation, vectorization, parallelization}\label{sec:parallel}

One of the benefits of \texttt{JAX} is that it provides just-in-time (\texttt{jit}) compilation for Python code. This allows the same code to be run on CPUs, GPUs and other accelerators without major restructuring. We have written our entire model to be \texttt{jit}-compatible. On modern CPUs, we find a $\sim10\times$ speed-up for single evaluations of our integrator function that returns ODE values given the current time, state variables and free parameters. In addition, \texttt{JAX} provides function transformations for vectorization (\texttt{vmap}) and parallelization (\texttt{shmap}). Combined with \texttt{jit}, these can provide orders of magnitude speed gains by solving large batches of ODE systems in parallel using vectorized inputs such as our interpolated merger tree matrices.

\begin{figure}
\centering
\includegraphics[width=0.75\hsize]{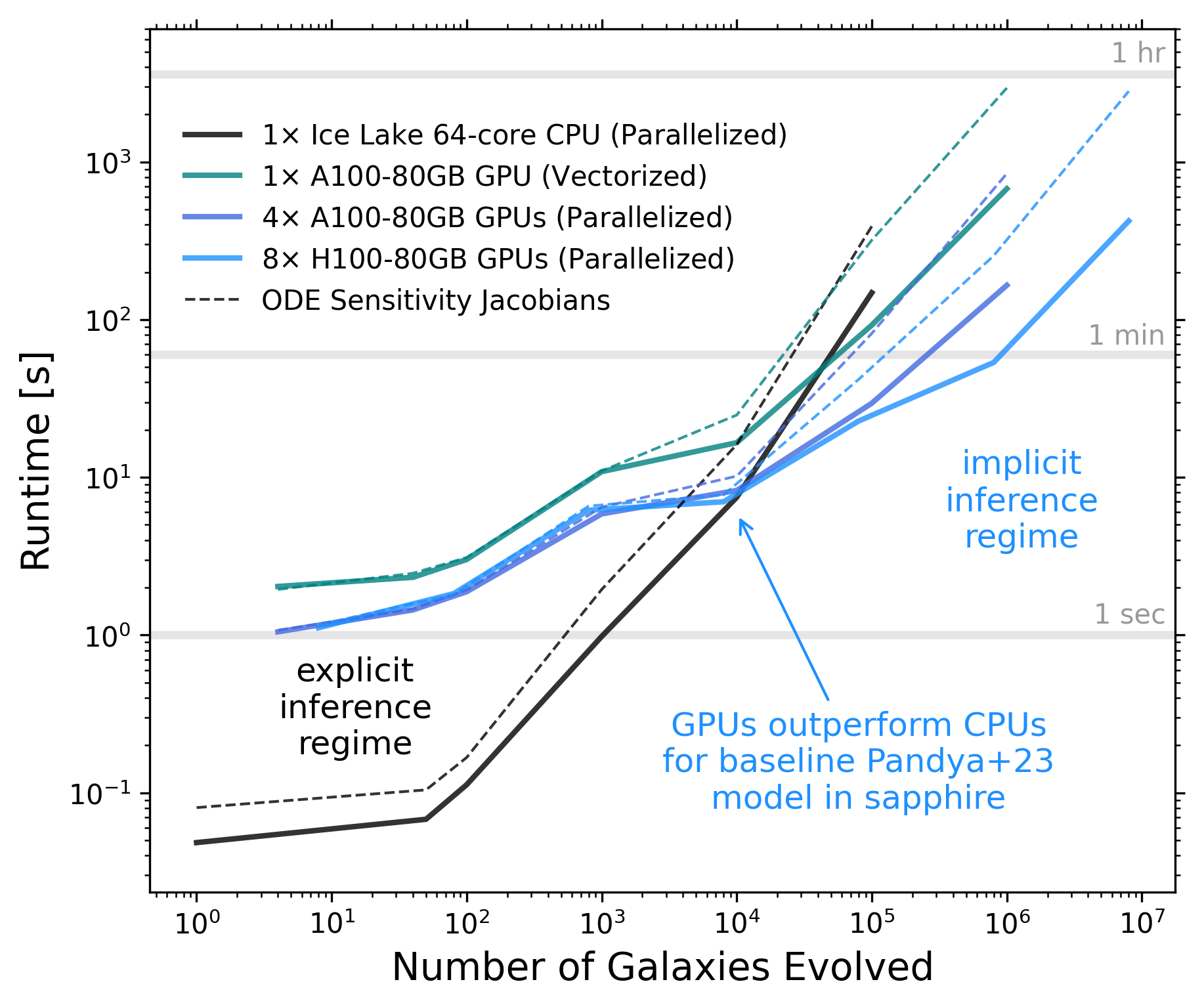}
\caption{Runtime for solving and auto-diffing through the ODE systems of different numbers of halos on a single 64-core Intel Ice Lake CPU node (black), single Nvidia A100-80GB GPU (teal), four A100-80GB GPUs (dark blue), and eight H100-80GB GPUs (light blue). Solid lines are for solving the ODE system whereas dashed lines are for computing the gradients (Jacobians) with respect to 9 parameters. For small batch sizes, GPUs are up to an order of magnitude slower than CPUs, so the latter are desirable for explicit inference with \texttt{adam} or HMC. For $\gtrsim10^5$ galaxies, GPUs out-perform CPUs, with runtime often dropping as $1/N_{\rm GPU}$. Gradients are at most only a factor of a few more expensive than solving the ODEs themselves.}
\label{fig:runtime}
\end{figure}

Figure \ref{fig:runtime} compares the runtime for solving the ODE systems of different numbers of halos on CPUs and GPUs. All runtimes exclude the initial \texttt{jit} compilation time. For small batch sizes, GPUs are $\sim10\times$ slower than CPUs. However, with vectorization alone, a single Nvidia A100-80GB GPU out-performs a single state-of-the-art Intel Ice Lake 64-core CPU node for batch sizes of $\gtrsim5\times10^4$ halos. On average, we can evolve 1000 independent galaxy ODE systems in $\sim1$ sec on a 64-core CPU and $\sim10$ sec on an A100-80GB GPU. For even larger batch sizes, multi-GPU parallelization becomes essential, with runtime often dropping as $1/N{\rm GPU}$. For example, at $10^5$ galaxies, CPU takes $\sim100$ sec, a single A100-80GB GPU is slightly quicker, and four A100-80GB GPUs are $\sim4\times$ faster still. At $10^6$ galaxies, four A100-80GB GPUs are $4\times$ faster than a single A100-80GB GPU, and we get another factor of $\sim4$ speed gain from parallelizing over eight H100-80GB GPUs. Finally, for $10^7$ galaxies, which is the largest batch size we test here, we can evolve these in $\sim420$ sec ($\sim7$ min) with eight H100-80GB GPUs. 

For the batch sizes tested here, we do not encounter GPU memory issues but that can be addressed with more sophisticated parallelization and batching strategies, including splitting work across multiple nodes. For the purposes of explicit inference with \texttt{adam} or HMC, solving ODEs and computing their gradients becomes too slow for more than a couple thousand halos. But at very large batch sizes, we can utilize multi-GPU parallelization for lightning-fast training set generation. Our computational efficiency will be key for future cosmological survey science with Roman, Rubin and Euclid, which will catalog $\sim10^9$ galaxies and which is motivating ever larger $\sim$Gpc-scale simulations \citep[e.g.,][]{heitmann21,ishiyama21}. For context, the flagship, highest resolution MillenniumTNG simulation provides $\sim10^8$ trees \citep{barrera23}. Despite the speed-up from JAX with \texttt{jit}, it will still be desirable to train emulators, but now with the added benefit of also having physically interpretable gradients from \texttt{sapphire}.

Beyond simply evolving different numbers of halos, benchmarking the full end-to-end inference pipeline runtime depends on a variety of factors such as convergence criteria. For our fiducial choice of evolving $\sim10^3$ halos for a single likelihood calculation in an eight-dimensional parameter space, our NUTS chains converge to the same posterior within a few hundred iterations at most. This takes $\sim30$ minutes when the NUTS chains are run in parallel on different 64-core Intel Ice Lake CPU nodes. In contrast, a single \adam\ trajectory hits our conservative convergence criteria (described in Subsection \ref{sec:map}) within a few minutes at most. We expect these gradient-based methods to scale more effectively than gradient-free methods as the number of parameters increases. For example, affine-invariant ensemble samplers require at least twice as many ``walkers'' as free parameters to ensure proper mixing and convergence \citep{aies10,emcee}. This is very expensive because a single walker would still need to evolve $\sim10^3$ halos to compute the loss, meaning a single step would take $\sim10^4$ ODE evaluations assuming $\sim10$ parameters. This is not feasible on a CPU node so we tested a GPU-native implementation of this method \citep[using \texttt{numpyro};][]{numpyro} and found that it took two hours (four times longer) to converge to the same NUTS posterior. All scripts and trace plots to validate this analysis are available on the public GitHub repository.

\subsection{Automatic differentiation}\label{sec:autodiff}
Perhaps most importantly, \texttt{JAX} gives us the ability to rapidly compute exact derivatives of our numerical ODE solutions with respect to input parameters, including through the internals of the adaptive ODE solver and some of its settings, via automatic differentiation (auto-diff). Briefly, auto-diff transforms a function into its derivative by ``tracing'' the inputs through the computation, accumulating partial derivatives with the chain rule \citep[see the review by][for more details]{baydin18}. The tracing can be run in ``forward-mode'' (perturbing the inputs) or ``reverse-mode'' (perturbing the outputs). There is no analytic solution for our ODE system so these derivatives are not computable symbolically. However, our code itself is still composed of a sequence of elementary mathematical operations, and auto-diff libraries like \texttt{JAX} know how to compute their partial derivatives. Alternatively, finite differentiation (finite-diff) can be applied to highly non-linear ODEs like ours but it will be both numerically error-prone and more expensive given that each parameter needs to be perturbed separately once or twice. In contrast, auto-diff is exact to machine precision and typically requires fewer function evaluations, of order the number of inputs (outputs) for forward-mode (reverse-mode) autodiff. However, care must be taken for control flow (if/else statements, hard-binning, nondifferentiable operations like absolute values). We will show below that comparing auto-diff to finite-diff is a useful consistency check. For both methods, accurately preserving gradient flow requires tightly controlling local ODE errors since they will otherwise accumulate. But the $a_{\rm tol}/r_{\rm tol}$ tolerance requirements are less stringent for auto-diff than finite-diff, further increasing the attractiveness of auto-diff. 

Figure \ref{fig:runtime} shows that the runtime of auto-diff is at most only a few times higher than that of solving the ODE system itself. The combined runtime of a batched ODE solve and gradient calculation determines the regime where inference is possible without emulators. For gradient descent and especially HMC, we generally want $\lesssim1$ sec total runtime per iteration and therefore prefer minibatches of size $\lesssim1000$ halos where CPUs win over GPUs. Note that calculating our likelihood function (i.e., loss; Subsection~\ref{sec:loss}) requires solving the full ODE system for a batch of halos, but the gradient may only be required for one or a few state variables (e.g., if SMHM is the only constraint, only $\partial \log M_*/\partial\vec{\theta}$ is needed). We find that the gradient of the loss alone is $\sim2-10\times$ faster than the full Jacobian depending on which constraints we use. In this paper, when trying to interpret the dynamics and sensitivities of our model, we calculate full Jacobians, whereas for inference purposes, we use only the gradient of the simpler scalar loss function. In Subsection~\ref{sec:fisher}, we will compute second-order gradients (Hessian matrix), which take at least $\sim10\times$ longer than Jacobians and scale with the square of the number of free parameters and/or state variables. Hessians are both computationally and memory-wise prohibitive to compute for large batches so we typically only calculate them once at the end of an inference loop for Fisher/Laplace forecasts.

In Section~\ref{sec:mocks}, we show that our Jacobian and Hessian matrices have interpretable, non-random structure that encodes the astrophysics of our dynamical model. This was surprising because we did not know a priori whether numerical errors would accumulate while solving our highly non-linear ODEs, resulting in statistically noisy matrices. To convince the reader (and ourselves) of the robustness of the trends we will show, we now compare the Jacobians from auto-diff to those from finite diff. The latter is generally more expensive because we require $2\times N_{\rm free}$ ODE system evaluations to fill out a single column of the Jacobian, and less reliable because finite differencing is prone to numerical errors that depend on an arbitrary parameter step size $\epsilon$. Nevertheless, auto-diff and finite-diff gradients should agree, though the latter may require much tighter ODE solver tolerances to reach the same accuracy. To test this, we generated 1000 parameter sets over a uniformly sampled latin hypercube (as described in Subsection~\ref{sec:lhs}). For a single random MW halo, we computed its auto-diff Jacobian for four different values of ODE solver tolerances $a_{\rm tol}=r_{\rm tol}=(10^{-6},10^{-8},10^{-10},10^{-12})$. We also computed finite-diff Jacobians using six different step sizes $\epsilon=(10^{-1},10^{-2},10^{-3},10^{-4},10^{-5},10^{-6})$ for each of the same four ODE tolerances. We use central finite differences:
\begin{equation}
\mathcal{J}(X_k,\theta_j) = \frac{f(\vec{\theta}_j+\epsilon \vec{e}_j)_k^{z=0} - f(\vec{\theta}_j-\epsilon \vec{e}_j)_k^{z=0}}{2\epsilon}
\end{equation}
where $X_k$ is the $k$th state variable, $\theta_j$ is the $j$th astrophysical parameter, $f$ represents our system of $k$ nonlinearly coupled ODEs, $f(\dots)_k^{z=0}$ denotes that we only take the $k$th state variable at $z=0$, and $\vec{e}_j$ is a ``one-hot'' vector (all zeros except for index $j$).

Next we define and compute the ``relative symmetric error'' between any two Jacobians as: 
\begin{equation}\label{eqn:relsymerr}
\Delta(\mathcal{J}_1,\mathcal{J}_2) \equiv \left\|\frac{\mathcal{J}_1 - \mathcal{J}_2}{0.5(\vert\mathcal{J}_1\vert+\vert \mathcal{J}_2\vert)}\right\| \;.
\end{equation}
The numerator is the residual matrix between any two Jacobians. The denominator is the average value of each Jacobian cell given that we do not consider either Jacobian as the ground truth. The closer these normalized errors are to one (or higher), the closer (or worse) the discrepancies are to the average magnitude of the gradient itself and thus more unreliable. As a simple summary statistic, we take the norm of this normalized residual matrix. 

\begin{figure}
\centering
\includegraphics[width=0.7\hsize]{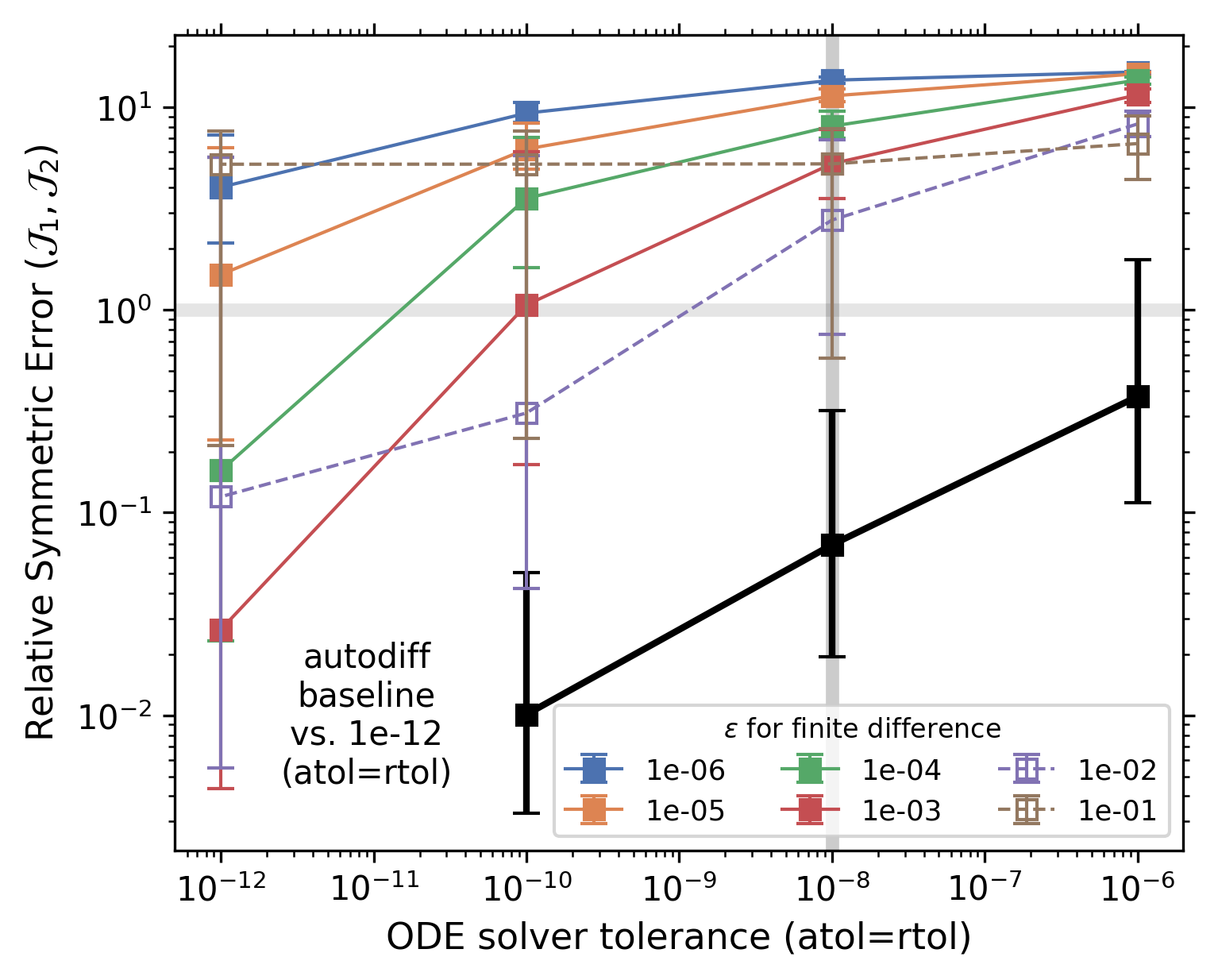}
\caption{Dependence of the relative symmetric error between autodiff and finite-difference Jacobians on ODE solver tolerances and the $\epsilon$ used for finite differencing. This is for a single random representative MW halo averaged over 2000 latin hypercube parameters with errorbars denoting 16-50-84 percentiles of the distribution of Jacobian residual norms. The black curve is a baseline from comparing autodiff Jacobians with various ODE tolerances to the autodiff Jacobian with our tightest and thus most accurate tolerances ($a_{\rm tol}=r_{\rm tol}=10^{-12}$). As expected, the autodiff Jacobian error decreases as we lower the ODE solver tolerances. The colored lines show that residuals between the finite-difference and autodiff Jacobians also decrease as we lower ODE solver tolerance (for a fixed $\epsilon$). Naturally, the finite difference residuals also decrease as we increase $\epsilon$ due to avoiding numerical round-off error except when $\epsilon\gtrsim10^{-2}$ where the step size becomes large enough to cause formula (extrapolation) errors. The horizontal gray band roughly marks where the residuals become of order the gradient magnitude itself. The vertical gray band denotes our fiducial choice of ODE solver tolerances to balance autodiff accuracy and speed.}
\label{fig:jacerrs}
\end{figure}

Figure \ref{fig:jacerrs} summarizes how $\Delta(\mathcal{J}_1,\mathcal{J}_2)$ depends on $a_{\rm tol}=r_{\rm tol}$ and $\epsilon$ for the single MW halo evaluated across parameter space. As a baseline, the black line compares each autodiff Jacobian to the ``true'' autodiff Jacobian with the tightest $a_{\rm tol}=r_{\rm tol}=10^{-12}$. Since autodiff Jacobians are not subject to finite-diff noise from an arbitrary $\epsilon$, this is a self-consistency check of the dependence of the numerical ODE solution on solver error tolerances alone. As expected, the autodiff residuals decrease as we decrease ODE solver tolerances. For our fiducial choice of $a_{\rm tol}=r_{\rm tol}=10^{-8}$ (chosen to balance accuracy and speed), the average autodiff Jacobian varies by $\lesssim10^{-2}$ compared to the lowest tolerance (most accurate) auto-diff Jacobian. The finite-difference residuals relative to auto-diff at a fixed ODE solver tolerance also decrease as we lower the tolerances. These finite-difference residuals also decrease as $\epsilon$ is increased which avoids numerical round-off error, but show a turn-around with larger residuals when $\epsilon\gtrsim10^{-2}$, which is likely too large of a step size and leads to non-linear extrapolation errors.

\begin{figure*}
\centering
\includegraphics[width=\hsize]{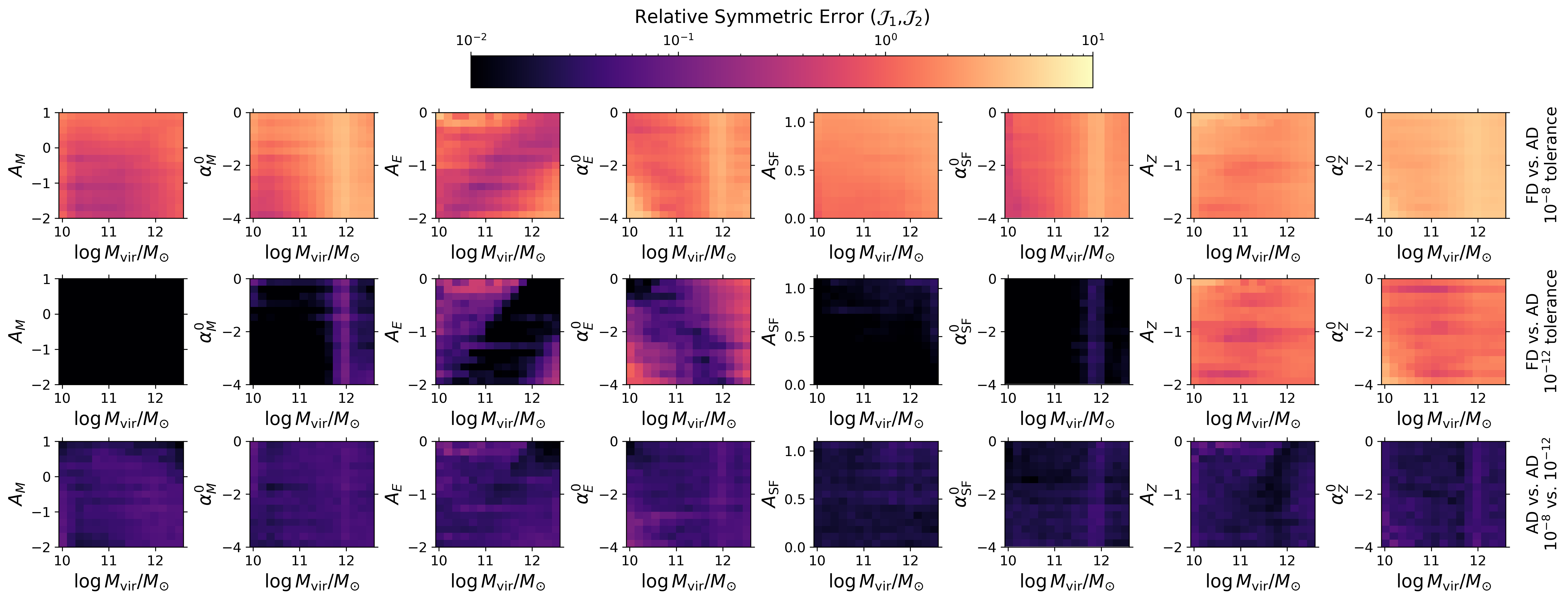}
\caption{Similar to Figure \ref{fig:jacerrs} but now across our latin hypercube parameter space as a function of halo mass. Brighter colors correspond to worse discrepancies. The top row shows the average relative symmetric error between finite-diff and auto-diff Jacobians evaluated with $a_{\rm tol}=r_{\rm tol}=10^{-8}$ and $\epsilon=10^{-4}$. These errors are typically of order the average gradient magnitude itself. The middle row is the same except we dropped the tolerances to $10^{-12}$ which improves agreement. The $A_Z$ and $\alpha_Z$ columns show large errors but only because auto-diff gives zero gradient for many halos whereas finite-difference is near but not exactly zero, causing the relative symmetric error to be large. The bottom row is a self-consistency check comparing the auto-diff Jacobian with $10^{-8}$ tolerances against $10^{-12}$ tolerances, revealing tight agreement. This demonstrates consistency with Figure \ref{fig:jacerrs} across parameter space for a range of halo masses (and formation histories at fixed mass).}
\label{fig:jacerrs2}
\end{figure*}

Figure \ref{fig:jacerrs2} expands the above to simultaneously consider variations across halo mass and parameter space. This is very expensive even with multiple GPUs, so we restrict ourselves to three illustrative cases. First, the average relative symmetric error between finite-diff and auto-diff Jacobians with $a_{\rm tol}=r_{\rm tol}=10^{-8}$ and $\epsilon=10^{-4}$ is of order the average Jacobian norm itself. This is mainly driven by numerical errors in the finite-diff Jacobian since dropping the ODE tolerances to $10^{-12}$ leads to much better agreement with the auto-diff Jacobians across parameter space for a range of halo masses (and formation histories at fixed halo mass). The autodiff Jacobians themselves are self-consistent as evidenced by the low errors between the $10^{-8}$ and $10^{-12}$ tolerance versions.

Lastly, we compute Hessians $\mathcal{H}$ in this paper using auto-diff for Fisher/Laplace analysis. It is useful to benchmark against finite-diff Hessians just as we did for Jacobians earlier. For finite-diff Hessians, we use the standard central finite-difference scheme \citep[e.g.,][]{nocedal06,press07} involving the log-likelihood so that the diagonals of the Hessian are given by 
\begin{equation}
\mathcal{H}_{ii} = \frac{\mathcal{L}(\vec{\theta}+\epsilon \vec{e}_i) - 2\mathcal{L}(\vec{\theta}) + \mathcal{L}(\vec{\theta}-\epsilon \vec{e}_i)}{\epsilon^2}
\end{equation}
and the off-diagonals are given by
\begin{equation}
\mathcal{H}_{ij} =
\frac{
\begin{array}{l}
\mathcal{L}(\vec{\theta}+\epsilon \vec{e}_i + \epsilon \vec{e}_j) - \mathcal{L}(\vec{\theta}+\epsilon \vec{e}_i -\epsilon \vec{e}_j) \\
-\mathcal{L}(\vec{\theta}-\epsilon \vec{e}_i + \epsilon \vec{e}_j) + \mathcal{L}(\vec{\theta}-\epsilon \vec{e}_i - \epsilon \vec{e}_j)
\end{array}
}{4\epsilon^2}
\end{equation}
where $\vec{e}_i$ and $\vec{e}_j$ are ``one-hot'' vectors (all zeros except for index $i$ and $j$, respectively). The Hessian is extremely expensive to evaluate with finite-diff since it requires $1+N+N(N-1)/2$ evaluations, which partly explains why this has not been done before for SAMs. For simplicity, we fix the finite-diff stepsize to $\epsilon=10^{-4}$ and ODE solver tolerances $a_{\rm tol}=r_{\rm tol}$ to either $10^{-8}$ (fiducial) or $10^{-12}$ (reference). Finite-diff Hessians can be numerically unstable but in practice we find that most of our Hessians are well behaved, likely because we are considering ensemble sensitivity which is more robust than single-halo sensitivity and thus do not require regularization (addition of a very small diagonal offset). For this exercise, we generate mocks at ten random points in parameter space, evaluate auto-diff and finite-diff Hessians of the loss at the true mock parameter values, and adapt Equation \ref{eqn:relsymerr} to compute the relative symmetric error between pairs of Hessians, taking the median error matrix across the ten mock parameter realizations.

\begin{figure}
\centering
\includegraphics[width=\hsize]{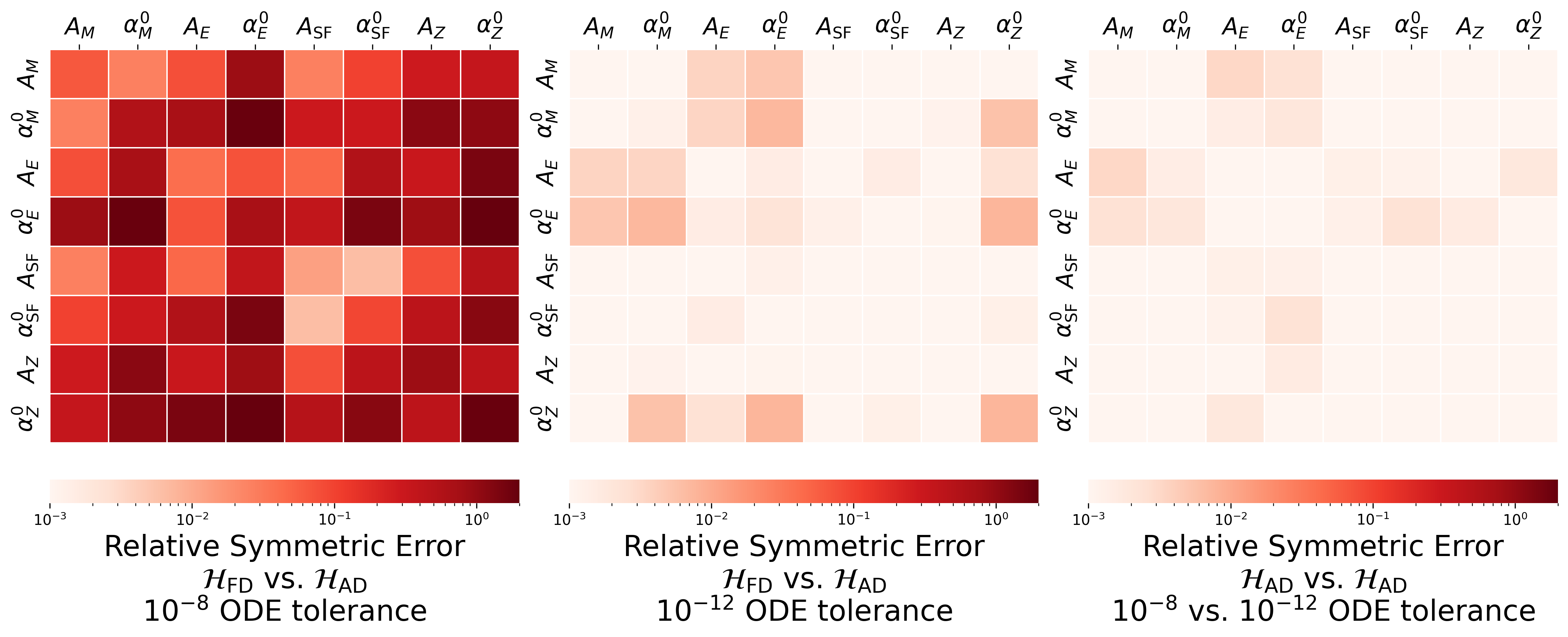}
\caption{Median relative symmetric error matrices between pairs of auto-diff and finite-diff Hessians at ten random points in parameter space. \textit{Left:} finite-diff and auto-diff Hessians show significant differences at our fiducial ODE tolerance of $r_{\rm tol}=a_{\rm tol}=10^{-8}$. \textit{Middle:} finite-diff and auto-diff Hessians agree much better when using tighter, more expensive $10^{-12}$ ODE tolerances. \textit{Right:} the $10^{-8}$ auto-diff Hessian agrees with the $10^{-12}$ auto-diff Hessian, demonstrating the self-consistency of the ODE solution and second-order gradients. This validates our usage of auto-diff Hessians which remain fast and accurate at $10^{-8}$ ODE tolerances.}
\label{fig:fishermethod}
\end{figure}

Figure \ref{fig:fishermethod} shows the median relative symmetric error matrix from comparing finite-diff and auto-diff Hessians at a fixed ODE tolerance. At $10^{-8}$ ODE tolerance, finite-diff disagrees significantly with auto-diff, with the discrepancies often of order the value of entries in either Hessian. The two converge at the tighter $10^{-12}$ ODE tolerance, with discrepancies dropping to $\lesssim10^{-2}$ of Hessian entry values but finite-diff Hessians become very expensive. Importantly, the $10^{-8}$ auto-diff Hessian agrees with the $10^{-12}$ auto-diff Hessian with relative symmetric errors $\lesssim10^{-2}$, demonstrating that auto-diff Hessians remain accurate and cheaper to compute at lower ODE tolerances. Note that small errors in Hessians will be compounded when performing matrix inversion to compute Fisher matrices, further increasing the appeal of auto-diff due to its self-consistency. This validates our use of auto-diff Hessian matrices which are $\sim5-10\times$ faster than finite-diff Hessians to compute while also being more numerically stable without any arbitrary dependence on $\epsilon$. In practice, when fitting actual observational data, we run full HMC to globally map the posterior, so these Hessian-related issues are not a concern for this paper.

The numerical exercises of this subsection increase our confidence in the auto-diff gradients used throughout this paper. Surprisingly, we are unable to find any other study in the $\gtrsim50$ year history of SAMs that attempted this kind of sensitivity analysis using finite-differences, likely because it is very expensive and requires multi-GPU parallelization. Nevertheless, even in the auto-diff era, we have found it tremendously useful to compute finite-diff gradients since they can help reveal where and how auto-diff gradients may not be flowing properly. The unprecedented speed and accuracy of our auto-diff Jacobians and Hessians will be very useful for unlocking previously inaccessible techniques for both global and local sensitivity analysis, parameter optimization and inference \citep[and for studying the equilibrium behavior of galaxies as prototyped by Figure 14 of][]{pandya23}. Indeed, the mock parameter recovery tests we present in Section~\ref{sec:mocks} are the ultimate arbiter of the correctness and practical utility of our gradients.

\section{Dependence of mock tests on observational errors}\label{sec:adamerrs}

\begin{figure*}
\centering
\includegraphics[width=\hsize]{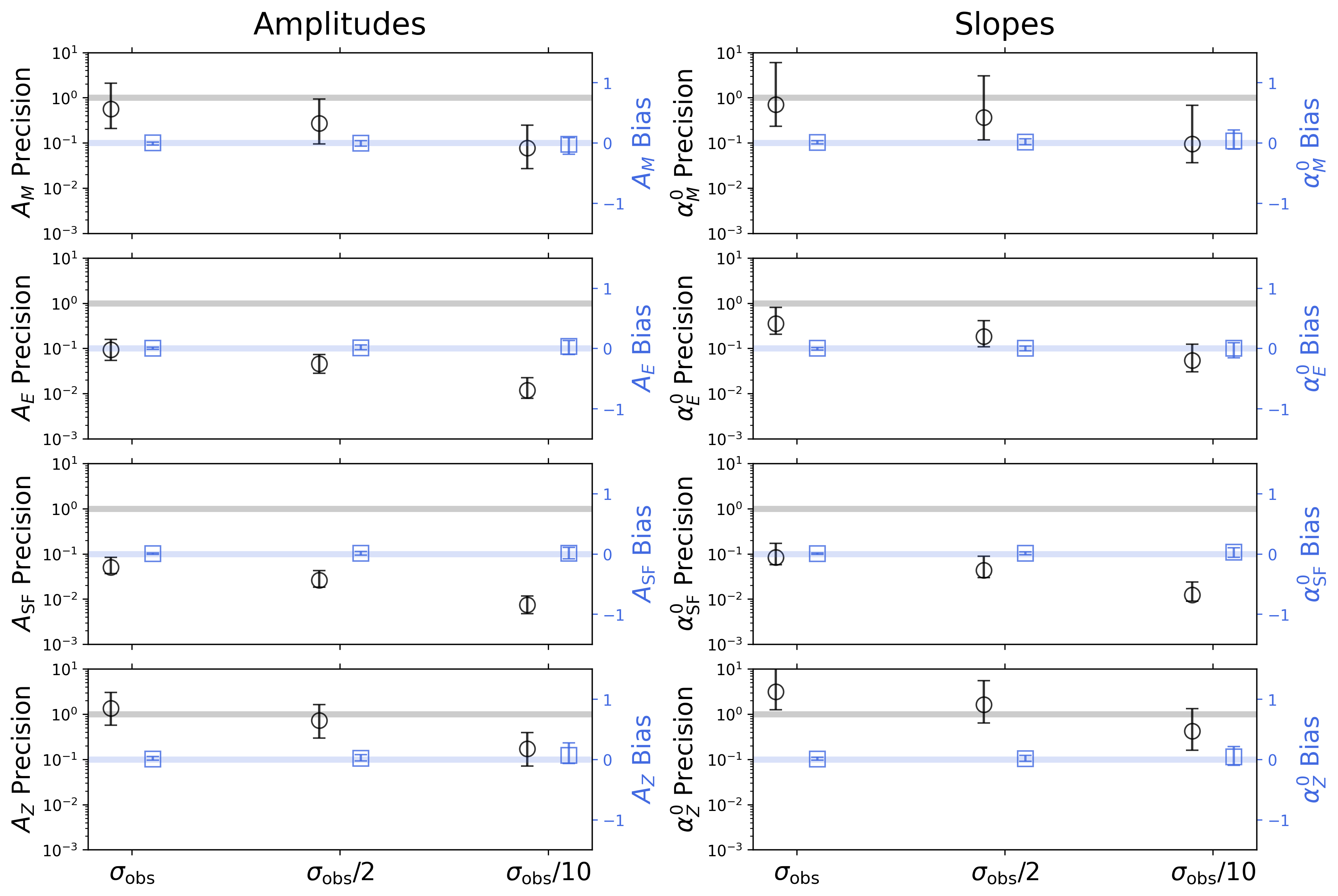}
\caption{Dependence of mock precision (black) and bias (blue) on assumed observational uncertainty. The left (right) column shows amplitude (slope) parameters. The fiducial observational uncertainty reflects our data as described in Section~\ref{sec:data} and Appendix \ref{sec:manga}. Here we always fit to all three $z=0$ scaling relations and the errorbars reflect the 16-84 percentile spread across 100 random mocks. The precision improves as the observational error decreases, following a roughly linear behavior. The bias, defined as in Figure \ref{fig:summaryfisher}, also remains close to zero. The larger spread in bias for smaller errors reflects the lack of tuning of \adam\ hyper-parameters to account for very steep loss landscapes, but that is unimportant for this paper.}
\label{fig:adamerrs}
\end{figure*}

In Subsections \ref{sec:adam} and \ref{sec:summarystats}, we assumed the best case scenario of no measurement uncertainties and systematics for mock tests of MAP parameter optimization with \texttt{adam}. Here we incorporate observational uncertainties reflective of our data as described in Section~\ref{sec:data} and Appendix \ref{sec:manga}. For simplicity, to each standard error from our Gaussian kernel regression, we add in quadrature a constant $0.1$ dex for the SMHM relation, $0.2$ dex for ISM gas fractions and $0.3$ dex for the ISM MZR. We then re-run our parameter recovery tests with \adam\ for the same 100 random mocks as before. We repeat this two more times using half and tenth the errors given above. For this subsection, we restrict ourselves to only fitting all three $z=0$ scaling relations together since the previous tests already revealed that using only one or two is insufficient for \adam.

Figure \ref{fig:adamerrs} summarizes the dependence of mock precision and bias on assumed error for each parameter. Importantly, the bias remains close to zero indicating that adding in these uncertainties does not cause \adam\ to systematically end too far from the true minimum. The spread in bias does get larger for smaller errors, but just like in Figure \ref{fig:summaryfisher}, this reflects the fact that we have not tuned the \adam\ hyper-parameters to take into account the changing steepness of the loss landscape. Naturally, the precision on every parameter improves as the observational error decreases. The behavior appears close to linear with a factor of two (ten) drop in observational error leading to a factor of two (ten) decrease in parameter uncertainty. This exercise suggests that we can achieve percent-level precision on many parameters with sufficiently small observational uncertainties, and is consistent with our finding applied to observational data in Subsection \ref{sec:forecasts}.

\end{document}